\documentclass[twocolumn, twocolappendix]{aastex631}

\usepackage{amsmath}
\usepackage{amsfonts}
\usepackage{color}
\usepackage{soul}
\usepackage{textcomp}
\usepackage{gensymb}
\usepackage{bm}
\usepackage{lmodern}
\renewcommand*{\thefootnote}{\alph{footnote}}

\newcommand{\um}{$\mu$m}
\newcommand{\kms}{km\,s$^{-1}$}
\newcommand{\msol}{M$_\odot$}
\newcommand{\msun}{M$_\odot$}
\newcommand{\lsun}{L$_\odot$}
\newcommand{\sfr}{M$_\odot$\,yr$^{-1}$}
\newcommand{\lfir}{$L_{\rm FIR}$}

\newcommand{\twidth}{.47\textwidth}
\newcommand{\width}{.31\textwidth}
\newcommand{\mycomment}[1]{}

\begin{document}

\title{A 100\,Mpc$^2$ structure traced by hyperluminous galaxies around a massive $z$\,=\,2.85 protocluster}

\correspondingauthor{George Wang}
\email{georgecpwang@phas.ubc.ca}

\author[0009-0003-4626-9777]{George C.P. Wang}
\affiliation{Department of Physics and Astronomy, University of British Columbia, 6225 Agricultural Road, Vancouver, V6T 1Z1, Canada}

\author[0000-0002-8487-3153]{Scott C. Chapman}
\affiliation{Department of Physics and Atmospheric Science, Dalhousie University, 6310 Coburg Road, B3H 4R2, Halifax, Canada}
\affiliation{National Research Council, Herzberg Astronomy and Astrophysics, 5071 West Saanich Road, Victoria, V9E 2E7, Canada}
\affiliation{Department of Physics and Astronomy, University of British Columbia, 6225 Agricultural Road, Vancouver, V6T 1Z1, Canada}
\affiliation{Eureka Scientific Inc, Oakland, CA 94602, USA}

\author[0000-0002-3187-1648]{Nikolaus Sulzenauer}
\affiliation{Max-Planck-Institut f\"{u}r Radioastronomie, Auf dem H\"{u}gel 69, Bonn, D-53121, Germany}

\author[0000-0002-1707-1775]{Frank Bertoldi}
\affiliation{Argelander-Institut f\"{u}r Astronomie, Auf dem H\"{u}gel 71, D-53121 Bonn, Germany}

\author[0000-0003-4073-3236]{Christopher C. Hayward}
\affiliation{Center for Computational Astrophysics, Flatiron Institute, 162 Fifth Avenue, New York, NY, 10010, USA}

\author[0009-0008-8718-0644]{Ryley Hill}
\affiliation{Department of Physics and Astronomy, University of British Columbia, 6225 Agricultural Road, Vancouver, V6T 1Z1, Canada}

\author[0000-0003-3214-9128]{Satoshi Kikuta}
\affiliation{National Astronomical Observatory of Japan, 2-21-1, Osawa, Mitaka, Tokyo 181-8588, Japan}

\author[0000-0003-1747-2891]{Yuichi Matsuda}
\affiliation{Graduate University for Advanced Studies (SOKENDAI), 2-21-1 Osawa, Mitaka, Tokyo 181-8588, Japan}
\affiliation{National Astronomical Observatory of Japan, 2-21-1, Osawa, Mitaka, Tokyo 181-8588, Japan}

\author[0000-0002-1619-8555]{Douglas Rennehan}
\affiliation{Center for Computational Astrophysics, Flatiron Institute, 162 Fifth Avenue, New York, NY, 10010, USA}

\author[0000-0002-6878-9840]{Douglas Scott}
\affiliation{Department of Physics and Astronomy, University of British Columbia, 6225 Agricultural Road, Vancouver, V6T 1Z1, Canada}

\author[0000-0003-3037-257X]{Ian Smail}
\affiliation{Centre for Extragalactic Astronomy, Department of Physics, Durham University, South Road, Durham DH1 3LE, UK}

\author[0000-0002-4834-7260]{Charles C. Steidel}
\affiliation{Cahill Center for Astronomy and Astrophysics, California Institute of Technology, MC249-17, Pasadena, CA 91125, USA}

\begin{abstract}
We present wide-field mapping at 850\,\um\ and 450\,\um\ of the $z$\,=\,2.85 protocluster in the HS\,1549$+$19 field using the Submillimetre Common User Bolometer Array 2 (SCUBA-2).  Spectroscopic follow-up of 18 bright sources selected at 850\,\um, using the Nothern Extended Millimeter Array (NOEMA) and Atacama Large Millimeter Array (ALMA), confirms the majority lies near $z$\,$\sim$\,2.85 and are likely members of the structure. Interpreting the spectroscopic redshifts as distance measurements, we find that the SMGs span 90\,Mpc$^2$ in the plane of the sky and demarcate a 4100\,Mpc$^3$ ``pancake''-shaped structure in three dimensions. We find that the high star-formation rates (SFRs) of these SMGs result in a total SFR of 20,000\,\sfr\ only from the brightest galaxies in the protocluster. These rapidly star-forming SMGs can be interpreted as massive galaxies growing rapidly at large cluster-centric distances before collapsing into a virialized structure. We find that the SMGs trace the Lyman-$\alpha$ surface density profile. Comparison with simulations suggests that HS\,1549$+$19 could be building a structure comparable to the most massive clusters in the present-day Universe.
\end{abstract}


\section{Introduction} \label{sec: intro}

Massive galaxy clusters, collections of gravitationally bound galaxies, are now found as early as 3 billion years after the Big Bang and contain stars that formed at even earlier epochs \citep{Stanford2012, Wang2016, Mantz2018}. The high-redshift progenitors of these galaxy clusters, termed ``protoclusters'', represent the highest dark matter overdensities at their epoch. While their observational signatures are less well-defined than the hot intra-cluster medium (ICM) of virialized clusters, protoclusters contain extremely massive galaxies that can be observed as luminous starbursts \citep[with star-formation rates $>$\,100\,\sfr;][]{Casey2015}. 

Galaxy evolution is known to be accelerated in regions with high overdensities -- many studies have demonstrated enhanced star-formation rates and a reversal of the star-formation-density relation (star-forming galaxies prefer low-density environments) at $z$\,$>$\,1.5 in galaxy clusters \citep{Elbaz2007, Tran2008, Elbaz2011, Cooke2019, Smail2024}. Exceptionally high levels of star formation have been found in the centres of clusters, reaching SFR surface densities (SFRDs) of more than 2000\,\sfr\,Mpc$^{-2}$ at $z$\,=\,2--3 \citep{Umehata2015}, and even higher at redshifts greater than four \citep{Oteo2018, Miller2018}. The reversal of the $z$\,$<$\,1.5 trend in SFRD suggests that massive cluster galaxies may form the bulk of their stars in protoclusters around redshift 2, prior to their virialization \citep{Smail2024}.

Performing surveys of the gas and dust in $z$\,$>$\,2 protoclusters is of emerging importance \citep{Casey2016}. Bright SMGs \citep[typically $S_{850}$\,$>$\,5\,mJy;][]{Chapman2005} are believed to be galaxies seen at a moment when they are rapidly building up their stellar mass \citep[e.g.,][]{smail2004, dudzeviciute2020}.



In the HS\,1549$+$19 field, a protocluster has been identified with significant galaxy overdensity traced by Lyman-$\alpha$ emitting galaxies (LAEs) and UV continuum-selected galaxies, revealing a filamentary structure in the early Universe at a redshift of $z$\,$\approx$\,2.85 when the Universe was 2.3\,Gyr old) \citep{Steidel2011, Lacaille2019, Kikuta2019}. 

The HS\,1549$+$19 (henceforth HS\,1549) survey field at redshift 2.85 contains the strongest galaxy overdensity found in the Keck Baryonic Structure Survey \citep{Rudie2012, Mostardi2013}, even exceeding well-studied protoclusters such as SSA22 at $z$\,$=$\,3.09 \citep{Steidel2000} or the Spiderweb protocluster at $z$\,$=$\,2.16 \citep{Dannerbauer2014}. Using Lyman--$\alpha$ (Ly$\alpha$) emitting galaxies imaged with Subaru's Hyper-suprime-cam (HSC), the extended environment of this protocluster has been mapped with unprecedented detail over a degree-diameter field, corresponding to projected distances of 30\,pMpc \footnote{cMpc or Mpc are comoving megaparsecs, the distance between objects factoring out the expansion of the Universe.}\,\footnote{pMpc are proper megaparsecs, the physical distance between objects, or cMpc/(1\,+\,z).} \citep{Kikuta2019}. This map covers the full density distribution of all the subregions that are likely to collapse and become a Coma-type rich galaxy cluster ($>$\,10$^{15}$\,\msol) at redshift zero \citep{Chiang2014}. 

In this paper, we report on 18 SMGs spectroscopically confirmed at $z$\,$\approx$\,2.85 with NOEMA band-1 and band-3 observations from initial SCUBA-2 850-\um\ observations. In Sec.~\ref{sec: obs} we describe the photometry and spectroscopy that facilitated the detection of galaxies in the HS\,1549 protocluster. In Sec.~\ref{sec: analysis}, we describe the spectral lines fitting, the SMGs membership and the derived properties from these lines. In Sec.~\ref{sec: results} we present our findings on the distribution of the SMGs across the protocluster and the resulting structural shape. Sec.~\ref{sec: discussion} discusses the star-formation rates in comparison with the LAEs and to other protocluster-SMG surveys and how HS\,1549 compares with dark matter simulations. We summarize and conclude the paper in Sec.~\ref{sec: conclusion}.

The paper assumes a standard $\Lambda$CDM model with cosmological parameters taken from \citet{Planck2018}. In cases where we need to model the spectral energy distribution (SED), we apply a modified blackbody function with a dust temperature of 40\,K and a median dust emissivity index of 2 \citep{Greve2012, swinbank2014, Cunha2015}.

\section{Observations} \label{sec: obs}

\begin{figure*}
 \includegraphics[width=\textwidth]{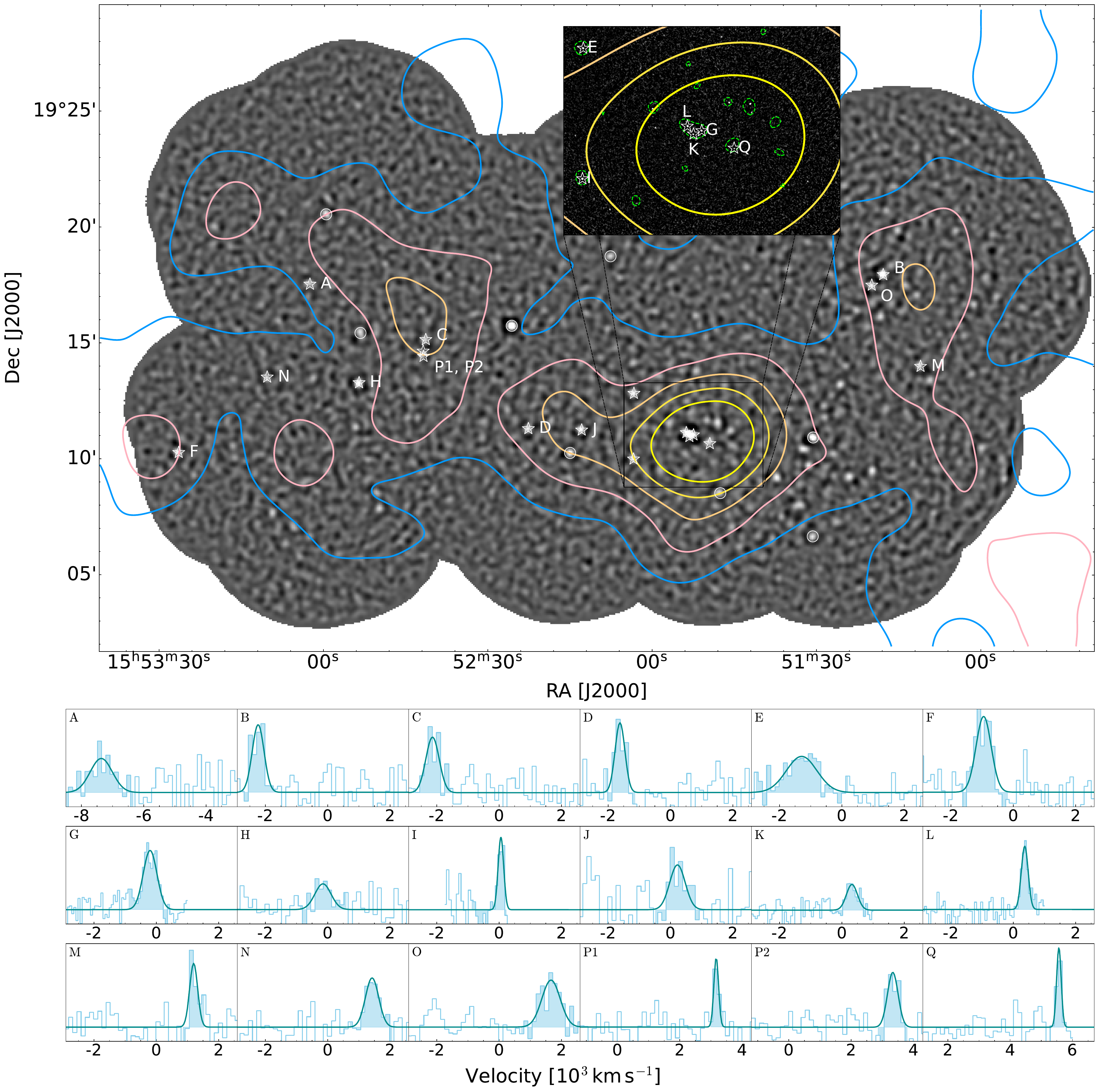}
 \caption{Overview of the protocluster HS\,1549 and individual galaxy members. \textit{Top left}: The 850-\um\ SCUBA-2 map with LAE surface density contours 
 \cite{Kikuta2019} overlaid (levels: 2300, 3400, 4400, 5500, and 6600 LAEs per pMpc$^2$). White circles are all bright submm sources observed in our survey without an observed spectral line. Sources spectroscopically confirmed to be protocluster members are plotted with a white star and have alphabetical labels. \textit{Top right}: The central region of HS\,1549 showing deep ALMA Band-6 (240\,GHz) imaging (PI: Kikuta). The same LAE contours \cite{Kikuta2019} are overlaid on the ALMA image, while SCUBA-2 contours from Ref.~\cite{Lacaille2019} are shown in green (10\,$\sigma$). Galaxy ``G'', the QSO, is in this central region. \textit{Bottom}: The CO (3--2 in most cases) transitions of 18 SMGs in the HS\,1549 protocluster. The galaxies are organized and labelled in ascending redshift order, starting with ``A''. The continuum levels of all galaxies are scaled to the same height; see Sec.~\ref{sec: spectral appendix} for the full detailed spectra.\label{fig: overview}}
\end{figure*}

\subsection{JCMT/SCUBA-2 observations} \label{sec: scuba2}
The James Clerk Maxwell Telescope's (JCMT) SCUBA-2 \citep{Holland2013} instrument was used to map the HS\,1549 field, encompassing the structure around the bright central quasi-stellar object (QSO). The project aimed to cover all regions of the highest overdensity in LAEs \citep{Kikuta2019}. The map covers a total area of 0.24\,deg$^2$ at 850\,\um\ (Fig.~\ref{fig: overview} and 0.18\,deg$^2$ at 450\,\um. SCUBA-2 \ previously observed the central core region \citep{Lacaille2019}, and these data were combined with our new map. There are a few outlying regions with comparable LAE overdensities that we have yet to map with SCUBA-2.

The SCUBA-2 observations were conducted in grade 2 weather conditions ($\tau_{\rm 225\,GHz}$ $<$\,0.08) over six nights between 23 May and 15 Oct.\ 2015, totalling 15 hr of on-sky integration in individual 30-min scans. Standard 3-arcmin diameter ``daisy'' mapping patterns were used, which kept the pointing centre on one of the four SCUBA-2 sub-arrays at all times during the exposure. The full map was constructed using a mosaic pattern of these daisy pointings. 

Data reduction followed standard recipes \citep{geach2017}, with individual 30-min scans reduced using the dynamic iterative map-maker of the SMURF package \citep{Jenness2009, Chapin2013}, flat-fielding, and then solving for model components assumed to make up the bolometer signals (atmospheric, astronomical, and noise terms). The signal from each bolometer’s time stream was then re-gridded onto a map, with the contribution to a given pixel weighted according to its time-domain variance. Since we are interested in extragalactic point sources, we applied a beam-matched filter to improve point source detectability, resulting in a map that is convolved with an estimate of the 850-$\mu$m beam. 

The 850-$\mu$m map has an average depth of 1.0$_{-0.2}^{+0.9}$\,mJy\,beam$^{-1}$ root-mean-square (RMS) outside the deep core region \citep{Lacaille2019} and the 450-\um\ map has a depth of 11.8$_{-6.1}^{+30.2}$\,mJy\,beam$^{-1}$ RMS. The smaller coverage and over 10 times shallower 450-\um\ map make these data less useful for characterizing the SMGs, which leads us to search for submm-bright sources at 850\,\um\, with 450-\um\ flux densities reported only when available.

\subsection{NOEMA observations} \label{sec: noema}

From the SCUBA-2 map, we selected a flux-limited sample of 25 sources with brightnesses greater than 8\,mJy at 850\,\um\ for spectroscopic follow-up with the IRAM NOEMA interferometer to identify SMGs lying at $z$\,$\approx$\,2.85 in the protocluster. The central core region was previously followed up with the Submillimetre Array (SMA) at 850\,\um\ and NOEMA at 3\,mm \citep{Lacaille2019}. These studies resolved two $S_{850}$\,$>$\,8\,mJy sources from the central blended SCUBA-2 source and detected CO(3--2) lines with NOEMA, both lying at $z$\,$\approx$\,2.85, along with two nearby $S_{850}$\,$\approx$\,5\,mJy sources. Our final NOEMA follow-up sample of S$_{850}$\,$>$\,8\,mJy sources in the wider field is thus reduced to 21.

Observations were obtained in two observing programmes using NOEMA/PolyFix in 2020 and 2021, with Project IDs W20DD (for the 3-mm follow-up of CO(3--2)) and W21DH (for the 1.4-mm follow-up of CO(7--6)). 21 targets were observed at 3\,mm, with spectral setups of two 8-GHz sidebands (26000\,\kms\ per sideband), and an integration time of 1 hour per setup using the combined ``CD'' configuration, which is suitable for low-resolution detection experiments. The 1.4-mm programme then targeted 14 sources with CO(3--2) line candidates found from the first programme. 

Reduction of the data was carried out using the standard Grenoble Image and Line Data Analysis Software ({\tt GILDAS})\footnote{\url{http://www.iram.fr/IRAMFR/GILDAS}}. The raw data were calibrated using standard pipelines, with bad visibilities flagged and removed in the process, resulting in calibrated $uv$ tables. The $uv$-visibilities were then imaged using the {\tt GILDAS} MAPPING pipeline. We used the H{\"o}gbom algorithm \citep{Hogbom1974} to clean the data products and deconvolve the telescope's beam from the image. The results are data cubes in the image plane, with frequency bins of 20\,MHz in width. 

The 1.4-mm and 3.3-mm observations have an average depth of 1.20\,mJy per channel and 0.69\,mJy per channel, respectively. The average continuum sensitivity at 1.4\,mm is 43\,$\mu$Jy and 25\,$\mu$Jy at 3.3\,mm. At 3.3\,mm the continuum emission falls to a level similar to that of our sensitivity (we quote the 3.3-mm continuum measurements in Table~\ref{tab: ctn} for completeness). The $z$\,$\approx$\,2.85 redshifted CO(3--2) and CO(7--6) spectral transitions are located in their receivers' upper and lower sidebands, respectively.

\subsection{ALMA observations} \label{sec: alma}

The Atacama Large Millimeter/ Submillimeter Array (ALMA) was used to map the HS\,1549 core region in Band 6 under a Cycle 6 program (2019.0.00236.T; PI: S.\ Kikuta) targeting a $6'\times4'$ region around the central QSO. Observations were obtained on 28 November 2019 in a 40-2 array configuration with baseline lengths of 15--459\,m, giving a naturally weighted synthesized beam size of 1.6$^{\prime \prime}$. There were 40 antennas available, with a total on-source integration time of 10 minutes per pointing. Ceres and J2357--5311 were used as flux and phase calibrators, respectively. The data were re-processed using CASA and the standard ALMA-supplied calibration, using natural beam weighting to maximize point-source sensitivity. The average 233\,GHz continuum sensitivity in the map is 66\,$\mu$Jy.

The frequency setting adopted for these observations covers the redshifted CO(8--7) transition around an observed frequency of 239\,GHz. Three of the bright SCUBA-2 sources found within the ALMA mosaic have strong CO(8--7) detections above 7\,$\sigma$. These include the two central sources described in Lacaille et al., 2019 \citep{Lacaille2019} (``K'' and ``L''), and an additional source in our bright SMG sample (``Q'' in Fig.~\ref{fig: overview}). Sources ``I'' and ``E'' also happen to fall within the field of view.

\begin{table*}
 \begin{tabular}{ccccccccc}
\hline
ID & RA & Dec & $z$ & $S_{450}$ & $S_{850}$ & $S_{1.4}$ & $S_{3.3}$\\
& [J2000] & [J2000] & & [mJy] & [mJy] & [mJy] & [$\mu$Jy] \\
\hline
A\phantom{$^\star$} & 15:53:02.6 & 19:17:33.0 & 2.758 & $--$ & \phantom{0}9.8\,$\pm$\,0.6 & $--$ & 60\,$\pm$\,30 \\
B\phantom{$^\star$} & 15:51:17.8 & 19:17:58.3 & 2.822 & 27.9\,$\pm$\,4.4\phantom{0} & 11.0\,$\pm$\,0.3 & 1.74\,$\pm$\,0.04\phantom{$^\mathrm{A}$} & 120\,$\pm$\,30 \\
C\phantom{$^\star$} & 15:52:41.4 & 19:15:10.0 & 2.823 & 37.5\,$\pm$\,11.8 & \phantom{0}8.1\,$\pm$\,0.4 & 0.75\,$\pm$\,0.04\phantom{$^\mathrm{A}$} & $<$\,80 \\
D\phantom{$^\star$} & 15:52:22.6 & 19:11:19.8 & 2.829 & 23.9\,$\pm$\,10.3 & \phantom{0}9.6\,$\pm$\,0.4 & 1.01\,$\pm$\,0.04\phantom{$^\mathrm{A}$} & 70\,$\pm$\,30 \\
E\phantom{$^\star$} & 15:52:03.3 & 19:12:51.3 & 2.834 & 17.9\,$\pm$\,5.0\phantom{0} & \phantom{0}8.3\,$\pm$\,0.3 & 1.45\,$\pm$\,0.04\phantom{$^\mathrm{A}$} & 60\,$\pm$\,30 \\
F\phantom{$^\star$} & 15:53:26.5 & 19:10:16.8 & 2.838 & $--$ & 17.3\,$\pm$\,0.7 & 0.92\,$\pm$\,0.05\phantom{$^\mathrm{A}$} & $--$ \\
F$^\star$ & 15:53:26.9 & 19:10:20.6 &  &  &  & 2.29\,$\pm$\,0.05\phantom{$^\mathrm{A}$} &  \\
G\phantom{$^\star$} & 15:51:52.4 & 19:11:03.8 & 2.847 & 29.3\,$\pm$\,3.7\phantom{0} & \phantom{0}8.8\,$\pm$\,1.0 & 3.40\,$\pm$\,0.07\phantom{$^\mathrm{A}$} & 110\,$\pm$\,30 \\
H\phantom{$^\star$} & 15:52:53.6 & 19:13:18.2 & 2.848 & 55.0\,$\pm$\,15.3 & 12.4\,$\pm$\,0.4 & 2.78\,$\pm$\,0.05\phantom{$^\mathrm{A}$} & 130\,$\pm$\,30 \\
I\phantom{$^\star$} & 15:52:03.4 & 19:10:01.3 & 2.851 & 24.1\,$\pm$\,5.6\phantom{0} & \phantom{0}9.0\,$\pm$\,0.4 & 1.99\,$\pm$\,0.05\phantom{$^\mathrm{A}$} & $<$\,70 \\
J\phantom{$^\star$} & 15:52:12.9 & 19:11:16.3 & 2.853 & 20.7\,$\pm$\,5.1\phantom{0} & 8.6\,$\pm$\,0.3 & 0.70\,$\pm$\,0.04\phantom{$^\mathrm{A}$} & 70\,$\pm$\,30 \\
J$^\star$ & 15:52:12.8 & 19:11:21.3 &  &  &  & 1.15\,$\pm$\,0.04\phantom{$^\mathrm{A}$} &  \\
K\phantom{$^\star$} & 15:51:53.2 & 19:10:59.5 & 2.854 & 29.3\,$\pm$\,3.7\phantom{0} & \phantom{0}5.6\,$\pm$\,1.1 & 1.34\,$\pm$\,0.07\phantom{$^\mathrm{A}$} & $<$\,70 \\
L\phantom{$^\star$} & 15:51:53.8 & 19:11:09.7 & 2.855 & 16.4\,$\pm$\,3.8\phantom{0} & \phantom{0}9.4\,$\pm$\,1.1 & 3.37\,$\pm$\,0.07\phantom{$^\mathrm{A}$} & 160\,$\pm$\,20 \\
M\phantom{$^\star$} & 15:51:10.1 & 19:14:00.8 & 2.865 & 14.8\,$\pm$\,5.9\phantom{0} & \phantom{0}8.0\,$\pm$\,0.4 & 1.39\,$\pm$\,0.05\phantom{$^\mathrm{A}$} & $<$\,100 \\
N\phantom{$^\star$} & 15:53:10.4 & 19:13:32.4 & 2.868 & $--$ & 11.4\,$\pm$\,0.6 & 2.26\,$\pm$\,0.05\phantom{$^\mathrm{A}$} & 50\,$\pm$\,30 \\
O\phantom{$^\star$} & 15:51:19.9 & 19:17:31.2 & 2.872 & 12.9\,$\pm$\,4.6\phantom{0} & \phantom{0}8.0\,$\pm$\,0.3 & 2.03\,$\pm$\,0.05\phantom{$^\mathrm{A}$} & 70\,$\pm$\,30 \\
P1\phantom{$^\star$} & 15:52:41.8 & 19:14:26.6 & 2.891 & 33.8\,$\pm$\,11.7 & 8.2\,$\pm$\,0.4 & 1.16\,$\pm$\,0.05\phantom{$^\mathrm{A}$} & 50\,$\pm$\,20 \\
P2$^\star$ & 15:52:41.9 & 19:14:40.2 & 2.893 &  &  & 0.45\,$\pm$\,0.05\phantom{$^\mathrm{A}$} & 40\,$\pm$\,30 \\
Q\phantom{$^\star$} & 15:51:49.5 & 19:10:41.1 & 2.923 & \phantom{0}8.6\,$\pm$\,3.8\phantom{0} & \phantom{0}4.9\,$\pm$\,0.2 & 0.57\,$\pm$\,0.09$^\mathrm{A}$ & 50\,$\pm$\,30 \\
\hline
\end{tabular}

 \caption{Observed properties of the SMGs ranked in ascending redshift order. The RA and Dec values are from the 1.4-mm continuum (3\,mm and 1.3\,mm for ``A'' and ``Q'', respectively), and $z$ is from the CO(3--2) transition. We give the total $S_{450}$ and $S_{850}$ flux densities of each SCUBA-2 source, and we quote all continuum flux densities that are resolved by NOEMA. ``A'', ``F'', and ``N'' were not observed in the 450-\um\ map. When the SCUBA-2 beam is resolved into multiple sources in the 1.4-mm continuum, we denote this by a $^\star$ (``F'', ``J'', and ``P2''). The 1.4-mm continuum of ``Q'' is measured from the ALMA Band-6 map (labelled with $^\mathrm{A}$). $S_{3.3}$ was not measured for ``F,'', and 3$\sigma$ upper limits are quoted for SMGs with a continuum smaller than their sensitivity.
 \label{tab: ctn}}
\end{table*}

\section{Analysis} \label{sec: analysis}

\subsection{Measuring spectral lines}

We ran the {\tt Python} package LineSeeker 
\citep{Gonzalez2017} on the CO(3--2) data cubes to find potential spectroscopic lines. The package outputs a list of potential lines, and we filtered this list to remove the outliers that possibly exist in the foreground or background compared to the protocluster. 

We correlated the locations of these potential lines with the corresponding SCUBA-2-identified submm bright source, and all lines outside of 1 SCUBA-2 beam (r\,$\approx$\,8$^{\prime \prime}$) were deemed unrelated to the source. We also removed all the line-of-sight (LOS) outliers. Then, a few sources with a low SNR were removed, resulting in 17 submm galaxies (SMGs) spectroscopically identified in CO(3--2). Source ``P'' has two components, both galaxies lying at a similar redshift. Here, we define multiplicity as resolved high-S/N ($>$\,5) sources in the SCUBA-2 beam detected at 1.3/1.4\,mm. The CO(7--6) program observed most (16) of the sample SMGs at a higher frequency (excluding ``A'' and ``Q''), introducing an additional SMG ``F'' to the sample, and bringing the total number of $z$\,$\approx$\,2.85 SMGs identified up to 18. 

We extracted the spectra at the locations of the peak pixels. The NOEMA beam is large with a full width at half maximum (FWHM) of 3.8$^{\prime\prime}$\,$\times$\,2.7$^{\prime\prime}$ at 3\,mm and 1.5$^{\prime\prime}$\,$\times$\,0.7$^{\prime\prime}$ at 1.4\,mm, leaving most galaxies unresolved or barely resolved, respectively. Therefore, we quote the higher SNR measurement of peak-pixel and aperture photometry. Here, the aperture used is scaled to the FWHM. The observed brightness of each galaxy in the submm is modelled by a combination of its emission from heated dust (modified blackbody radiation) and emission lines from the spectral transitions. Our model for each spectrum is thus
\begin{equation} 
{\rm Line}(\nu) = B(\nu) + {\rm Gauss}_i(\nu, \sigma_\mathrm{FWHM}, \mu_i, A_i).
\end{equation}
Here, $B(\nu)$ is the dust continuum emission (which remains constant for the narrow frequency range of our observation window). We model each emission line with a Gaussian function, where the width of the transition, $\sigma_\mathrm{FWHM}$ is the FWHM of the line, $\mu_i$ is the frequency at the peak of the $i$-th line, and $A_i$ is the maximum brightness of the $i$-th line. We introduce a new Gaussian term to the emission function for each transition within the observation window with one value for the FWHM and redshift across all lines. We have tried modelling each emission line with a double Gaussian model, but we did not find an improvement in the fit. In Table~\ref{tab: derived}, we quote the physical parameters of each galaxy derived from their spectral properties: dynamical mass, $M_{\rm dyn}$; cosmological redshift, $z$; and gas mass, $M_{\rm gas}$. The galaxies are organized and labelled in increasing redshift in the table. The protocluster field at 850\,\um\ and the central core \citep{Lacaille2019} are shown in Fig.~\ref{fig: overview}, with LAE contours \citep{Kikuta2019} overlaid. The bottom of the figure shows the individual spectral lines, in all cases CO(3--2) except for ``I'' where we show the higher SNR CO(8--7) line from ALMA, revealing the shift in velocity across the sample. The redshift distribution is shown in Fig.~\ref{fig: distribution}. Individual continuum maps and spectral lines for each SMG are shown in Sec.~\ref{sec: spectral appendix}.

\subsection{Redshift estimates}
Spectroscopy gives precise measurements of the redshift of each galaxy. By knowing which transition we are observing, we can calculate the redshift of photons using
\begin{equation}
z = \frac{\nu_\mathrm{rest}}{\nu_\mathrm{obs}} - 1.
\end{equation}
Here, $\nu_\mathrm{rest}$ and $\nu_\mathrm{obs}$ are the rest and observed frequencies of the spectral transition, respectively. We can rule out interloper lines in the SCUBA-2 beam by observing different spectral transitions from the same galaxy. We consider galaxies as protocluster members if we have detected both their CO(3--2) and CO(7--6) or CI(2--1) emission lines, otherwise we require the spectral line to have high S/N$>$\,5.5 (above the LineSeeker significance threshold) for galaxies with only a single line detection (namely the ALMA source ``I'' SNR$_{CO8-7}$=10.2, and sources ``A'', SNR$_{CO3-2}$=5.5 and ``F'', SNR$_{CO7-6}$=8.2). The redshifts of all the galaxies are shown in the second column of Table~\ref{tab: derived}.

\subsection{SMGs in HS\,1549}
We used NOEMA to search for the $^{12}$CO(3--2) transition near the median protocluster redshift of 2.85, revealing 16 $z$\,$\approx$\,2.85 line emitters at a high signal-to-noise ratio (SNR $>$\,5) from 15 unresolved SCUBA-2 sources (source P breaks up into two SMGs: ``P1'' and ``P2''). Together with two previously discovered $S_{850}$\,$\sim$\,5\,mJy sources with CO(3--2) redshifts \citep{Lacaille2019, Chapman2023} forms a sample of 18 bright SMGs in the HS\,1549 protocluster. Follow-up observations with NOEMA identified that 14 of the SMGs also detected CO(7--6) or/and CI(2--1) line detection(s). All of the galaxies are individually detected in continuum at $>$\,5\,$\sigma$, either through NOEMA Band-3 maps or the ALMA Band-6 map (also reported as 1.4\,mm, with average continuum RMS of 66\,\um), with the continuum flux density of SMGs ranging from 0.4 to 2.1\,mJy at 1.4\,mm (Table~\ref{tab: ctn}). Excluding the protocluster SMGs, we find the number counts (not in the HS\,1549 structure) is around 30\,deg$^{-2}$, which is comparable to typical number counts of the brightest SCUBA-2 sources in the blank field \citep{Weiss2009, geach2017}.

Most of the SCUBA-2 sources outside the core region are resolved as a single bright object with NOEMA, with only three sources breaking into two galaxies of roughly equal flux density, thus representing a low source multiplicity (3/15 or 20\%) for such bright SMGs. HyLIRG SMGs in the field are relatively rare, although the bright source density varies between different fields. \citet{Hill2018} selected bright field SMGs ($S_{850}\,>\,10\,$mJy) and find a similar source multiplicity (15\%). This is in contrast with the field SMGs from \citet{Stach2018} and \citet{simpson2020}, where 30--53\% of bright 850-\um\ sources break up into multiples at similar depths to our data ($\approx$\,1\,mJy at 850\,\um).

\subsection{Gas mass estimates}
In Table~\ref{tab: derived}, we give the gas mass of each galaxy. The gas mass measures the size of the gas reservoir, which is predominantly H$_2$. We apply the relation
\begin{equation}
M_{\rm gas} = \alpha_{\rm CO(1-0)}\,L^{\prime}_{\rm CO(1-0)}
\end{equation}
to scale the line strength, $L^{\prime}_{\rm CO}$, to the gas mass. Since we measure the higher $J$ transitions ($J$\,=\,3 and $J$\,=\,7), we first scale the observed line strengths to the ground state, $J$\,=\,1, by applying standard $r_{3,1}$ ($0.52$\,$\pm$\,0.09) and $r_{7,1}$ ($0.18$\,$\pm\,$0.04) scaling ratios \citep{Bothwell2013}. We then assume an $\alpha_{\rm CO(1-0)}$ of 1\,M$_\odot$\,(K\,km\,s$^{-1}$\,pc$^2$)$^{-1}$ to derive the gas mass. It has been shown \citep{Sulzenauer2021} that the $r_{i,1}$ depends on the sample of galaxies selected. However, we simply quote the gas mass using the ratios from Bothwell et al., 2013 \citep{Bothwell2013} for easier comparisons with other protocluster samples. The weighted average of $r_{37}$\,($L^\prime_\mathrm{CO(3-2)}/L^\prime_\mathrm{CO(7-6)}$) for HS\,1549 is 3.06\,$\pm$\,0.22, which is comparable to 2.89\,$\pm$\,0.55 found in \citet{Bothwell2013}.

\subsection{Dynamical mass estimates}
The dynamical mass of a galaxy, $M_{\rm dyn}$, is the mass calculated by assuming virial equilibrium and measures the mass required to keep a test particle in a circular orbit with a measured velocity at a given radius. The difficulty of modelling the dynamics of the galaxies requires us to assume some underlying physical distribution of matter in three-dimensional space. We approach quantifying the dynamical mass using the equation \citep{Gnerucci2011}:
\begin{equation}
M_{\rm dyn}\sin^2(i) = f\frac{\sigma_\mathrm{FWHM}^2 r}{G},
\end{equation}
where $i$ is the inclination angle, $f$ is the correction fraction for a disk galaxy, $\sigma_\mathrm{FWHM}$ is the FWHM of the line, $r$ is the galaxy's half-light radius, and $G$ is the gravitational constant. We assume an average $\langle \sin(i) \rangle$\,=\,$\frac{\pi}{4}$ \citep[Appendix A]{Law2009}, and an $f$ of 1.54 \citep{Bothwell2013}. Typically, $r$ is calculated by fitting a disk profile to the continuum image of the galaxy or fitting a three-dimensional ring profile to the data cube. However, because we lack the spatial resolution required to fully resolve the galaxy, we assume an average $r$ value of 7\,kpc \citep{Ivison2011}. The list of $M_\mathrm{dyn}$ values can be found in the last column of Table~\ref{tab: derived}, ranging from 10$^{10}$ to 10$^{12}$\,\msol.

\subsection{Star-formation rate estimates} \label{sec: sfr}

The far-infrared (FIR) brightness for high-redshift galaxies correlates with the instantaneous SFR because star formation produces dust. Dust naturally absorbs optical starlight and thermally re-radiates it at FIR wavelengths. The FIR photons are then redshifted into the mm/submm regime. The ``total'' thermal emission of the dust, known as the total FIR luminosity, or \lfir, is found by fitting FIR photometry to a spectral energy distribution (SED) and integrating, conventionally, from 42 to 500\,\um. We assume that the SED is well-modelled by a modified blackbody. The \lfir\ is then converted to a SFR using the functional form \citep{Kennicutt1998} assuming a Chabrier IMF \citep{Chabrier2003}
\begin{equation}
{\rm SFR}[M_\odot\,{\rm yr}^{-1}] = 0.95 \times 10^{-10} {\rm L}_{\rm FIR} [{\rm L}_\odot].
\end{equation}

Our observations cover four photometry measurements in the FIR to mm regime, namely 3.3\,mm, 1.4\,mm (1.3\,mm for ALMA), 850\,\um\, and 450\,\um\ (we mainly focus on 1.4\,mm and 850\,\um). The first two measurements trace the Rayleigh-Jeans tail, and at redshifts of around 2.85, the 850- and 450-\um\ observations begin to trace the peak of the blackbody (more so at 450\,\um\ than at 850\,\um). However, the 450-\um\ SCUBA-2 map generally has worse effective sensitivity to these galaxies than at 850-\um\, with a variable noise across the map due to differing atmospheric transparency through the nights the cluster was observed, resulting in low constraining power of the SED shape from the 450-\um\ photometry.
Our SMGs have an average $S_{850}$/$S_{1.4}$\,=\,5.4\,$\pm$\,2.2, which is expected of typical spectral energy distributions (SED) of SMGs (see \citep{swinbank2014}). 

To more easily compare with other protocluster samples, we scale SFRs from the 850-\um\ photometry of our HS\,1549 SMGs, using a modified blackbody with a dust temperature of 40\,K and an emissivity index of 2 \citep{swinbank2014, Cunha2015} assuming optically thick dust. In the cases where our NOEMA data resolves the SCUBA-2 source into multiple sources in the 1.4-mm continuum (``F'', ``J'', and ``P''), we scale the 850\,\um\ flux densities by their 1.4-mm flux density fractions. 

Integrating the modified blackbody function gives the total \lfir, which we convert to SFR \citep{Kennicutt1998}. The uncertainty in our quoted SFR is based purely on the 850-\um\ photometry, but we note that SFR is also highly dependent on the significant uncertainty in dust temperatures (i.e., a decrease of 5\,K reduces the SFR by 50\%). However, we adopt the same modified blackbody to all galaxies, limiting the impact of this uncertainty when comparing SFRs to other protoclusters.

\section{Results} \label{sec: results}

\begin{table}[!ht]
 \centering
  \begin{tabular}{cccccc}
\hline
ID & $R_{\mathrm{LOS}}$ & SFR & $M_\mathrm{gas}$ & $M_\mathrm{dyn}$ \\
& [pMpc] & [M$_\odot$\,yr$^{-1}$] & [10$^{10}$\,M$_\odot$] & [10$^{10}$\,M$_\odot$]\\
\hline
A & $-$25.08 & 1430\,$\pm$\,80\phantom{0} & \phantom{0}7.0\,$\pm$\,1.8 & \phantom{0}67.0\,$\pm$\,39.9 \\
B & $-$7.54 & 1580\,$\pm$\,50\phantom{0} & \phantom{0}7.9\,$\pm$\,1.7 & \phantom{0}19.8\,$\pm$\,7.7\phantom{0} \\
C & $-$7.27 & 1170\,$\pm$\,60\phantom{0} & \phantom{0}7.0\,$\pm$\,1.5 & \phantom{0}25.3\,$\pm$\,9.3\phantom{0} \\
D & $-$5.65 & 1390\,$\pm$\,60\phantom{0} & \phantom{0}6.2\,$\pm$\,1.4 & \phantom{0}13.6\,$\pm$\,5.7\phantom{0} \\
E & $-$4.30 & 1200\,$\pm$\,40\phantom{0} & 10.5\,$\pm$\,2.3 & 133.0\,$\pm$\,50.0 \\
F & $-$3.22 & \phantom{0}540\,$\pm$\,70\phantom{0} & \phantom{0}7.9\,$\pm$\,2.0 & \phantom{0}36.0\,$\pm$\,9.7\phantom{0} \\
G & $-$0.80 & 1270\,$\pm$\,150 & \phantom{0}7.8\,$\pm$\,1.5 & \phantom{0}27.1\,$\pm$\,5.3\phantom{0} \\
H & $-$0.54 & 1790\,$\pm$\,60\phantom{0} & \phantom{0}3.6\,$\pm$\,1.3 & \phantom{0}37.4\,$\pm$\,30.4 \\
I & $+$0.27 & 1290\,$\pm$\,50\phantom{0} & \phantom{0}3.3\,$\pm$\,1.0 & \phantom{0}\phantom{0}2.2\,$\pm$\,0.5\phantom{0} \\
J & $+$0.80 & \phantom{0}510\,$\pm$\,30\phantom{0} & \phantom{0}3.9\,$\pm$\,1.2 & \phantom{0}34.3\,$\pm$\,21.7 \\
K & $+$1.07 & \phantom{0}810\,$\pm$\,150 & \phantom{0}2.6\,$\pm$\,0.7 & \phantom{0}17.4\,$\pm$\,8.5\phantom{0} \\
L & $+$1.34 & 1360\,$\pm$\,150 & \phantom{0}4.1\,$\pm$\,0.8 & \phantom{0}\phantom{0}6.6\,$\pm$\,1.8\phantom{0} \\
M & $+$4.01 & 1140\,$\pm$\,50\phantom{0} & \phantom{0}4.8\,$\pm$\,1.2 & \phantom{0}\phantom{0}9.2\,$\pm$\,4.5\phantom{0} \\
N & $+$4.81 & 1640\,$\pm$\,80\phantom{0} & \phantom{0}6.6\,$\pm$\,1.6 & \phantom{0}25.6\,$\pm$\,12.0 \\
O & $+$5.87 & 1140\,$\pm$\,50\phantom{0} & \phantom{0}9.0\,$\pm$\,1.9 & \phantom{0}48.6\,$\pm$\,17.9 \\
P1 & $+$10.90 & \phantom{0}830\,$\pm$\,40\phantom{0} & \phantom{0}3.3\,$\pm$\,0.8 & \phantom{0}\phantom{0}3.2\,$\pm$\,1.6\phantom{0} \\
P2 & $+$11.43 & \phantom{0}340\,$\pm$\,60\phantom{0} & \phantom{0}6.6\,$\pm$\,1.5 & \phantom{0}16.5\,$\pm$\,7.0\phantom{0} \\
Q & $+$19.29 & \phantom{0}690\,$\pm$\,30\phantom{0} & \phantom{0}4.3\,$\pm$\,0.9 & \phantom{0}\phantom{0}3.1\,$\pm$\,2.0\phantom{0} \\
\hline
\end{tabular}

 \caption{Derived properties of the SMGs in HS\,1549 ranked in ascending redshift order. Here $z$ is measured from spectroscopy, $R_\mathrm{LOS}$ is the line-of-sight offset (in units of pMpc) from $z$\,=\,2.85, SFR is scaled from the $S_{850}$ flux density from Table~\ref{tab: ctn}, $M_\mathrm{gas}$ is scaled from CO line strength (Table~\ref{tab: lines}) $M_\mathrm{dyn}$ is scaled from the CO line width (Table~\ref{tab: lines}).
 \label{tab: derived}}
\end{table}

\subsection{Velocity dispersion of the protocluster}

\begin{figure*}
 \centering
 \includegraphics{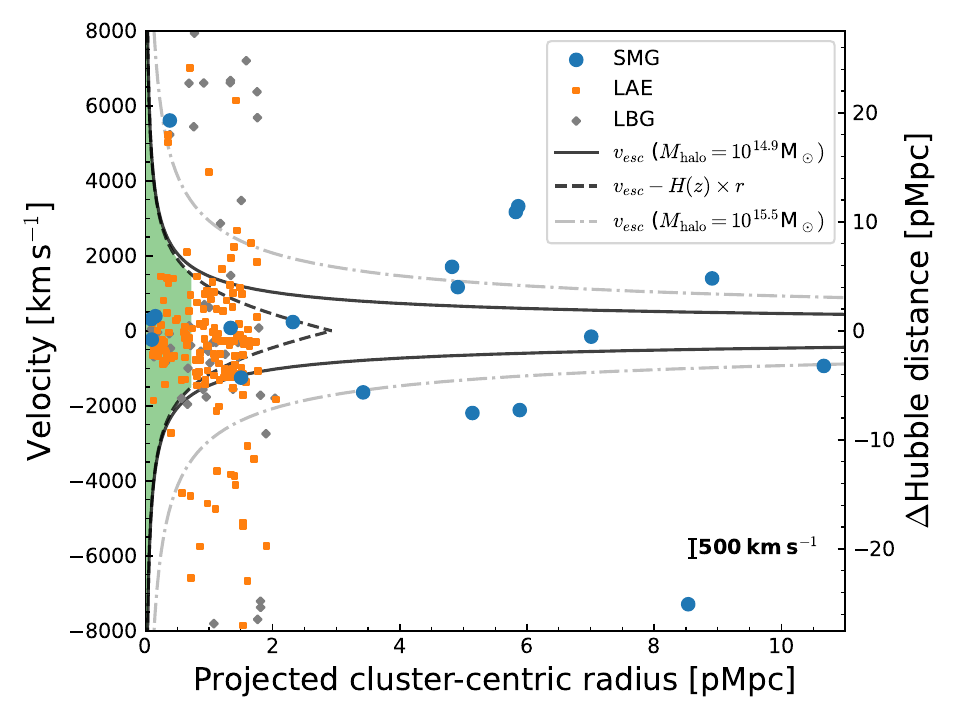}
 \caption{LOS velocity as a function of projected on-sky cluster-centric radius. The left axis shows the LOS velocity centred at $z$\,=\,2.85, and the right axis shows the cosmological distance in proper megaparsecs. The blue circles are the spectroscopically confirmed SMG members of the protocluster. The orange squares are LAEs, and the red diamonds are Lyman-break galaxies (Steidel et al., in prep.). The solid black curve shows the escape velocity of a dark matter halo of mass equal to 7.4\,$\times$\,10$^{14}$\,\msol\ ($v_{200}$\,=\,2100\,\kms) and defines a region where objects are likely gravitationally bound to the protocluster. The dashed curve includes the correction from cosmological expansion. The dash-dotted curve represents the escape velocity of the most massive halos (3\,$\times$\,10$^{15}$\,\msol\ with $v_{200}$\,=\,3300\,\kms, e.g., the Phoenix cluster \citep{Gao2012}). These escape velocity curves have been reduced by a factor of $\sqrt{3}$ so that they can be treated as LOS velocities. The green-shaded region represents the central area within $R_{200}$; the velocities of these galaxies should be dominated by their peculiar motions.\label{fig: velocity}}
\end{figure*}

We calculate LOS velocities relative to the central redshift using
\begin{equation}
\begin{split}
v_i &= H(z_\mathrm{cluster}) d_\mathrm{P}, \\ 
d_\mathrm{P} &= \frac{d_\mathrm{C}(z) - d_\mathrm{C}(z_\mathrm{cluster})}{1 + z_\mathrm{cluster}},
\end{split}
\end{equation}
where galaxies moving away from us have a positive velocity. Here $H$ is the Hubble parameter, $d_\mathrm{P}$ is the proper distance (see Table~\ref{tab: derived}), $d_\mathrm{C}$ is the comoving distance, $z_\mathrm{cluster}$ is the median cluster redshift (2.85), and $z$ is the redshift. However, what we measure ($z_\mathrm{obs}$) is the product of the peculiar and cosmological redshifts, $(1 + z_\mathrm{obs})\,=\,(1 + z_\mathrm{pec})(1 + z_\mathrm{cosmo})$. We interpret $z_\mathrm{cosmo}$ as a measure of distance and $z_\mathrm{pec}$ as peculiar velocities. In comoving units (cMpc), the cosmological distance is the difference between the SMG and the ``centre'', defined as barycentre of the structure. Combining the RA, Dec, and cosmological distance allows us to define a three-dimensional coordinate system.

Within a {\it core} region surrounding the QSO (SMGs ``G'', ``K'', and ``L'' fall within this region of phase space), the cluster-centric redshift of galaxies is likely to be dominated by their peculiar motions, resulting in their LOS cluster-centric distances being uncertain. By contrast, at large cluster-centric radii, where even a halo with a mass of $\sim$\,10$^{14}$\,\msun\ has a small escape velocity, the $z_\mathrm{cosmo}$ from the Hubble flow dwarfs the expected $\sim$\,500\,\kms\ peculiar velocities, and their redshifts can be interpreted as distances. The galaxies with the most extreme radii (i.e., ``A'', and ``Q'') may not necessarily be gravitationally bound to the virialized cluster by redshift zero, although they are still affected by the gravitational potential of the growing protocluster. 

In Fig.~\ref{fig: velocity}, we plot the LOS velocities of the SMGs as a function of their cluster-centric distances. The radial distance is determined by the straight line between the cluster's centre and each SMG, with the on-sky separation scaled to a length using the angular diameter distance. However, the Hubble flow will not affect some galaxies, i.e.\ they are already bound to the cluster. We determine the subset of these galaxies using the criteria that they must fall within an assumed virial radius of the core, which we define as $R_{200}$, the radius at which the density of the cluster falls to 200 times the critical density of the Universe at the given redshift. The critical density is given by
\bigskip
\begin{equation}
\begin{split}
\rho_{\rm crit} & = \frac{3H^2(z_\mathrm{cluster})}{8\pi G}, \\
R_{200} & =\left(\frac{3M_{200}}{800 \pi \rho_{\rm crit}} \right)^{1/3}. 
\end{split}
\end{equation}
We assume an upper bound to the mass, $M_{200}$\,=\,7.4\,$\times$\,10$^{14}$\,$h^{-1}$\,\msol\ to be the dark matter halo mass of the protocluster (Steidel et al., in prep.), defined by an escape velocity profile that contains 95\% of the UV-selected galaxies within a broad $\pm$\,3000\,\kms\ envelope.

The region of the cluster within $R_{200}$ is shaded in light green in Fig.~\ref{fig: velocity}, with SMGs ``G'', ``K'', and ``L'' lying in this region. Assuming the halo mass, $M_{200}$, we calculate an envelope within which galaxies would be gravitationally bound. We reduce the escape velocity by a factor of $\sqrt{3}$ to be effectively LOS values and then correct for the on-sky Hubble expansion. All the other bright SMGs have observed velocities greater than the escape velocity.

For the central region of the protocluster, there are spectroscopic observations of the LAEs and Lyman-break galaxies (LBGs, Steidel et al., in prep.). We also observe substructure across the protocluster \citep{Trainor2012} with typical LOS velocities of 250--300\,km\,s$^{-1}$ ($\leq$\,10$^{13}$\,\msun). We estimate their true velocities by multiplying by $\sqrt{3}$. The LAEs and LBGs are displayed in Fig.~\ref{fig: velocity}, where most of the population falls within the gravitational envelope of the central dark matter halo. The dashed region removes the cosmic expansion of the Universe from the gravitational envelope.

\subsection{3D Structure of the protocluster} \label{sec: structure}

\begin{figure*}
 \centering
 \includegraphics[width=12cm]{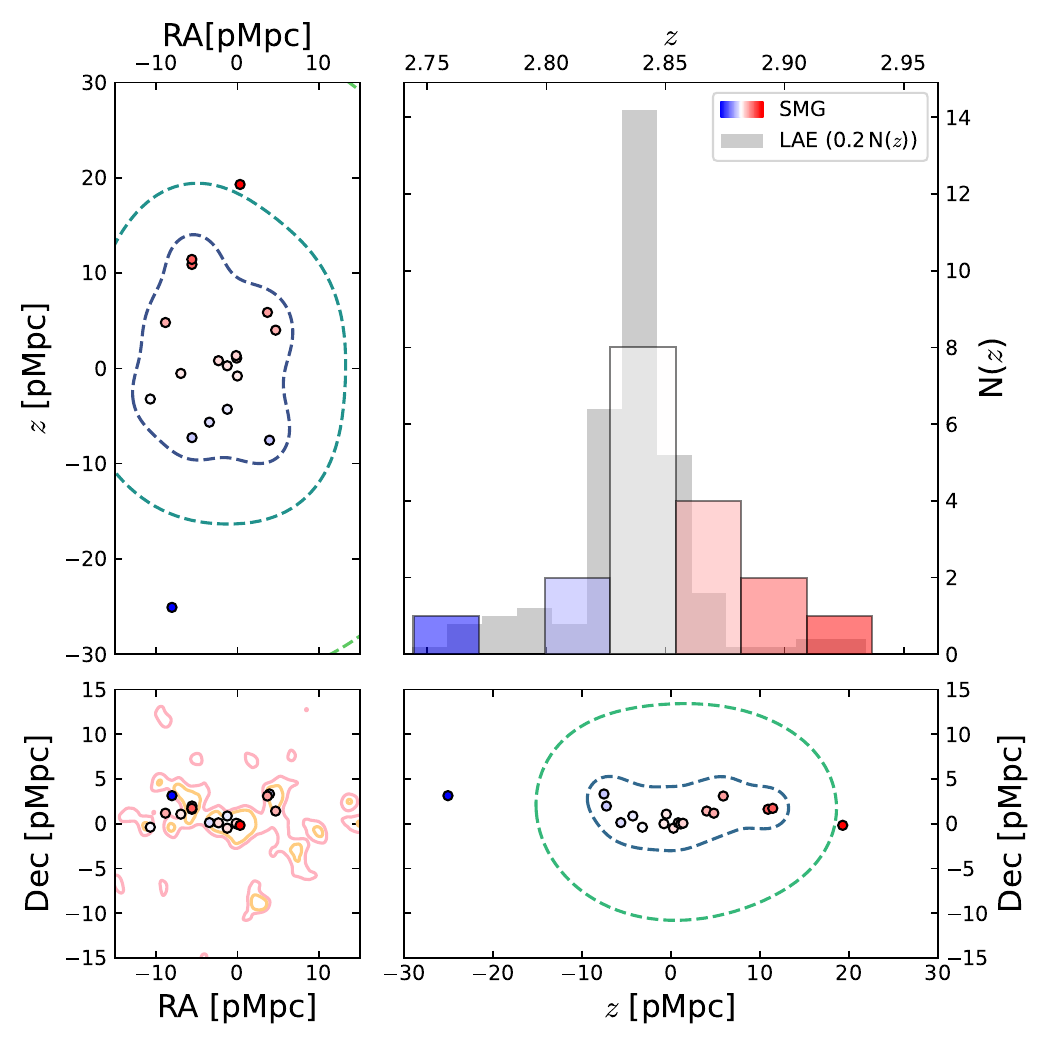}
 \vskip-0.5cm
 \caption{Four different projections of the three-dimensional structure of HS\,1549. \textit{Bottom left}: The on-sky projection of the 18 spectroscopically confirmed submm galaxies in HS\,1549, plotted in RA and Dec. The contours shown are from the LAE surface density distribution \cite{Kikuta2019, Lacaille2019}, and we find the SMGs situated in the highest-density regions. \textit{Top left}: The projection of HS\,1549 as seen from the ``top'' perspective, in RA and redshift. The contours are drawn from the distribution of the SMGs, showing the structure's shape. \textit{Bottom right}: The projection of HS\,1549 as seen from the ``side'' perspective, in Dec and redshift. The SMGs trace a structure extended in the redshift axis and compressed along the Dec axis. \textit{Top right}: The one-dimensional histogram of the spectroscopic redshifts. The distribution peaks around $z$\,$\approx$\,2.85, which we use to define the mean redshift of HS\,1549. We compare with the LAE redshift distribution (scaled by a factor of 0.2) and find that SMGs and LAEs peak around the same redshift ($z$\,=\,2.852\,$\pm$\,0.025). In all panels, SMGs are colour-coded by their redshifts; redder galaxies have higher redshifts. \label{fig: distribution}}
\end{figure*}

Fig.~\ref{fig: distribution} shows three different 2D projections of the 3D protocluster. For each projection, we collapse the structure along one of the three axes: $z$, declination, and right ascension, forming the on-sky, top and side perspectives, respectively. The SMGs trace a structure that is compressed along the Dec axis, comparable to the LAEs, and elongated along the redshift axis, assuming redshifts are purely cosmological. We find, somewhat unexpectedly, that the SMGs reside relatively uniformly in projection across the dense LAE region of the protocluster.

The distance measurements become more secure as we move further away from the median redshift, since the expansion of the Universe increasingly dwarfs the peculiar motions of individual galaxies. However, the redshift outliers (``A'' and ``Q'') could potentially be non-protocluster members, suggesting that there may be less elongation of the structure along the redshift axis than we measure from the full SMG sample. However, the SMGs with the most extreme redshifts (``A'' and ``Q'') are still likely to be attracted to the protocluster due to the strong gravitational pull of the massive dark matter halo, even if they retain some non-random velocity in the cluster by $z$\,=\,0.

To assess the likelihood of SMG membership at the extremes of the redshift distribution, we compare with simulations of overdense structures at a similar redshift \citep{Muldrew2015}. We compare the fractional number of SMGs in a given 3D radius with the typical size of a protocluster that will ultimately form a cluster at the current epoch. A typical protocluster found in simulations (with descendant $M_\mathrm{halo}$\,$>$\,10$^{15}$\,$h^{-1}$\,M$_\odot$) \citep{Muldrew2015} has a radius of (6.8\,$\pm$\,1.5) pMpc. We compare this with our protocluster sample and estimate the radius at which we recover 90\% of our SMGs under the approximation that the SMGs in our sample have the same stellar mass. This radius is tightly correlated with the radius derived from the dark matter mass. We find that HS\,1549 is larger than 99.99\% of all protoclusters found in the simulation, and even if we exclude ``A'' and ``Q'', HS\,1549 is still larger than 99.78\% of all protoclusters. It is unclear what is happening at the locations of these outliers. It could be the case that these SMGs are separating from the protocluster core through the Hubble flow or that they are situated in a filament of the cosmic web with some streaming velocity that will eventually collapse at a node (i.e., where the protocluster core is located). More likely than not, these sources will not fall into HS\,1549's gravitational potential. However, the gravitational effects these galaxies experience from HS\,1549 are larger than the average peculiar motion of a galaxy, meaning that they are not typical field galaxies, regardless of their endpoint at $z$\,=\,0 (similar to LBGs with redshifts on the outskirts of the HS1549 structure). We consider the restricted (and full) sample cases in the following analysis and discussion.

\subsection{Shape of the protocluster} \label{sec: shape}

With the SMGs' redshift information providing our third spatial dimension, we attempt to characterize the protocluster's structure to compare with clusters found in simulations. We model the structure by approximating the protocluster as an ellipsoid and constructing its moment-of-inertia tensor. Each element of the tensor is defined by
\begin{equation} 
I_{i,j} = \sum^{N_{\rm SMG}}_{p = 1}w_px_{p, i}x_{p, j} ,
\end{equation}
where $x_{p,i}$ defines the position of galaxy $p$ along the $i$th axis, and $w_p$ is the weight associated with each galaxy. Intensity \citep{Harvey2021} and mass \citep{Velliscig2015} are common choices for the weight; however, we use a uniform weight for simplicity because our $S_{850}$ measurements are relatively uniform across the field and to account for filamentary structure \citep{lovell2018}. Solving the eigenvalue problem of the moment-of-inertia tensor, we can approximate the major, intermediate, and minor axes of an ellipsoid with the square root of the eigenvalues, 
\begin{equation}
\begin{split}
a & = \sqrt{\lambda_1}, \\
b & = \sqrt{\lambda_2}, \\
c & = \sqrt{\lambda_3}, \\
\lambda_1 & \geq \lambda_2 \geq \lambda_3.
\end{split}
\end{equation}
The sphericity and triaxiality \citep{Velliscig2015, lovell2018} parameters are defined as
\begin{equation}
\begin{split}
S & = \frac{c}{a}, \\ 
T & = \frac{a^2 - b^2}{a^2 - c^2}.
\end{split}
\end{equation}
A unitary sphericity, $S$, is a perfect sphere and the triaxiality, $T$, would then be undefined. As $S$ approaches zero, the protocluster becomes more disk-like. $T$ is a measure of oblateness/prolateness, with a high $T$ representing a more prolate structure.

\begin{figure}
 \centering
 \includegraphics[width=\twidth]{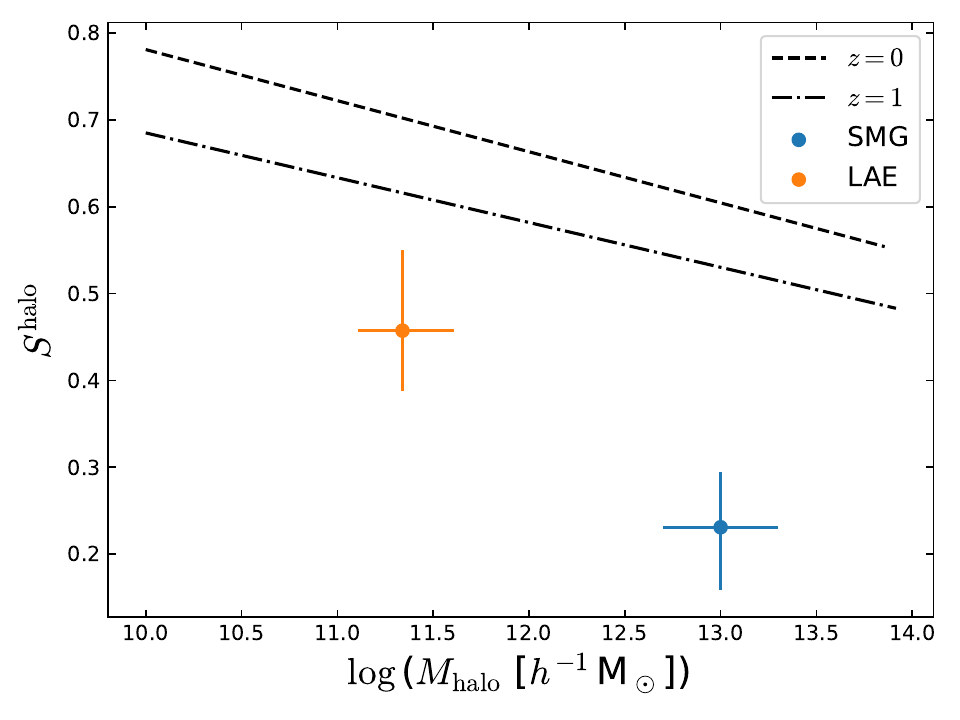}
 \vskip-0.5cm
 \caption{Sphericity of the protocluster compared to sub-halo mass. The HS\,1549 points are created by solving the moment-of-inertia tensor for the SMGs ($\approx$\,10$^{13}$\,$h^{-1}$\,M$_\odot$, in blue) and the LAEs ($\approx$\,10$^{11.3}$\,$h^{-1}$\,M$_\odot$, in orange) with bootstrapped uncertainties. The two dotted black curves are calculated from the inertia tensors in simulated clusters at two different epochs \citep{Velliscig2015}; higher redshift structures are difficult to identify in simulations of restricted sizes. The HS\,1549 protocluster is more flattened than typical lower redshift clusters.\label{fig: sphericity}}
\end{figure}

In Fig.~\ref{fig: sphericity}, we compare the sphericity of HS\,1549 to clusters found in dark matter simulations at different epochs \citep{Velliscig2015}. The simulations show a downward trend in sphericity at larger sub-halo masses (the halo mass of each galaxy), which we replicate by studying the LAEs, with an average halo mass of $\log$($M_{\rm halo}$\,[$h^{-1}$\,M$_\odot$])\,=\,$11.34_{-0.27}^{+0.23}$ \citep[see clustering analysis on LAEs,][]{Alonso2023} and the SMGs, with $\log$($M_{\rm halo}$\,[$h^{-1}$\,M$_\odot$])\,=\,13.0\,$\pm$\,0.3 \citep[see clustering analysis on SMGs,][]{Stach2021}. We do not have spectroscopic data across the entire LAE distribution. Therefore, we calculate the eccentricity (the ratio of the minor to major axis) of the LAEs in the SCUBA-2 map and scale it to the sphericity by calculating the ratio of the eccentricity to sphericity using the SMGs. The uncertainty in sphericity is bootstrapped 10,000 times from the SMGs to account for outliers and propagated to the LAE eccentricity. HS\,1549 follows the trend of structures being more disk-like at higher redshifts, and the protocluster is flatter than lower redshift clusters at the same sub-halo mass. This suggests a sheet-like collapse of the extended protocluster \citep{Casey2016}, which is expected early in galaxy cluster formation \citep[see Zeldovich pancakes][]{Zeldovich1970} during which matter is streaming in along filaments \citep{Bond1996}.

\section{Discussion} \label{sec: discussion}

\subsection{Radial profile of SFR in the protocluster} \label{sec: lae}

\begin{figure}
 \centering
 \includegraphics[width=\twidth]{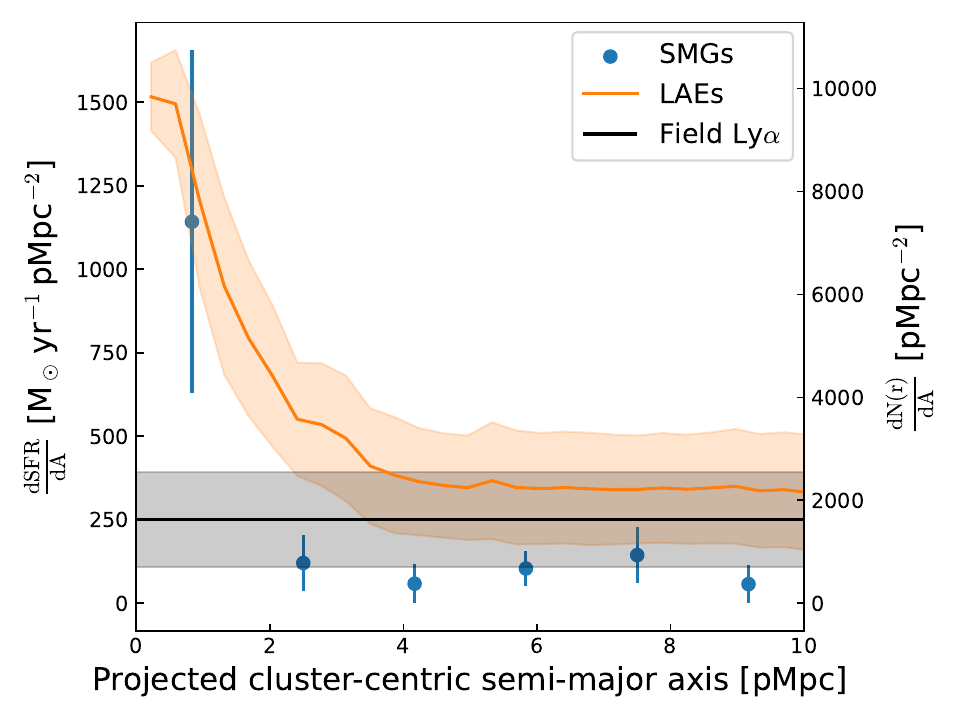}
\vskip-0.5cm
\caption{Differential SFR surface density of the protocluster as a function of projected cluster-centric radius. The SMGs are in blue, and the LAEs \citep{Kikuta2019} are in orange. The horizontal black region shows the field LAE surface density. The LAEs trace the protocluster with a profile comparable to the SMGs.\label{fig: diff sfr}}
\end{figure}

Given that the SMGs at $z$\,$=$\,2.85 are found within the region of highest LAE overdensity in the protocluster, understanding how the distributions of SMGs and LAEs compare is of interest. In Fig.~\ref{fig: diff sfr}, we show the radial distribution of differential SFR per unit area for the SMGs and the number density of the LAEs. We derive the number density of LAEs by taking the counts \citep{Kikuta2019} of the LAEs in each radial bin and dividing by the area of an ellipsoidal shell (with eccentricity and position angle derived from solving the moment-of-inertia tensor). The number density can be scaled to a SFR density using an average SFR for LAEs, (2.6\,$\pm$\,0.8) \sfr\ \citep{Alonso2023}. 
We follow a similar procedure for calculating the SFR density of SMGs but truncate the area to overlap with our SCUBA-2 maps. We find that the SMGs trace the LAEs' profile out to the structure's edges. The LAE profile also rapidly falls to the average field level of LAEs ($\approx$\,4\,pMpc), determined by averaging the LAE density at the edges of the Subaru Hyper-Suprime Cam 1-degree field. 

We then assess the local LAE overdensity around each SMG in HS\,1549. The average LAE density in a 1.7-arcmin radius around the protocluster SMGs is 4300\,pMpc$^{-2}$ (3900\,pMpc$^{-2}$ if we exclude the central 4 SMGs: ``G'', ``K'', ``L'', and ``Q''). This corresponds to a 1.6 (1.5) $\sigma$ overdensity which is twice the overdensity found around the non-protocluster SMGs (0.8\,$\sigma$, with a number density of 3100\,pMpc$^{-2}$). In the case of HS\,1549, we usually find SMGs in overdense regions of the Ly$\alpha$ surface distribution.

\subsection{Protocluster SFR comparison sample}

\begin{figure*}
 \centering
 \includegraphics[width=12cm]{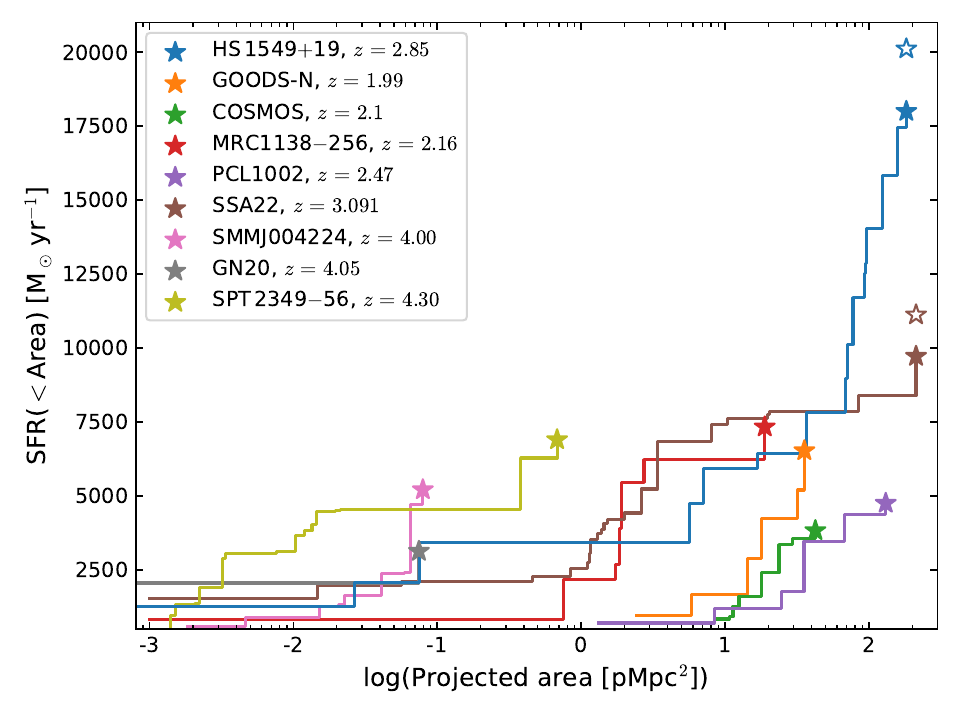}
 \caption{Comparison of $z$\,$>$\,2 protoclusters: the cumulative SFR from SMGs (selected by SCUBA-2) as a function of the projected on-sky area. For HS\,1549, we define the centre as the QSO position, with the star representing the total SFR and the hollow star representing the total without the two SMGs with the most discrepant redshifts. Here SFR is scaled from the 850-\um\ flux density of the spectroscopically confirmed SMGs in all the protoclusters shown, assuming $T_\mathrm{dust}$\,=\,40\,K. We show the integrated total SFRs. Protoclusters found in the literature lie in the range $z$\,=\,2--4.3: GOODS-N \cite{Chapman2005, Cowie2017}; COSMOS \cite{Yuan2014}; MRC1138$-$256 \cite[the Spiderweb,][]{Dannerbauer2014}; PCL1002 \cite{Casey2015}; SSA22 \cite[][, Chapman et al., in prep.]{Umehata2015}; SMMJ\,004224 \cite[Distant Red Core,][]{Oteo2018}; GN20 \cite{Daddi2009, Hodge2013, Cowie2017}; and SPT\,2349$-$56 \cite{Miller2018, Hill2020}. \label{fig: growth}}
\end{figure*}

To place HS\,1549 in context and compare to other systems claimed to be protoclusters, we assemble from the literature various SMG-rich overdensities at $2$\,$<$\,$z$\,$<$\,5. Although a direct comparison of the number counts (number deg$^{-2}$) of SMG-overdense systems can be performed, it involves making somewhat arbitrary choices of enclosed areas and redshift boundaries. 

In Fig.~\ref{fig: growth}, we show instead the cumulative 850-\um\ flux density translated to SFR in HS\,1549 compared to other published protoclusters found in the literature \citep[for more details, see Appendix~\ref{sec: comparison sample}]{Chapman2009, Yuan2014, Dannerbauer2014, Casey2015, Umehata2015, Oteo2018, Daddi2009, Hodge2013, Miller2018, Hill2020}. We assume the same dust temperature of 40\,K for all protoclusters. The comparison protocluster data are drawn partially from a recent compilation \citep[see references in][]{Casey2016}, and partially from primary references. Only galaxies confirmed to be protocluster members via spectroscopic redshifts are considered. 

To create the cumulative submm flux distributions for the comparison protoclusters in Fig~\ref{fig: growth}, the centre of each protocluster is defined by computing the median RA and Dec of all submm sources. We checked that adjusting the centres for the curves randomly by $\sim$\,$1^{\prime}$ did not change the curves by more than 10\%, demonstrating that the curves for the SMG overdensities are fairly insensitive to the adopted centre.

Similar to other protoclusters, HS\,1549 has an extreme core, but the integrated SFR rises steeply above all other high-redshift protoclusters beyond the 2.4--4.8\,pMpc inner regions, exceeding a total SFR of 1.8\,$\times$\,10$^4$\,\sfr. Furthermore, no other protocluster contains nearly as many bright $S_{850}$\,$>$\,8\,mJy SMGs, and its uniqueness becomes more dramatic if we limit the comparison to only the most luminous SMGs. HS\,1549 is also one of the largest protoclusters in the sky. The excess of bright SMGs highlights how hyper-luminous infrared galaxies (HyLIRGs) trace the large-scale structure in HS\,1549, presumably indicating very rapid galaxy growth.

\subsection{Simulations} \label{sec: simulations}

To understand how rare the observed density of bright SMGs in HS\,1549 is, we used the Big MultiDark Planck (BigMDPL) simulation \citep{prada2012} at the closest available redshift ($z$\,=\,2.89) to examine the neighbouring regions of massive halos. BigMDPL is a dark-matter-only simulation containing 3840$^3$ $N$-body particles within a volume 50\,cGpc$^3$, giving a mass resolution of $3.5$\,$\times$\,10$^{10}$\msun. This resolution is sufficient to resolve halos of mass $M_\mathrm{vir}$\,=\,3.5\,$\times$\,10$^{12}$\,\msun\ and the volume is large enough that it should provide hundreds of massive group-sized structures at $z$\,$\sim$\,3. 

Based on the estimated central halo mass of HS\,1549, $M_\mathrm{vir,\,est}$\,$\sim$\,10$^{14}$\,\msun, we searched BigMDPL for all halos more massive than $\log(M_\mathrm{vir,\,limit})$\,=\,13.8. We found 435 halos at $z$\,=\,2.89 that satisfy this mass cut. For each massive halo, we searched three separate spherical regions with radii $R$\,=\,$\{20,\,30,\,40\}$\,cMpc for neighbouring halos. In each spherical region, we ranked all neighbouring halos by virial mass and determined the mass of the 10th most massive halo. Ten is chosen as a fiducial baseline, since at least three of the 18 SMGs in HS\,1549 lie within the central halo, and at least two of the outlying SMGs likely probe the same halo (e.g., P1/P2 in Table~\ref{tab: derived}). This gives us a minimum mass, $M_{\mathrm{vir,}\,10}$, above which at least 10 neighbours are equally or more massive. 

\begin{figure}
 \centering
 \includegraphics[width=\twidth]{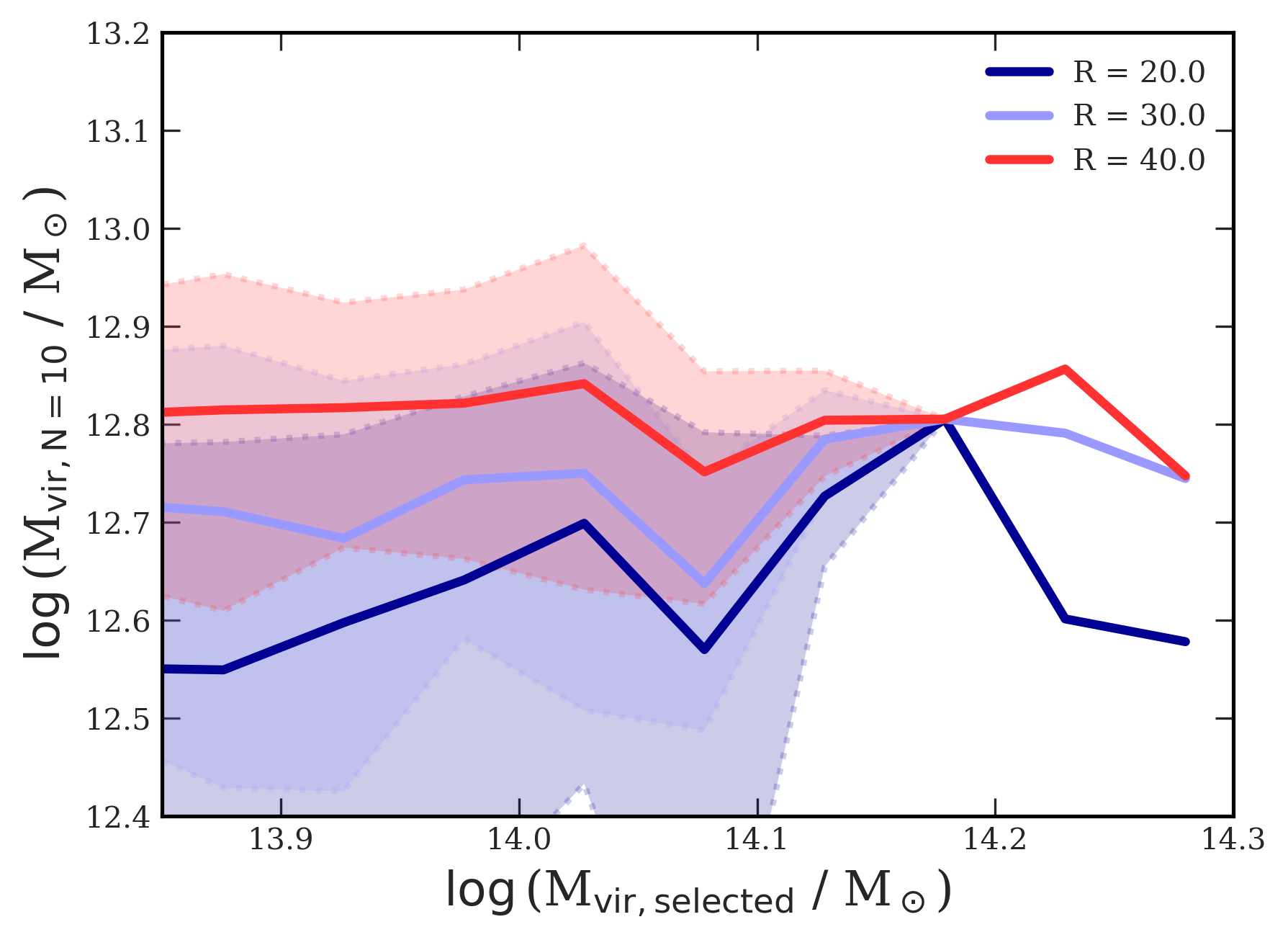}
 \caption{From the BigMDPL simulation (see text), the median mass (vertical axis) of the 10th most massive nearby halo in bins of halo virial mass for all halos above $\log(M_\mathrm{vir})$\,=\,13.8\,\msun\ (horizontal axis) at $z$\,=\,2.89. The curves show the mass at which at least $10$ halos have equal or greater virial mass within $R$, for a given selected massive halo $M_\mathrm{vir,\,selected}$. Each coloured line shows a different spherical radius in cMpc that was used to find neighbouring halos. The shaded regions show a $\pm$\,1\,$\sigma$ spread around the median for each radius. \label{fig: rennehan}}
\end{figure}

Fig.~\ref{fig: rennehan} shows the result of our calculation for $M_{\mathrm{vir,}\,10}$. We binned the result by the virial mass of the most massive halos in the simulation at $z$\,=\,2.89 and found the median value for each massive halo in the neighbourhood, given by $R$. The shaded regions show the uncertainty for each massive halo. The shading disappears at the high mass end, where the number of central halos diminishes exponentially until only one halo with greater mass than 10$^{14.1}$\,\msun\ remains. 

From clustering analysis of SMGs \citep{Stach2021}, the typical halo mass of $z=2-3$ SMGs is inferred to be $\approx$\,10$^{13}$\,\msun. Strictly interpreted, none of the simulated central haloes have 10 satellite haloes with masses $>10^{13}$\,\msun, even out to the largest radii considered (40\,cMpc). The closest system being a single $10^{14}$\,\msun\ central halo whose tenth most massive satellite is $9\times10^{12}$\,\msun.
However, any of these simulated central haloes do have at least 10 massive satellites within R\,$<$\,40\,cMpc, which are massive enough within uncertainties to be viable hosts of SMGs.

For an explicit example, at an R\,=\,20\,cMpc, which would enclose exactly 10 of the HS\,1549 SMGs, the figure shows that the most massive central halos in the simulation have a tenth most massive satellite on average $4\times10^{12}$\,\msun. 
Again, while the satellite haloes are, on average, a factor of two lower than the mass required to support the SMG halo mass (from field SMG clustering analysis), the masses of satellite haloes are not completely out of line with requirements for HS\,1549.

The figure demonstrates that very few of the most massive central haloes have enough neighbouring halos of sufficiently high mass to host the 15 SMGs observed outside the core region of HS\,1549. However, the complication in simulating the high density of HyLIRGs in HS\,1549 is unlikely the result of the number of available massive haloes in simulations, but instead, the star formation prescriptions, feedback recipes, and baryonic physics adopted in simulations.

\subsection{HS\,1549 in context of cluster galaxy evolution}
The cores of present-day galaxy clusters contain massive elliptical galaxies with old stellar populations \citep{renzini2006}, and SMGs are believed to be the high-redshift progenitors of these ellipticals \citep{smail2004, dudzeviciute2020, simpson2020}. Therefore, the 18 bright SMGs within a volume of 4100\,pMpc$^3$ are probably set to rapidly become massive elliptical galaxies located within the same structure.

Assuming the individual galaxies' halo masses are characterized by their generally broad central CO(3--2) line widths (Table~\ref{tab: derived}), the implied masses are already comparable to cluster ellipticals (with typical masses 10$^{10}$--10$^{11}$\msun\ by $z$\,$\sim$\,1) \citep{saracco2014}. Thus, a key consequence of our finding is that massive elliptical galaxies can form rapidly at large cluster-centric distances. This conclusion has been found to a lesser extent in other clusters, with fainter SMGs appearing at smaller cluster-centric radii near the core (Fig.~\ref{fig: growth}). Recent semi-analytic models \citep{lovell2018} have indicated that the number of massive galaxies in the wider field correlates strongly with the final descendent cluster mass. In HS\,1549, the formation of massive galaxies with high SFRs (up to 1100\,\sfr) over enormous scales (0.24\,deg$^2$) suggests that HS\,1549 may be the progenitor of a very massive galaxy cluster. 

Theoretical studies have shown that at $z$\,$\sim$\,3, the progenitors of galaxy clusters should span 5\,pMpc (5000\,kpc\,$\approx$\,0.2\,deg\,$\approx$\,625$^{\prime\prime}$) \citep{Chiang2014, onorbe2014} to 14\,pMpc (0.5\,deg) \citep{Muldrew2015}, which is consistent with the observations of LAEs in HS\,1549. Observing so many HyLIRGs over these large scales in HS\,1549 suggests that it is in an evolutionary phase where cold gas flows \citep{Dekel2009} into the highest density nodes replenishing the fuel reservoirs for the many short-lived HyLIRGs being observed simultaneously \citep{Casey2015, Umehata2019}. For some of the protoclusters from the literature, it is still being determined whether an overdensity of bright SMGs extends over such a large scale. However, evidence generally suggests that it does not exist in most systems (as shown in Fig.~\ref{fig: growth}). While the core region of HS\,1549 is comparable in its SMG-inferred cumulative SFR to other protoclusters, in the projected region beyond 10\,pMpc$^2$, the SFR of the HS\,1549 protocluster is more than double that of any other known protocluster (and many more times if only HyLIRGs are considered). Probing the HyLIRGs over the wide field in protoclusters provides a unique opportunity to study galaxy cluster formation at a crucial evolutionary epoch and may become a well-calibrated metric of descendent cluster mass.

\section{Conclusion} \label{sec: conclusion}

We have mapped the high-density LAE regions of HS\,1549 at 850\,\um\ with SCUBA-2 \citep{Holland2013} to explore how SMGs are distributed throughout the protocluster. We show the SCUBA-2 map covering most of the central Ly$\alpha$ overdensity, including 25 sources detected at 850\,\um\ with flux densities brighter than 8\,mJy over roughly 35$^{\prime}$\,$\times$\,18$^{\prime}$ (0.19 deg$^2$) of the map, or 137 sources per deg$^2$. The surface density in bright SMGs represents about 3$\times$ the blank field SCUBA-2 number counts \citep{geach2017}. Previously, the central core of HS\,1549 was studied with deep SCUBA-2 maps \citep{Lacaille2019}, which identified 6$\times$ the number density of bright SMGs compared to blank field surveys. We performed a blind spectral line survey of all 25 850-\um\ sources with $S_{850}$\,$>$\,8\,mJy and identified 18 SMGs at the protocluster redshift. Our findings are as follows:
\begin{itemize}
  \item The brightest sources have SFRs $>$\,1000\,\sfr, while the total SFR, including all 18 sources, is $>$\,2.0\,$\times\,10^4$\,\sfr. The SMGs have gas masses ($M_\mathrm{gas}$) around 10$^{10}$\msun\ individually, with dynamical masses ($M_\mathrm{dyn}$) being around $10^{11}$\msun\ (Table~\ref{tab: derived}). 
  
  \item Redshift distributions of SMGs allow us to measure the line-of-sight distances and the 3D distribution of bright galaxies. We find an almost uniform distribution of bright SMGs across the HS\,1549 protocluster.
  
  \item SMGs trace a highly compressed structure along the declination axis and a more elongated structure along the right ascension and redshift axes, forming an apparent {\it pancake}-like structure. 

  \item Bright SMGs in HS\,1549 and the LAE population (Fig.~\ref{fig: diff sfr}) traces a similar surface density profile. 

  \item Protocluster SMGs are located in more LAE overdense regions compared to the non-protocluster SMGs \citep[2$\times$ more overdense; see also][]{Tamura2009}.

  \item The shape of the protocluster can be characterized by a sphericity of around 0.25. The implied sheet-like collapse of the extended protocluster \citep{Casey2016} may be evidence that we are observing the early formation of a galaxy cluster \citep[see Zeldovich pancakes][]{Zeldovich1970, Kravtsov2012}) and the in-falling of matter along filaments \citep{Bond1996}.

  \item The integrated SFR of HS\,1549 rises steeply above all other high-redshift protoclusters beyond the inner regions, exceeding a total SFR of 2\,$\times$\,10$^4$\,\sfr. Furthermore, no other protocluster contains nearly as many bright $S_{850}$\,$>$\,8\,mJy SMGs, and its uniqueness becomes more dramatic if we limit the comparison to only the most luminous SMGs.

  \item We searched the simulations for central halos ($M_\mathrm{selected}$\,$>$\,6\,$\times$\,10$^{13}$\,\msun) approaching the inferred mass of HS\,1549. For each central halo, we looked at the ten most massive satellite halos within a volume of 50\,cMpc$^3$ comparable to HS\,1549. There is a deficit of halos of the typical clustering mass ($\approx$\,10$^{13}$\,M$_\odot$) to host the SMGs in HS\,1549. 

  \item That HS\,1549 is larger than 99.99\% of all protoclusters found at $z$\,=\,2.85 when compared to simulated protoclusters \citep{Muldrew2015}. 
\end{itemize}

The discovery of the 18 bright SMGs across the $\sim$\,0.2\,deg$^2$ field represents a new opportunity to study and understand galaxy cluster evolution. Individually, the SMGs in HS\,1549 can evolve into giant elliptical galaxies, which makes HS\,1549 one of the most massive galaxy clusters found to date. The discovery of this system also represents a new opportunity to look for these extended protoclusters, as current studies are focused more on the central region/core of the systems. In the future, more comprehensive multi-wavelength surveys will help us better understand the astrophysics of HS\,1549.

\section{Acknowledgments}

%
%
This paper makes use of the following ALMA data: ADS/JAO.ALMA\#2019.0.00236.T; PI: S.\ Kikuta.
ALMA is a partnership of ESO (representing its member states), NSF (USA) and NINS (Japan), together with NRC (Canada), MOST and ASIAA (Taiwan), and KASI (Republic of Korea), in cooperation with the Republic of Chile. The Joint ALMA Observatory is operated by ESO, AUI/NRAO and NAOJ.
The National Radio Astronomy Observatory is a facility of the National Science Foundation operated under cooperative agreement by Associated Universities, Inc.
G.W., S.C., D.S., \ gratefully acknowledge support for this research from NSERC.
Part of the analysis in this work was made possible by SciNet and the Niagara supercomputing cluster. D.R. is supported by the Simons Foundation.

\vspace{5mm}
\facilities{JCMT(SCUBA-2), NOEMA(Band 1 and 3), ALMA(Band 6)}




\appendix

\section{Spectral lines of each SMG} 
\label{sec: spectral appendix}
Here, we show the spectral line of all the SMGs (CO(3--2), CO(7--6)/[CI]2--1) and their corresponding continuum maps. Spectra are extracted at the peak pixel in the continuum map. For clarity, we have removed the continuum emission from the spectral lines in the figures.

\begin{table*}
 \centering
 \begin{tabular}{ccccc}
\hline
 ID & Line & FWHM & $I_{\nu}$ & $L^\prime$\\
  & & [km\,s$^{-1}$] & [Jy\,km\,s$^{-1}$] & [10$^{10}$\,K\,km\,s$^{-1}$\,pc$^2$]\\
\hline
A & CO(3--2) & \phantom{0}820\,$\pm$\,240 & 0.90\,$\pm$\,0.18 & 3.62\,$\pm$\,0.73 \\
B & CO(3--2) & \phantom{0}440\,$\pm$\,90\phantom{0} & 0.98\,$\pm$\,0.13 & 4.08\,$\pm$\,0.55 \\
 & CO(7--6) &  & 1.68\,$\pm$\,0.12 & 1.28\,$\pm$\,0.09 \\
 & CI(2--1) &  & 1.09\,$\pm$\,0.12 & 0.83\,$\pm$\,0.09 \\
C & CO(3--2) & \phantom{0}500\,$\pm$\,90\phantom{0} & 0.88\,$\pm$\,0.11 & 3.65\,$\pm$\,0.48 \\
 & CO(7--6) &  & 0.83\,$\pm$\,0.12 & 0.63\,$\pm$\,0.09 \\
 & CI(2--1) &  & 0.59\,$\pm$\,0.12 & 0.45\,$\pm$\,0.09 \\
D & CO(3--2) & \phantom{0}370\,$\pm$\,80\phantom{0} & 0.77\,$\pm$\,0.12 & 3.20\,$\pm$\,0.50 \\
 & CO(7--6) &  & 0.97\,$\pm$\,0.10 & 0.75\,$\pm$\,0.08 \\
 & CI(2--1) &  & 0.62\,$\pm$\,0.10 & 0.47\,$\pm$\,0.08 \\
E & CO(3--2) & 1150\,$\pm$\,220 & 1.31\,$\pm$\,0.17 & 5.48\,$\pm$\,0.71 \\
 & CO(7--6) &  & 0.87\,$\pm$\,0.18 & 0.67\,$\pm$\,0.14 \\
 & CI(2--1) &  & 0.88\,$\pm$\,0.18 & 0.67\,$\pm$\,0.14 \\
F & CO(7--6) & \phantom{0}600\,$\pm$\,80\phantom{0} & 1.83\,$\pm$\,0.22 & 1.42\,$\pm$\,0.17 \\
G & CO(3--2) & \phantom{0}520\,$\pm$\,50\phantom{0} & 0.97\,$\pm$\,0.07 & 4.08\,$\pm$\,0.29 \\
 & CO(7--6) &  & 2.30\,$\pm$\,0.30 & 1.78\,$\pm$\,0.23 \\
 & CI(2--1) &  & 1.03\,$\pm$\,0.27 & 0.79\,$\pm$\,0.21 \\
H & CO(3--2) & \phantom{0}610\,$\pm$\,250 & 0.44\,$\pm$\,0.14 & 1.87\,$\pm$\,0.59 \\
 & CO(7--6) &  & 0.28\,$\pm$\,0.16 & 0.22\,$\pm$\,0.13 \\
I & CO(8--7) & \phantom{0}150\,$\pm$\,20\phantom{0} & 1.01\,$\pm$\,0.08 & 0.6\,$\pm$\,0.05 \\
J & CO(3--2) & \phantom{0}580\,$\pm$\,190 & 0.48\,$\pm$\,0.12 & 2.03\,$\pm$\,0.50 \\
 & CI(2--1) &  & 0.45\,$\pm$\,0.13 & 0.34\,$\pm$\,0.10 \\
K & CO(3--2) & \phantom{0}420\,$\pm$\,100 & 0.32\,$\pm$\,0.06 & 1.34\,$\pm$\,0.27 \\
 & CO(7--6) &  & 0.73\,$\pm$\,0.24 & 0.57\,$\pm$\,0.19 \\
L & CO(3--2) & \phantom{0}260\,$\pm$\,40\phantom{0} & 0.51\,$\pm$\,0.05 & 2.16\,$\pm$\,0.23 \\
 & CO(7--6) &  & 0.62\,$\pm$\,0.15 & 0.48\,$\pm$\,0.12 \\
 & CI(2--1) &  & 0.53\,$\pm$\,0.13 & 0.41\,$\pm$\,0.10 \\
M & CO(3--2) & \phantom{0}300\,$\pm$\,70\phantom{0} & 0.59\,$\pm$\,0.10 & 2.50\,$\pm$\,0.42 \\
 & CI(2--1) &  & 0.56\,$\pm$\,0.12 & 0.43\,$\pm$\,0.10 \\
N & CO(3--2) & \phantom{0}500\,$\pm$\,120 & 0.80\,$\pm$\,0.14 & 3.42\,$\pm$\,0.59 \\
 & CO(7--6) &  & 1.23\,$\pm$\,0.45 & 0.97\,$\pm$\,0.35 \\
 & CI(2--1) &  & 0.88\,$\pm$\,0.15 & 0.69\,$\pm$\,0.12 \\
O & CO(3--2) & \phantom{0}700\,$\pm$\,130 & 1.09\,$\pm$\,0.14 & 4.68\,$\pm$\,0.59 \\
 & CI(2--1) &  & 0.85\,$\pm$\,0.19 & 0.66\,$\pm$\,0.15 \\
P1 & CO(3--2) & \phantom{0}180\,$\pm$\,50\phantom{0} & 0.40\,$\pm$\,0.07 & 1.73\,$\pm$\,0.30 \\
 & CO(7--6) &  & 0.65\,$\pm$\,0.11 & 0.52\,$\pm$\,0.09 \\
 & CI(2--1) &  & 0.35\,$\pm$\,0.10 & 0.28\,$\pm$\,0.08 \\
P2 & CO(3--2) & \phantom{0}410\,$\pm$\,90\phantom{0} & 0.80\,$\pm$\,0.11 & 3.45\,$\pm$\,0.48 \\
 & CI(2--1) &  & 0.32\,$\pm$\,0.15 & 0.26\,$\pm$\,0.12 \\
Q & CO(3--2) & \phantom{0}180\,$\pm$\,60\phantom{0} & 0.50\,$\pm$\,0.06 & 2.21\,$\pm$\,0.28 \\
\hline
\end{tabular}

 \caption{Observed line properties of the SMGs ranked in ascending redshift order.\label{tab: lines}}
\end{table*}

\begin{figure*}
    \centering
    \begin{tabular}{cc}
    \includegraphics[width=\width]{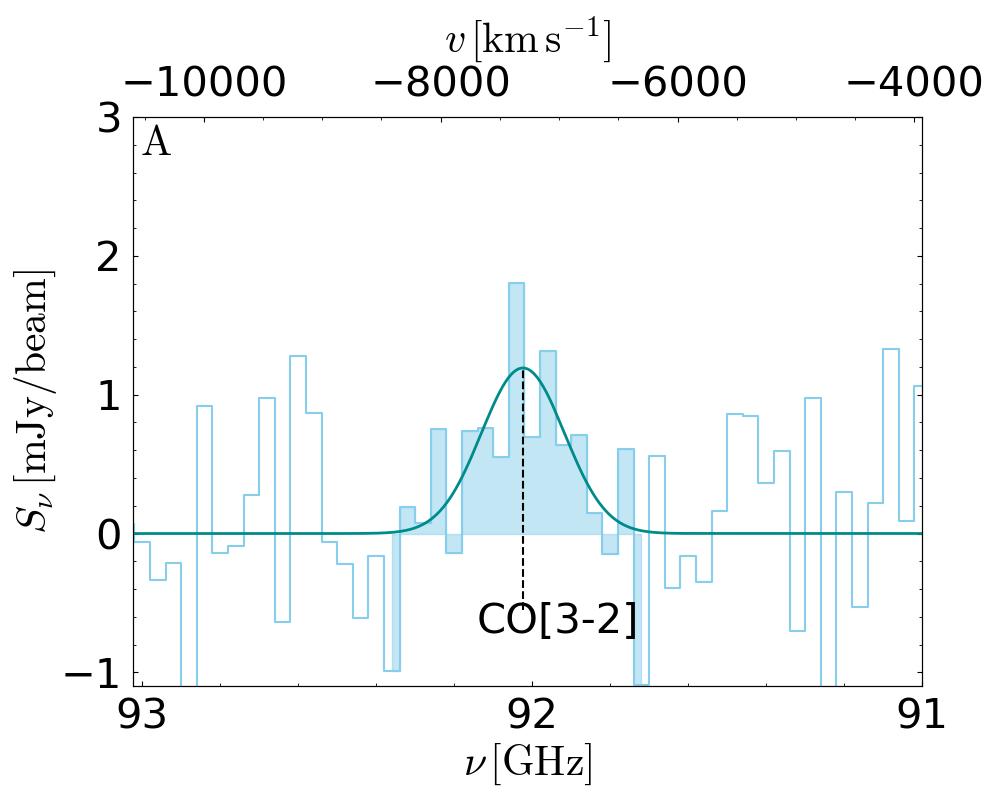} &
    \includegraphics[width=\width]{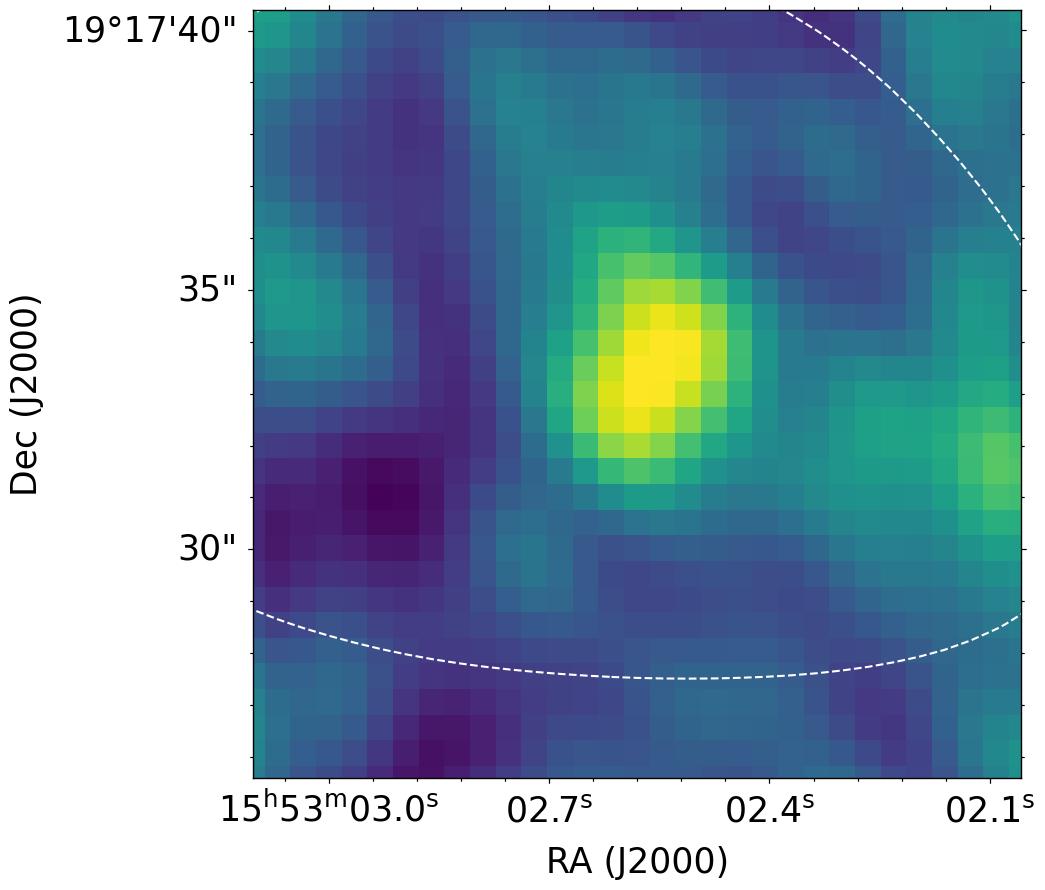} \\
    \includegraphics[width=\width]{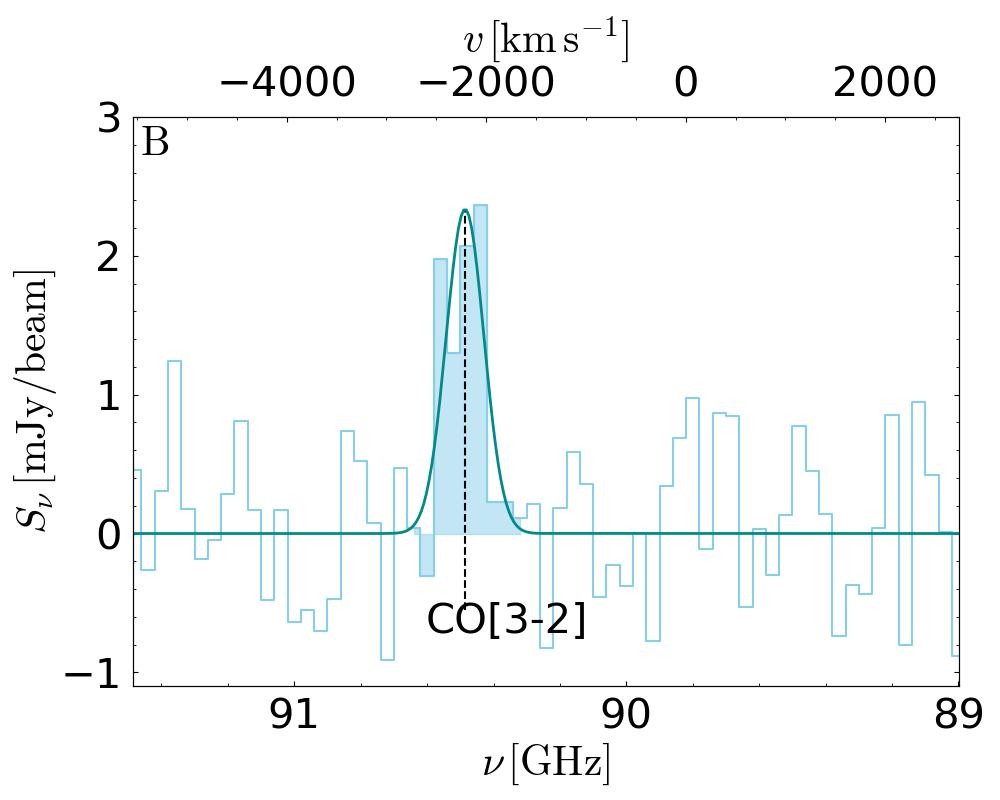} & 
    \includegraphics[width=\width]{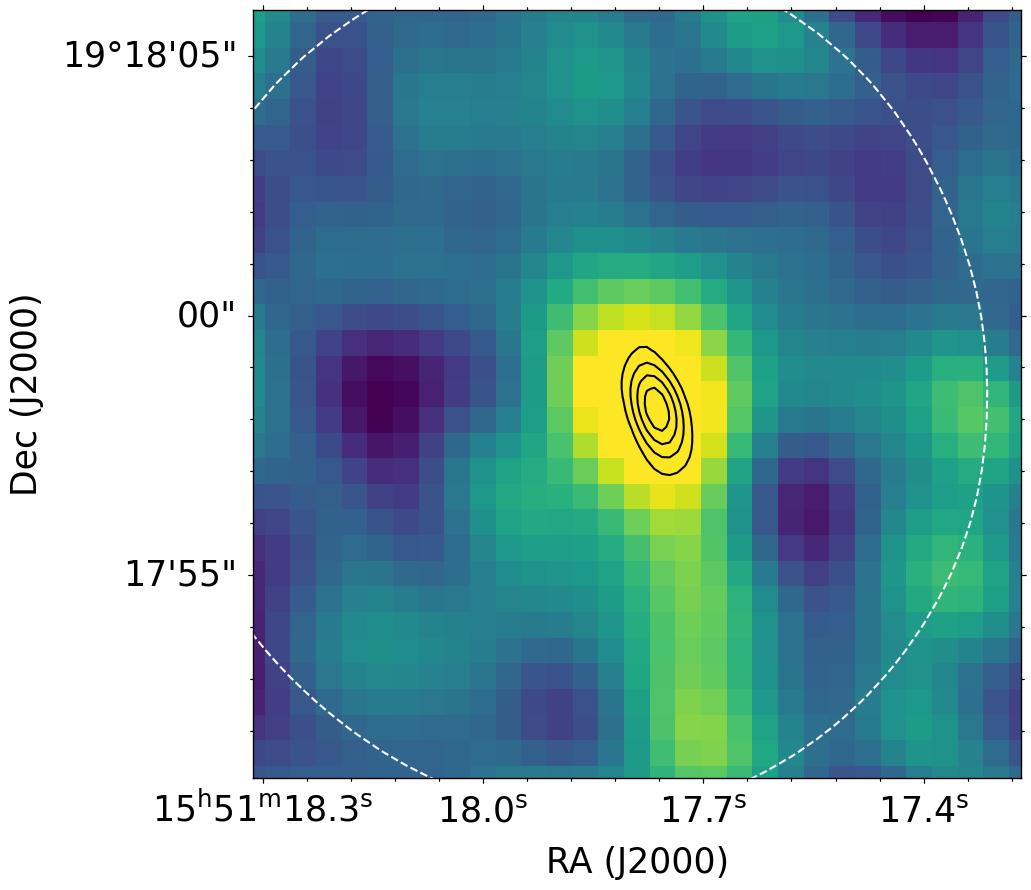} \\
    \includegraphics[width=\width]{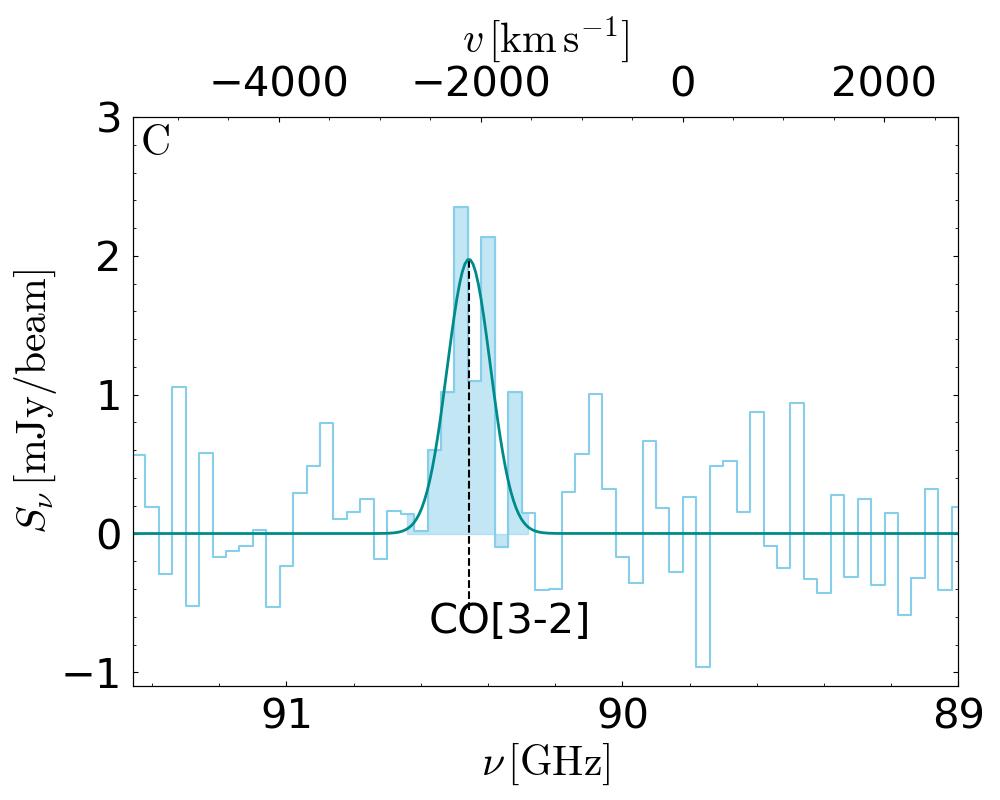} & 
    \includegraphics[width=\width]{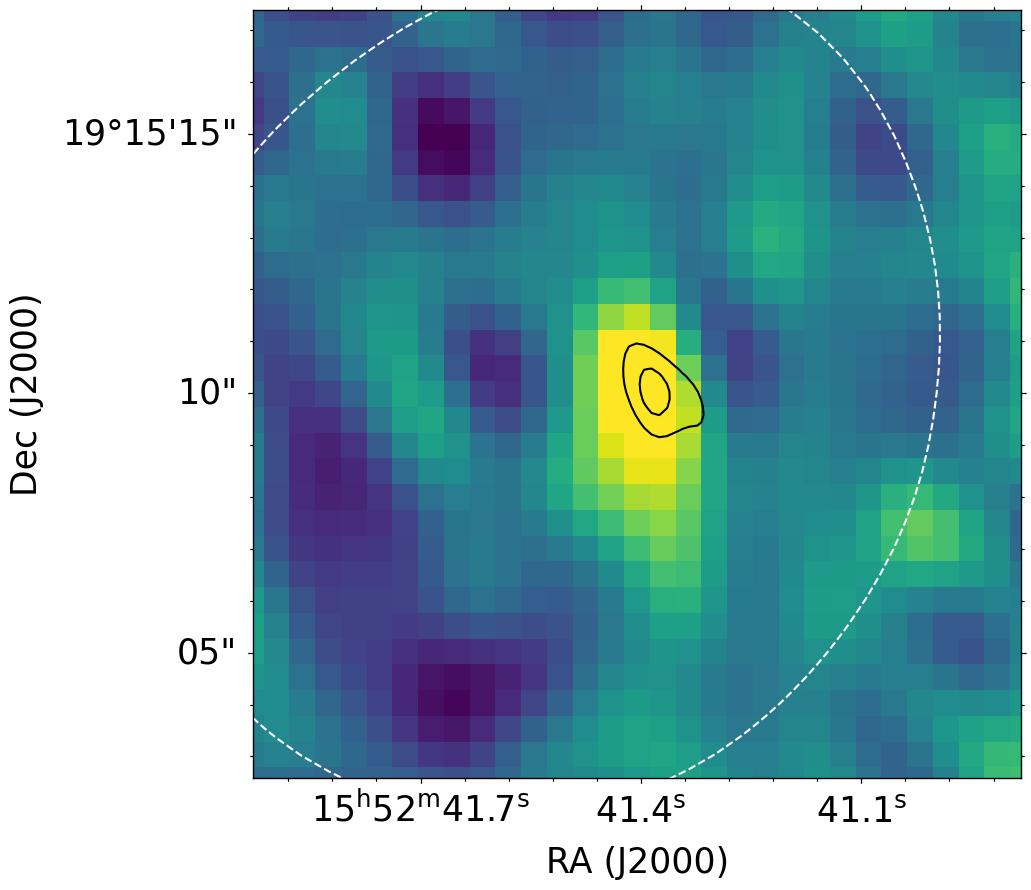} \\
    \includegraphics[width=\width]{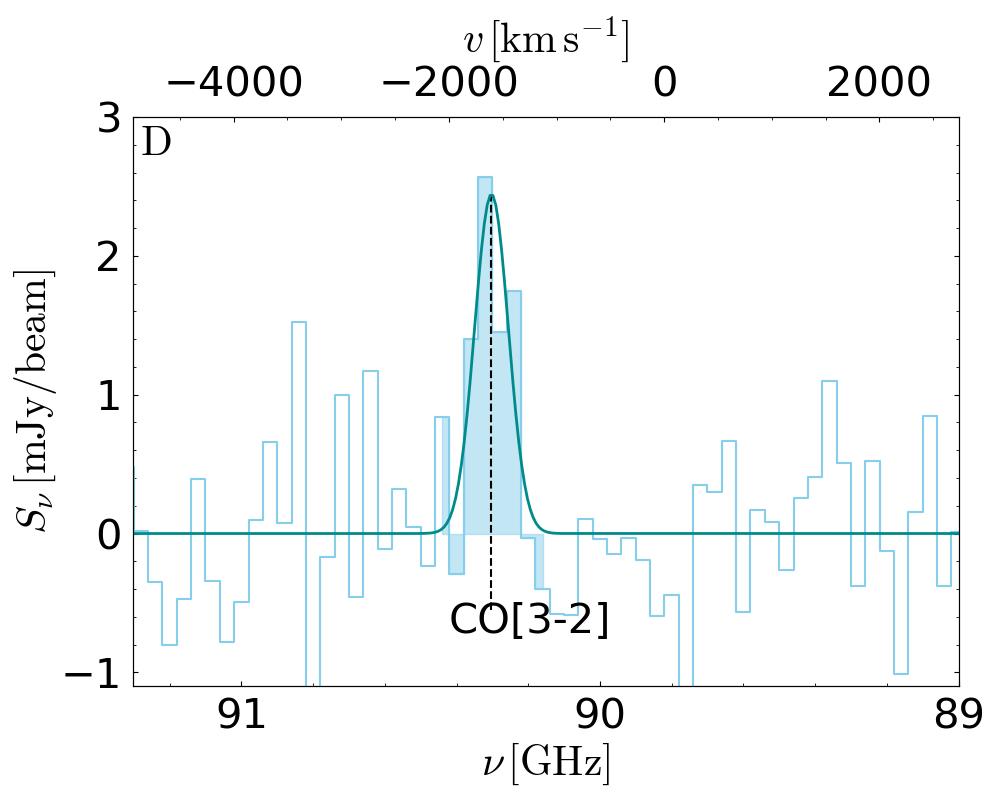} & 
    \includegraphics[width=\width]{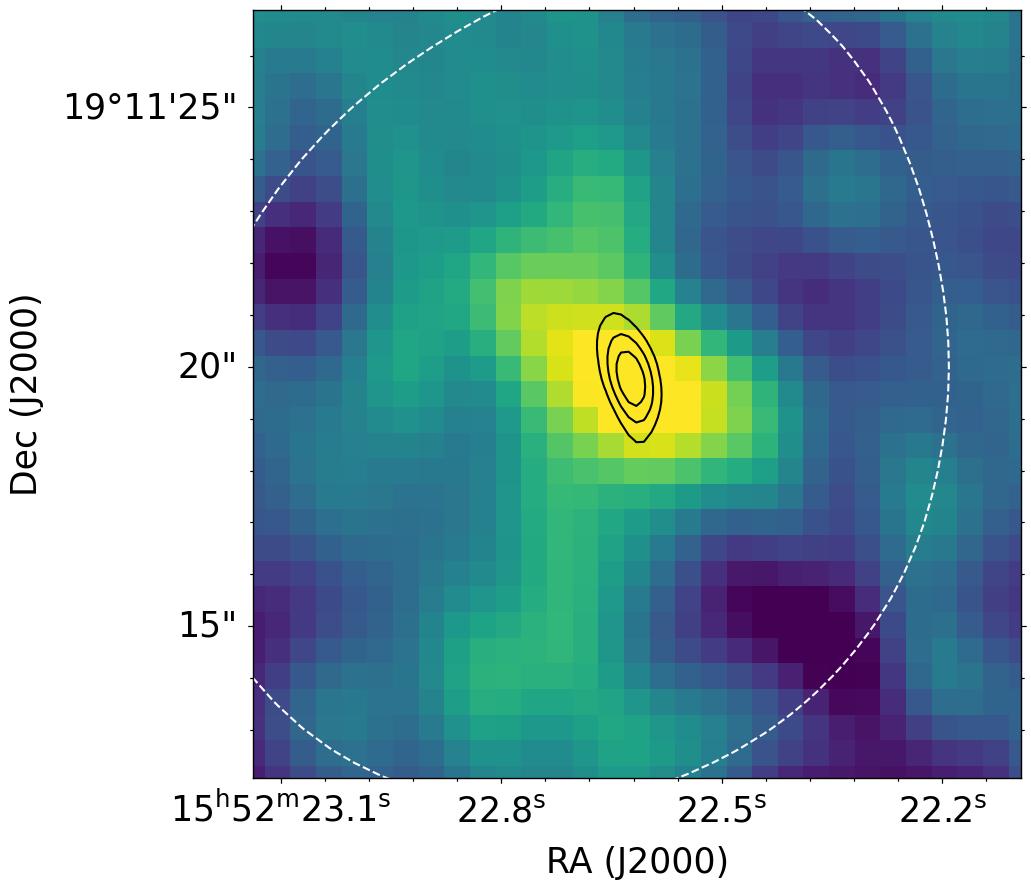} \\
    \end{tabular}
    \caption{CO(3--2) spectra and channel maps. The 850-\um\ SCUBA-2 emission is shown by white contours (70\% of the peak flux of each source). Identifications within the SCUBA-2 beam are shown with NOEMA 1.4-mm continuum contours in blue (``A'' and ``Q'' were not observed at 1.4\,mm). Spectra are extracted at the peak pixel of the channel map constructed from the identified CO(3--2) line. The spectral resolution is 60\,\kms. Gaussian fits to the line profiles are overlaid. \label{fig: CO3-2}}
\end{figure*}

\addtocounter{figure}{-1}
\begin{figure*}
    \centering
    \begin{tabular}{cc}
    \includegraphics[width=\width]{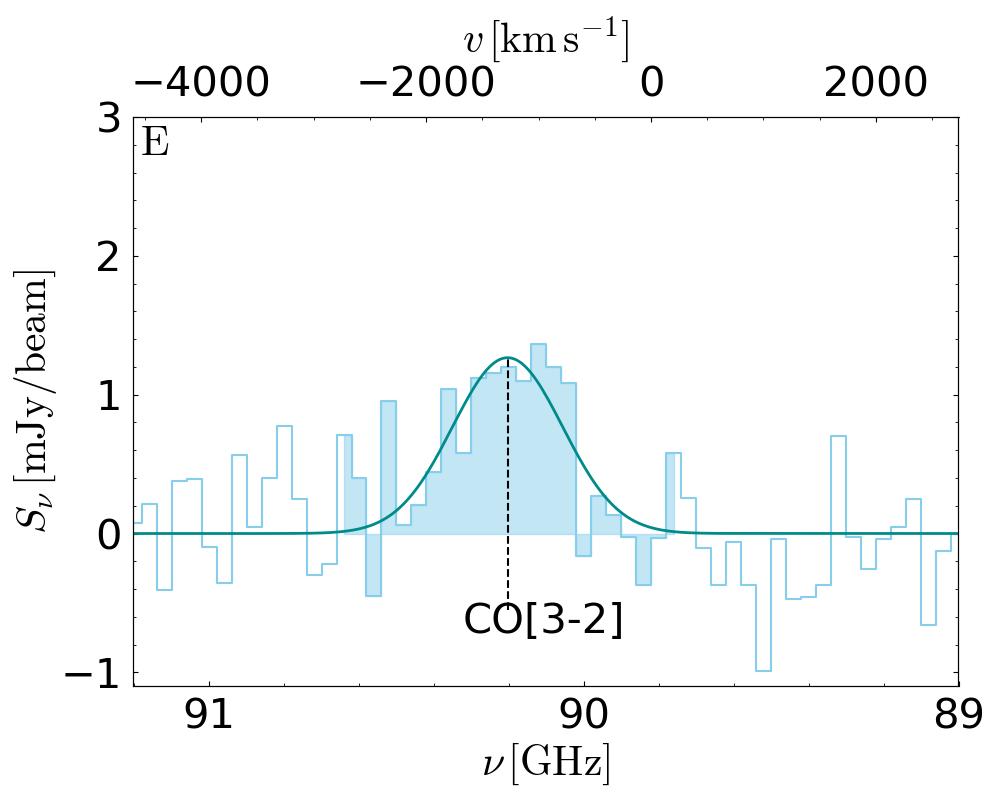} & 
    \includegraphics[width=\width]{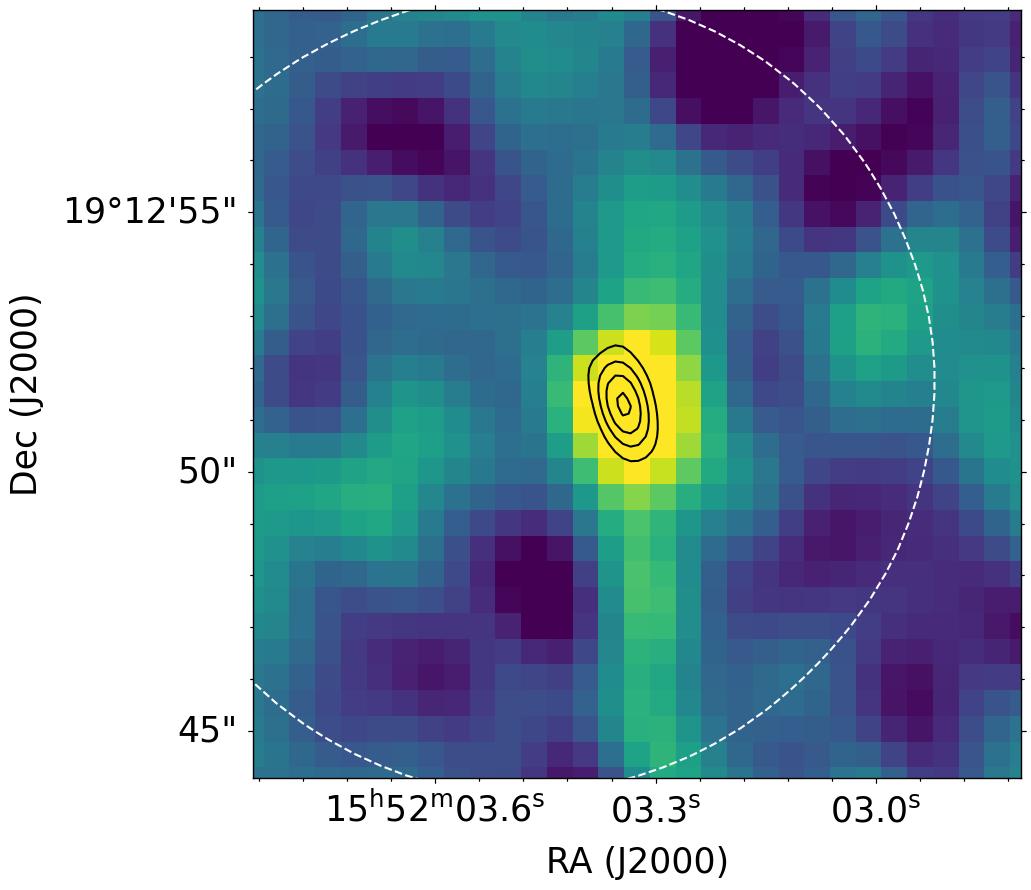} \\
    \includegraphics[width=\width]{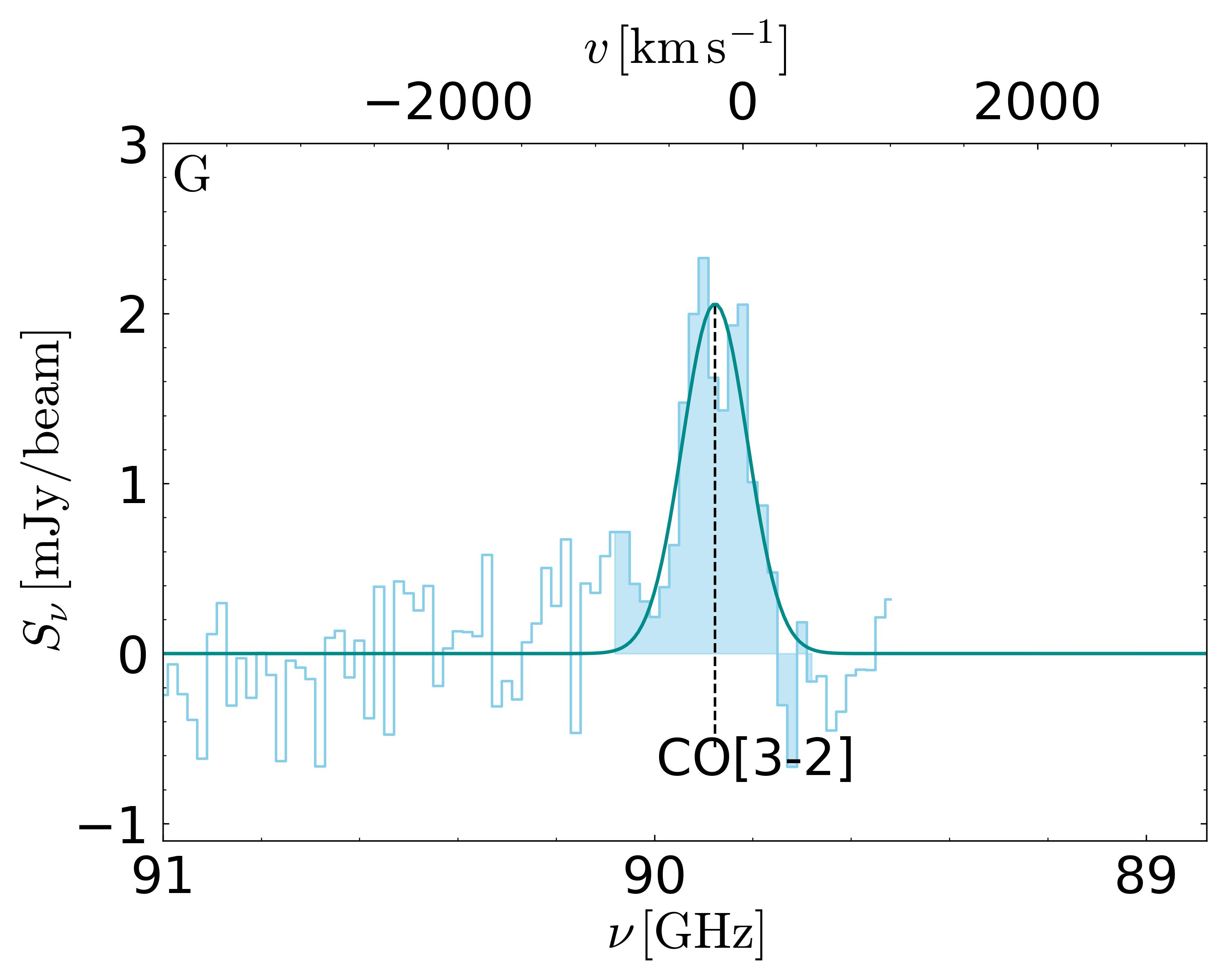} & 
    \includegraphics[width=\width]{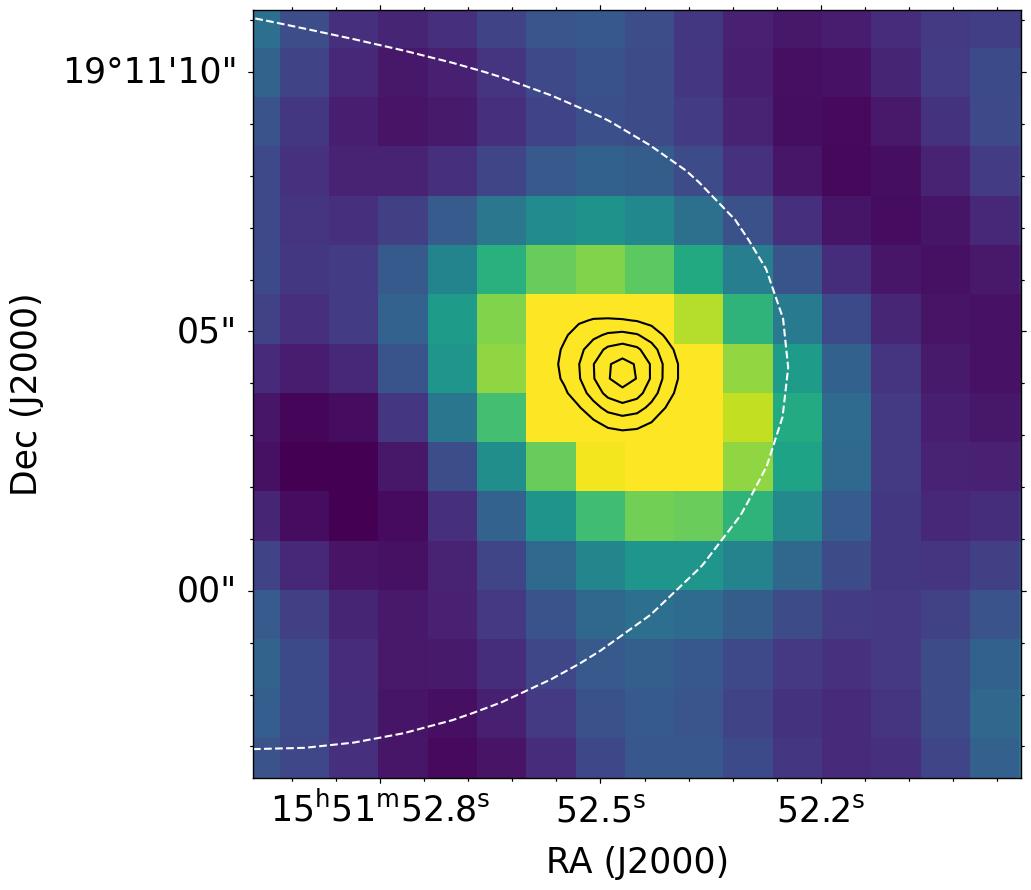} \\ 
    \includegraphics[width=\width]{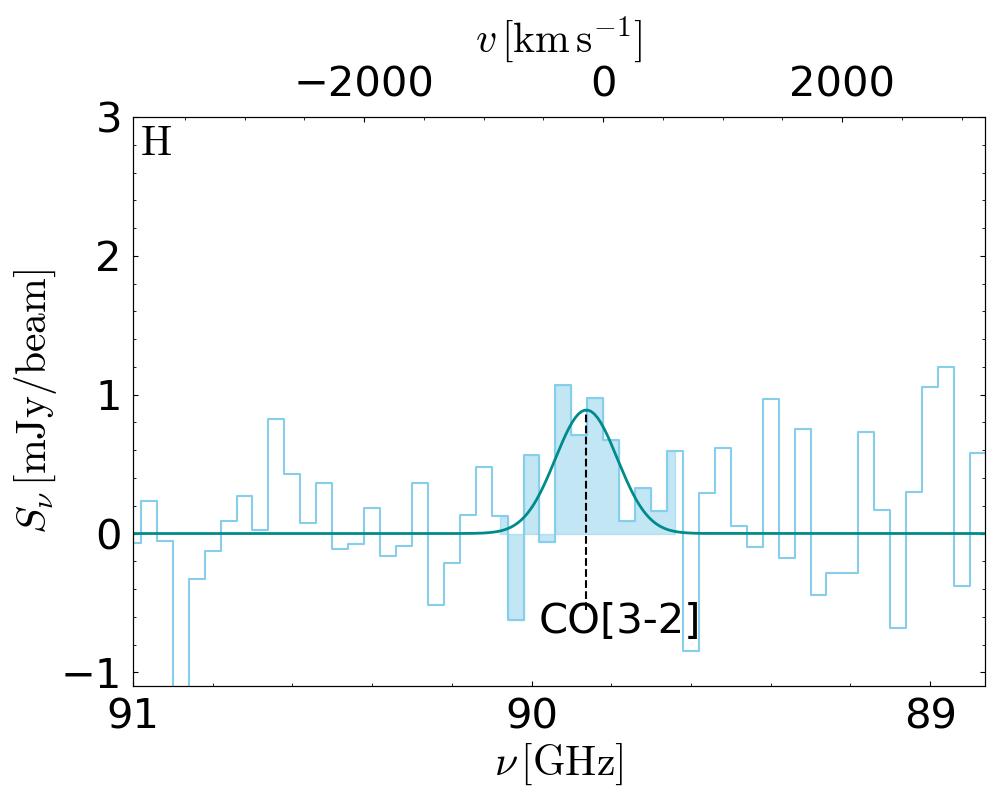} & 
    \includegraphics[width=\width]{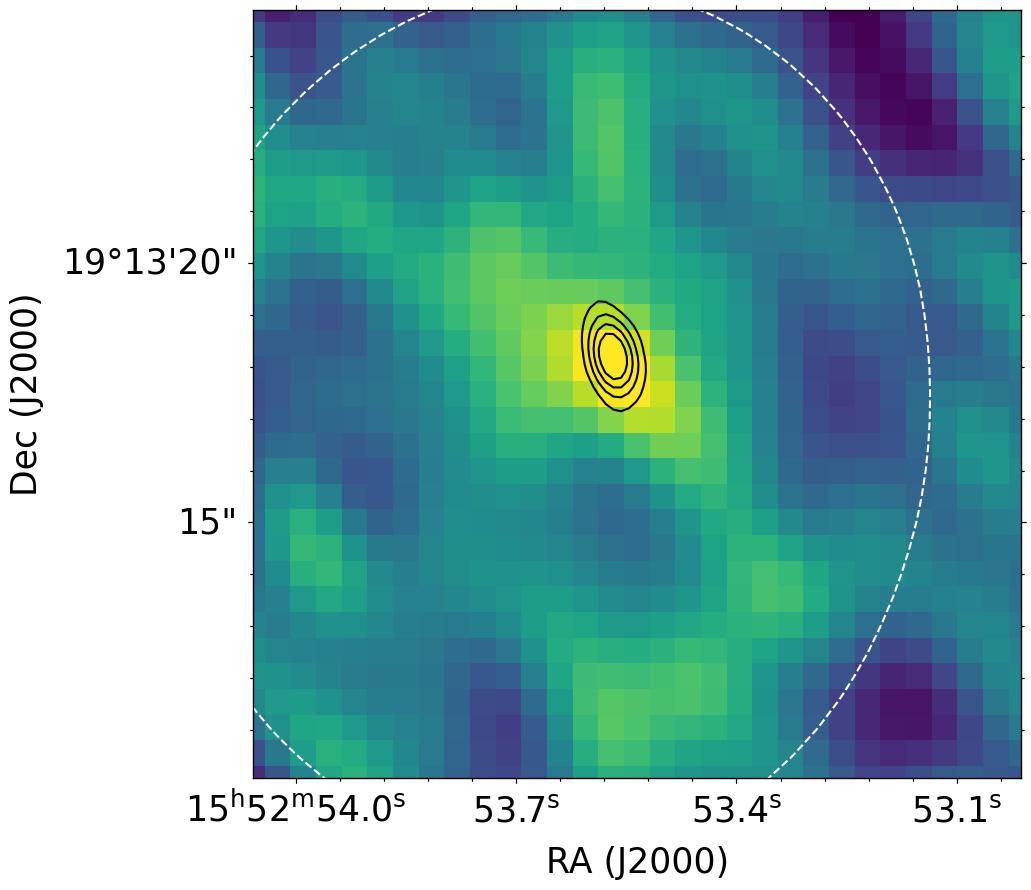} \\ 
    \includegraphics[width=\width]{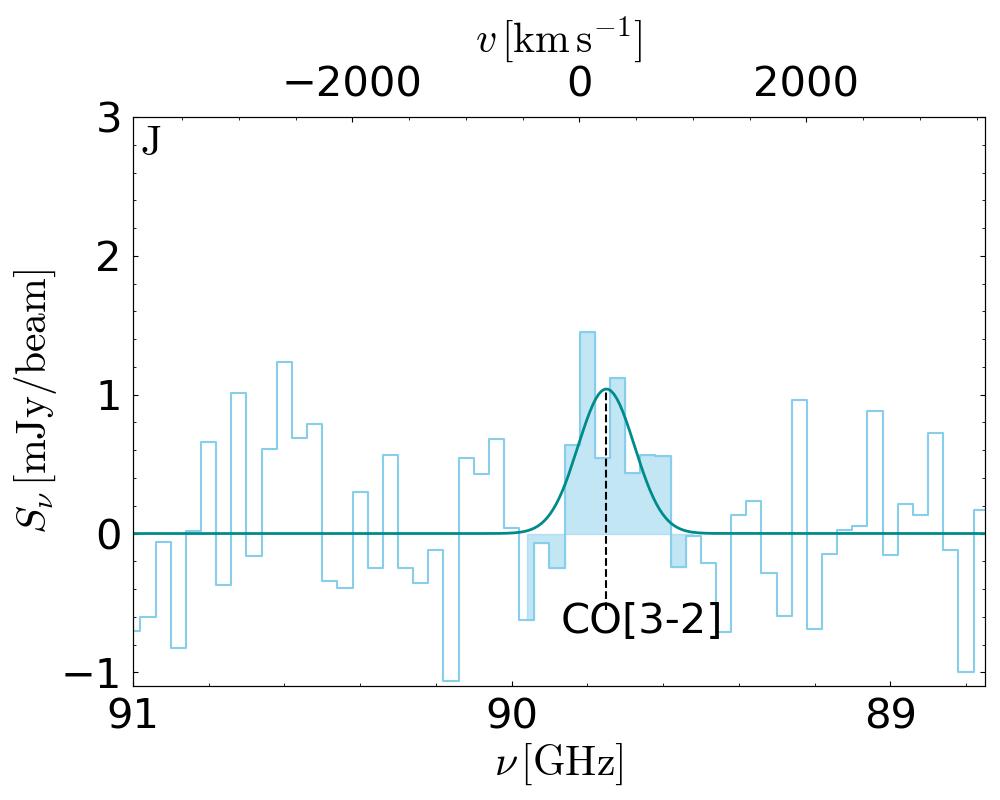} & 
    \includegraphics[width=\width]{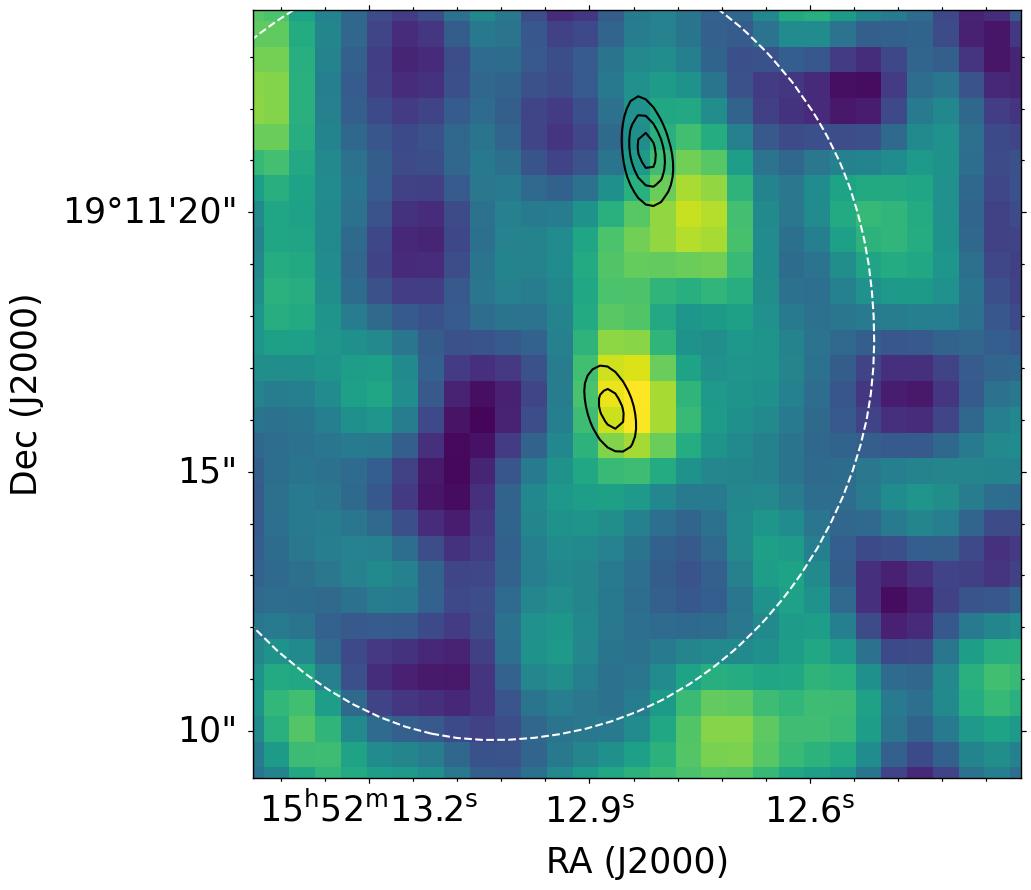} \\
    \end{tabular}
    \caption{CO(3--2) spectra and channel maps, continued.}
\end{figure*}

\addtocounter{figure}{-1}
\begin{figure*}
    \centering
    \begin{tabular}{cc}
    \includegraphics[width=\width]{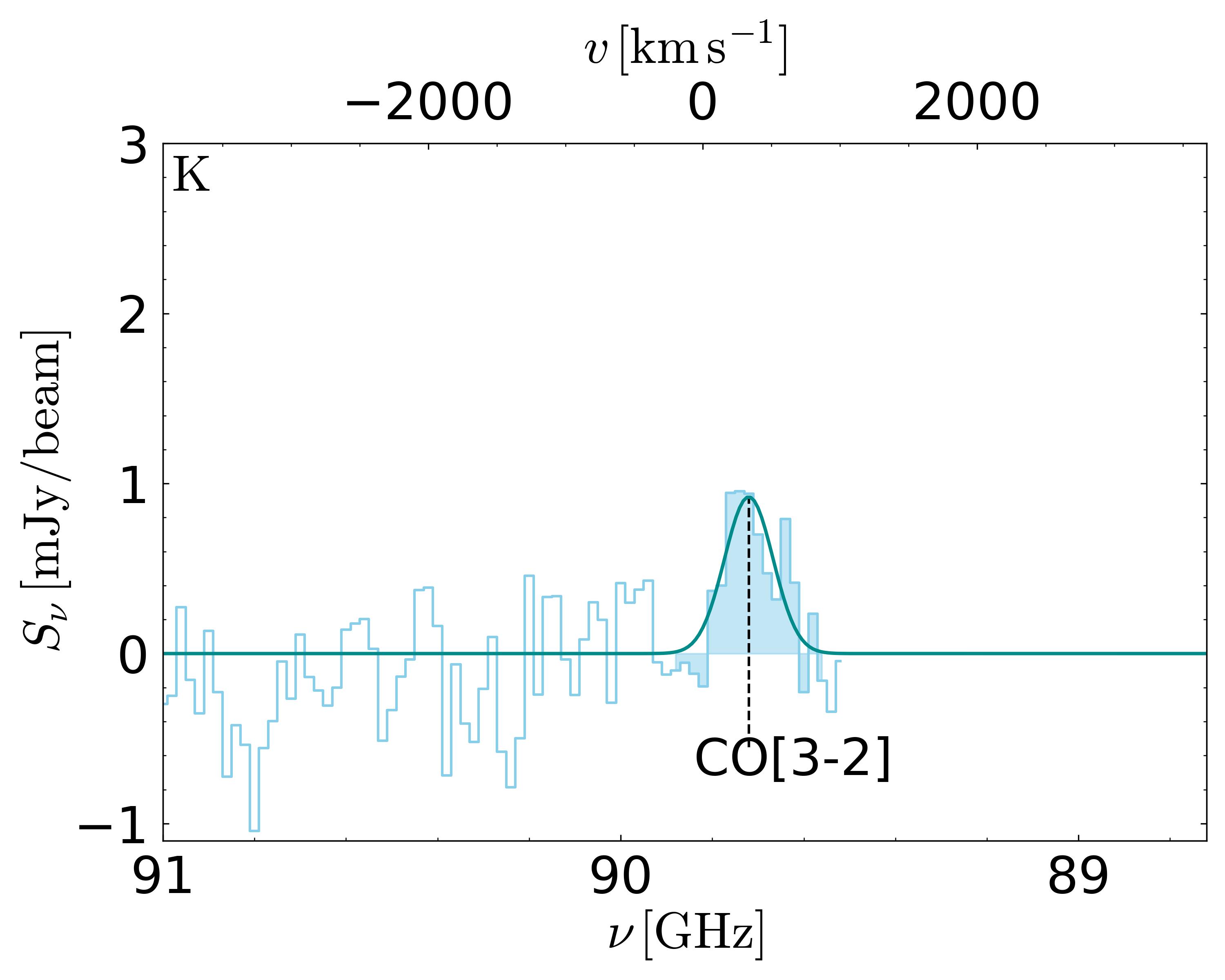} & 
    \includegraphics[width=\width]{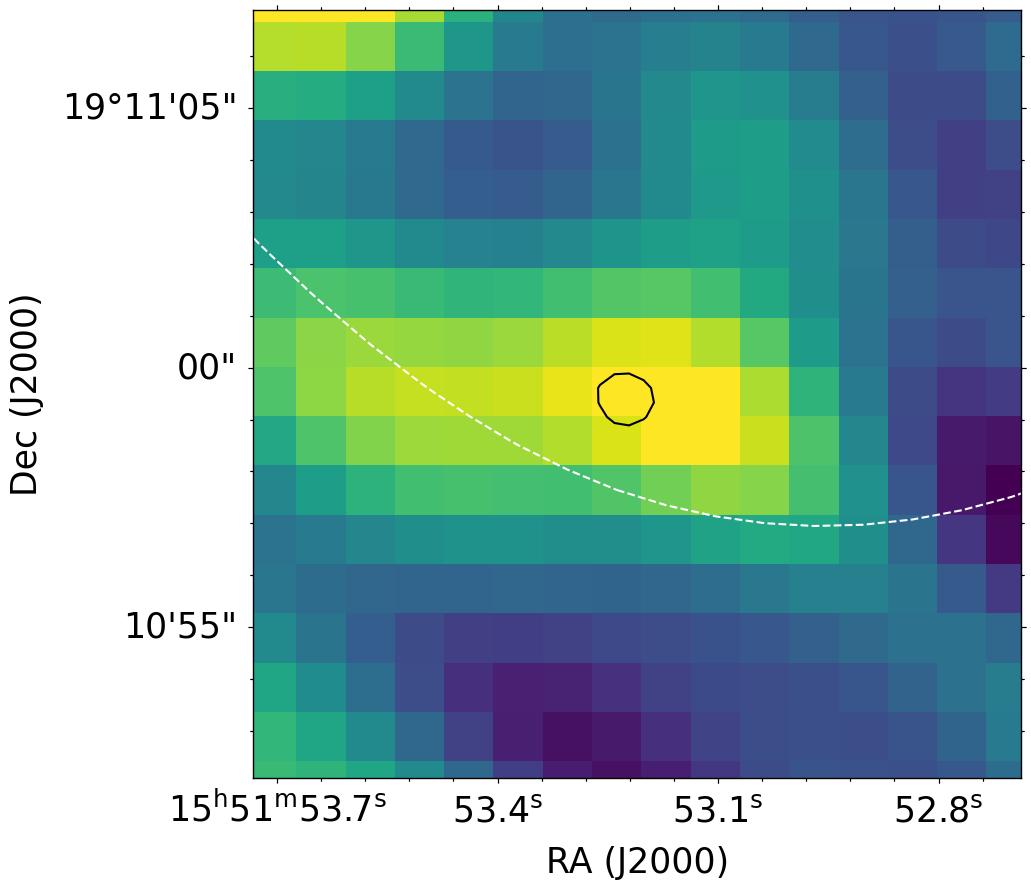} \\
    \includegraphics[width=\width]{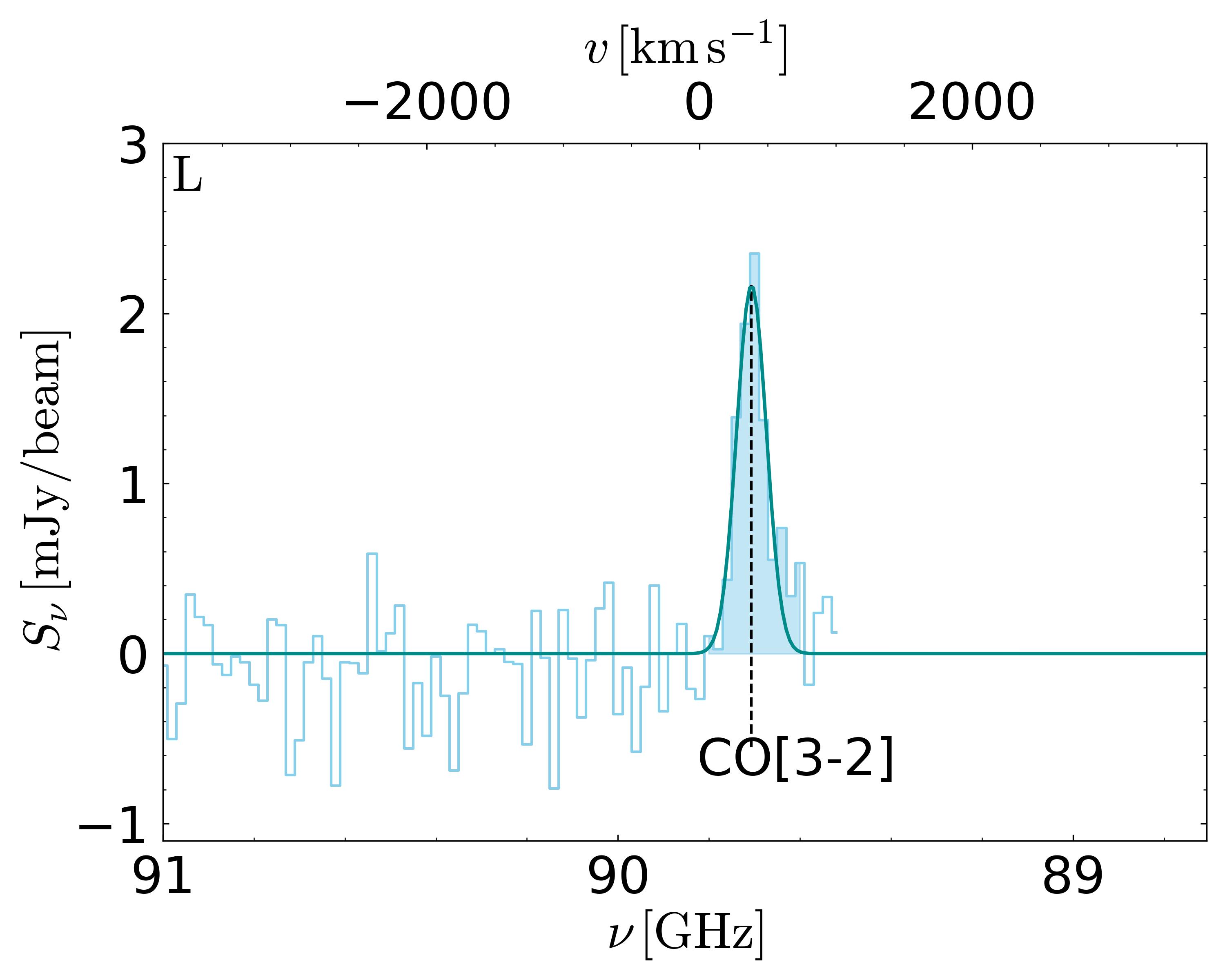} & 
    \includegraphics[width=\width]{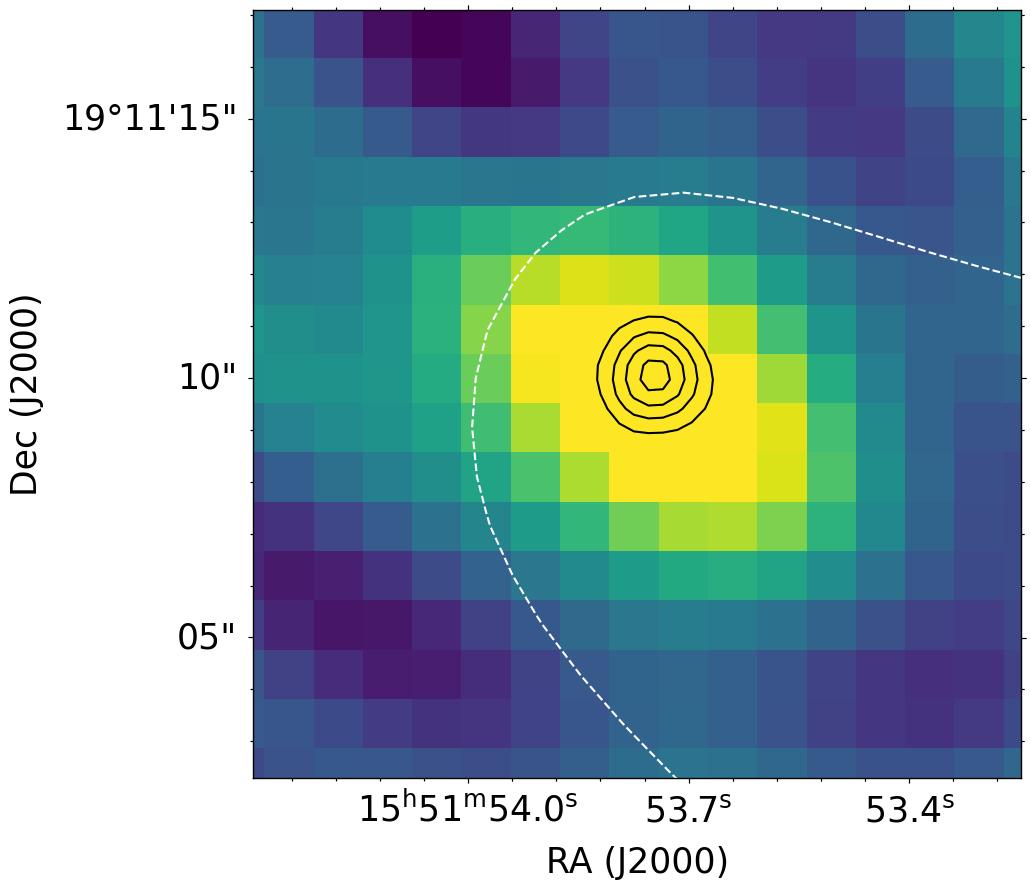} \\ 
    \includegraphics[width=\width]{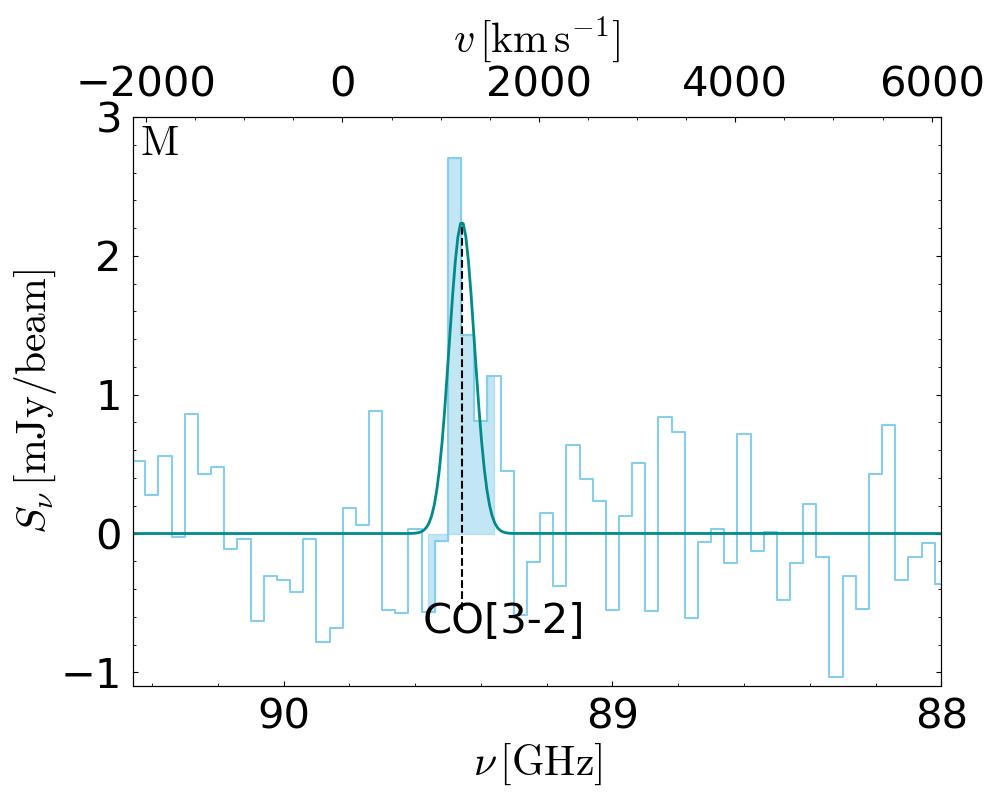} & 
    \includegraphics[width=\width]{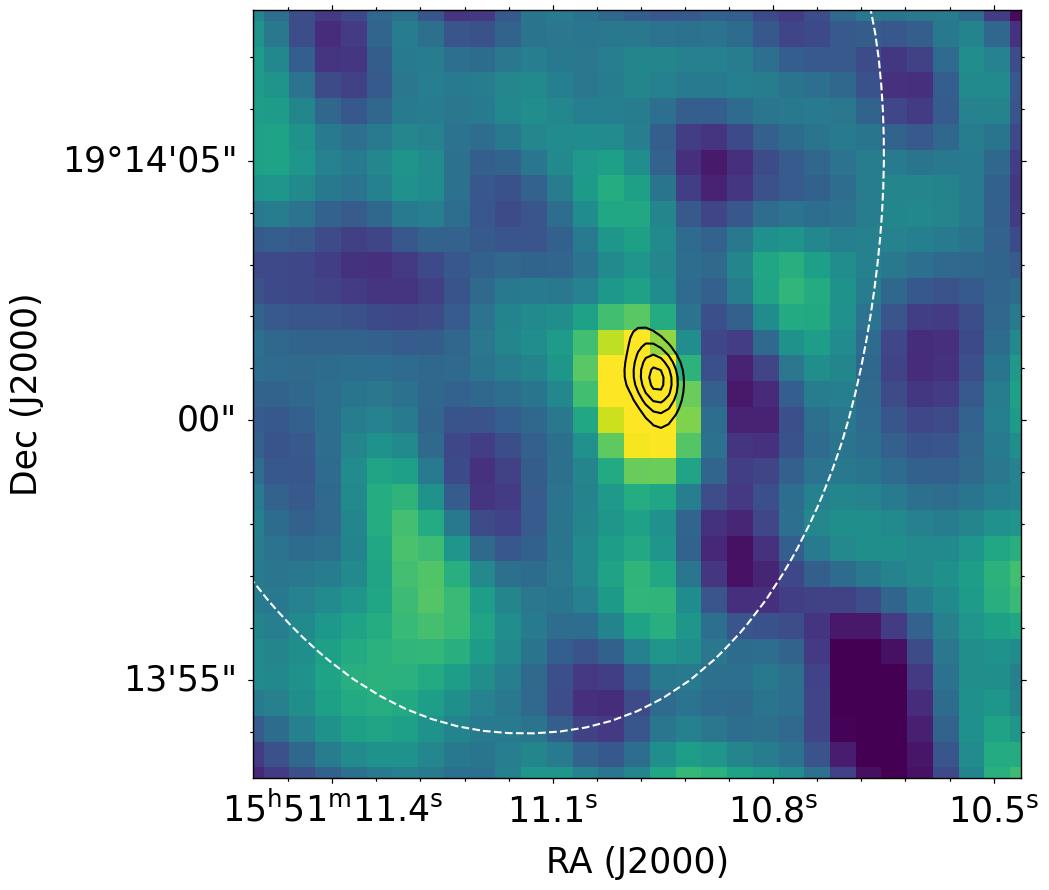} \\ 
    \includegraphics[width=\width]{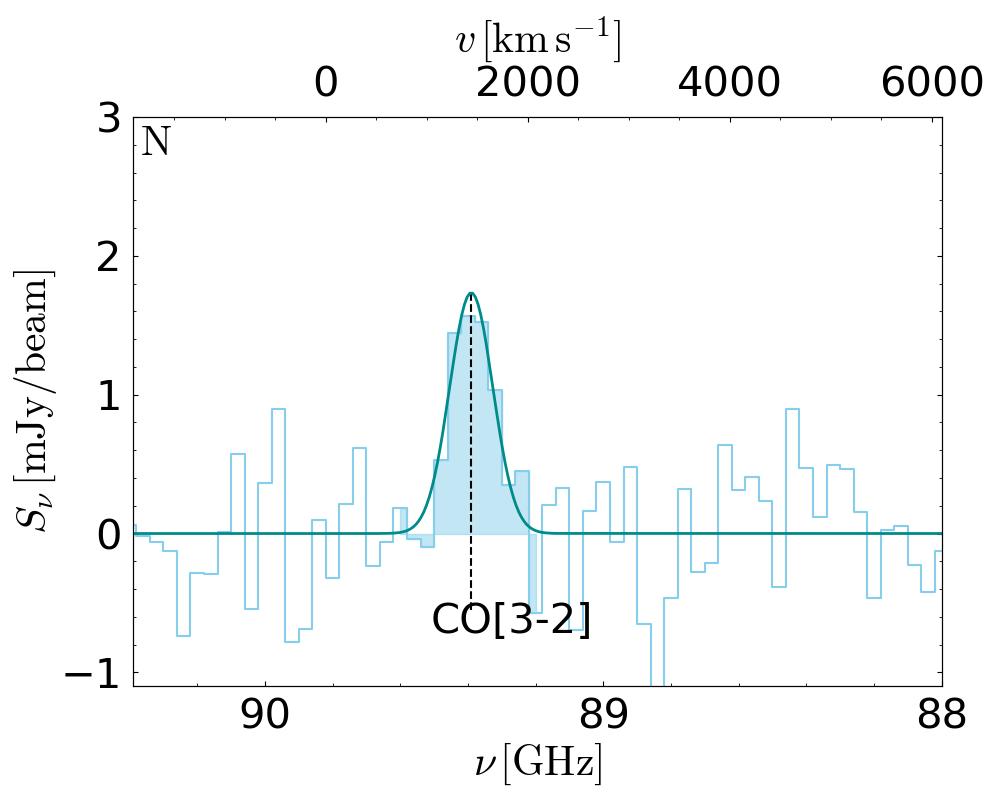} & 
    \includegraphics[width=\width]{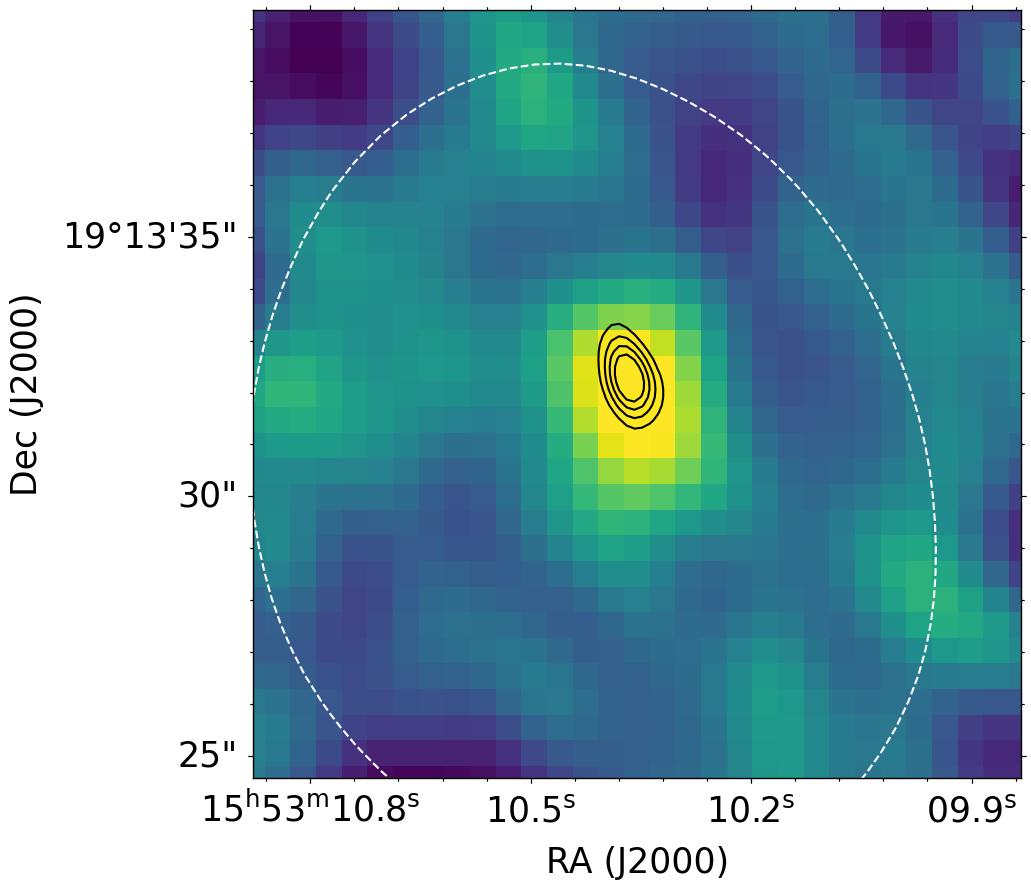} \\
    \end{tabular}
    \caption{CO(3--2) spectra and channel maps, continued.}
\end{figure*}

\addtocounter{figure}{-1}
\begin{figure*}
    \centering
    \begin{tabular}{cc}
    \includegraphics[width=\width]{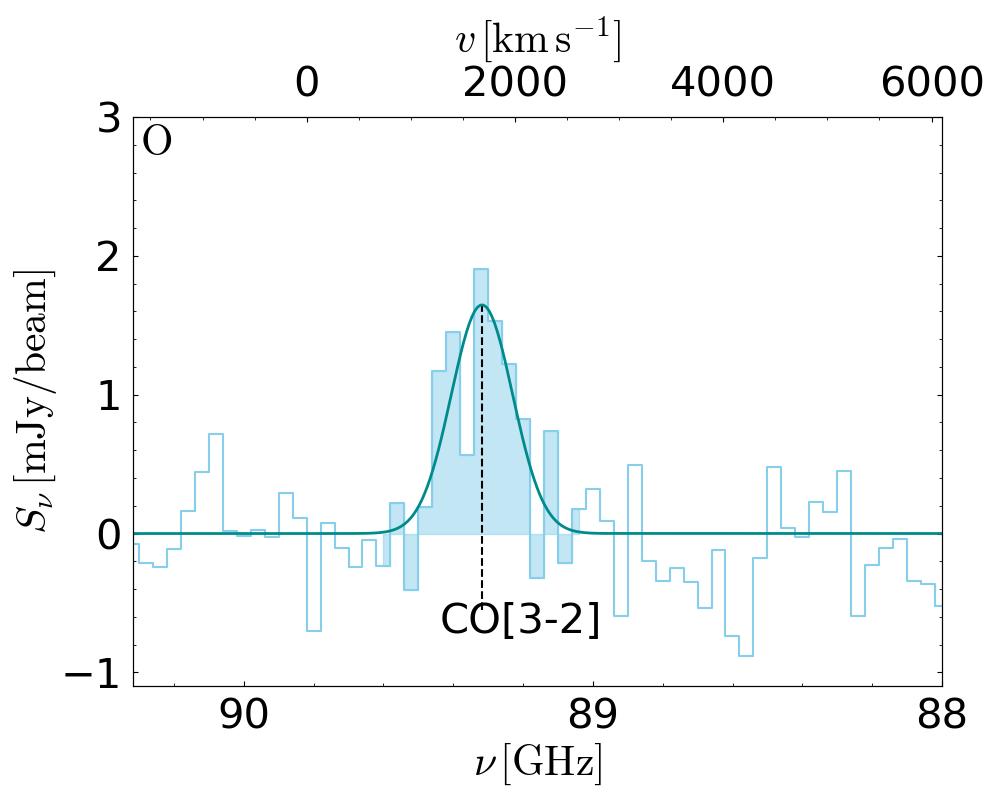} & 
    \includegraphics[width=\width]{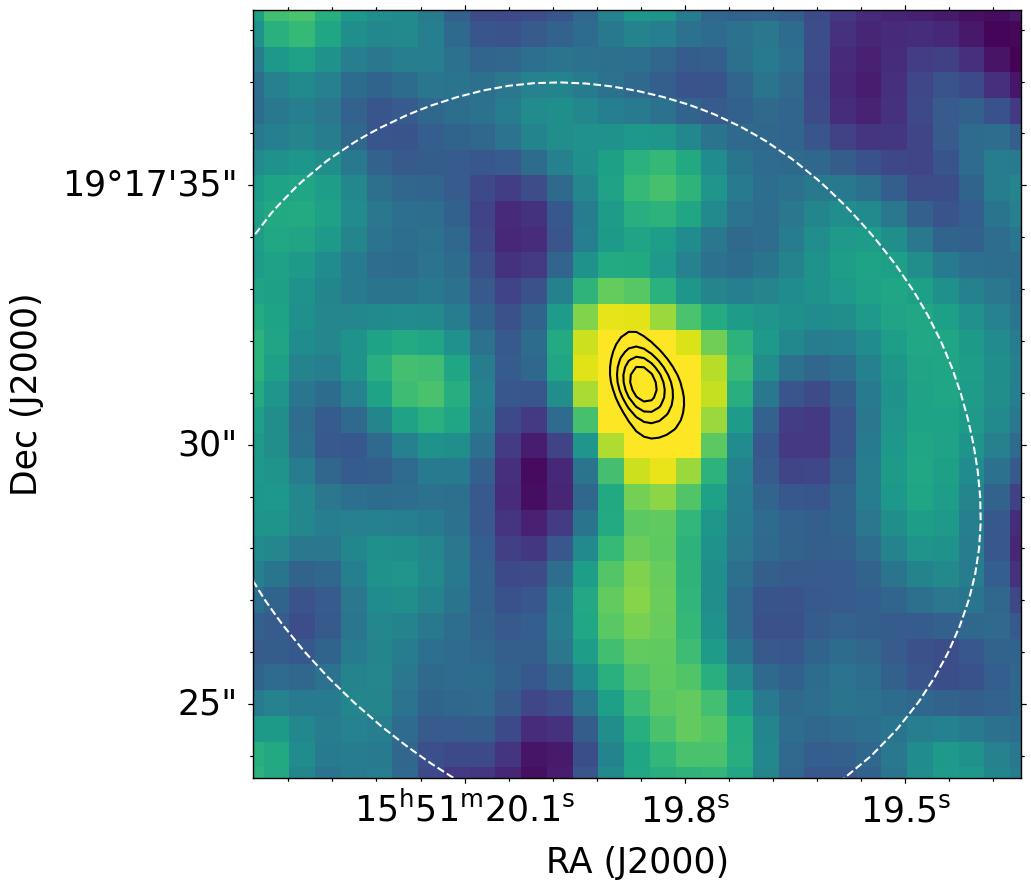} \\
    \includegraphics[width=\width]{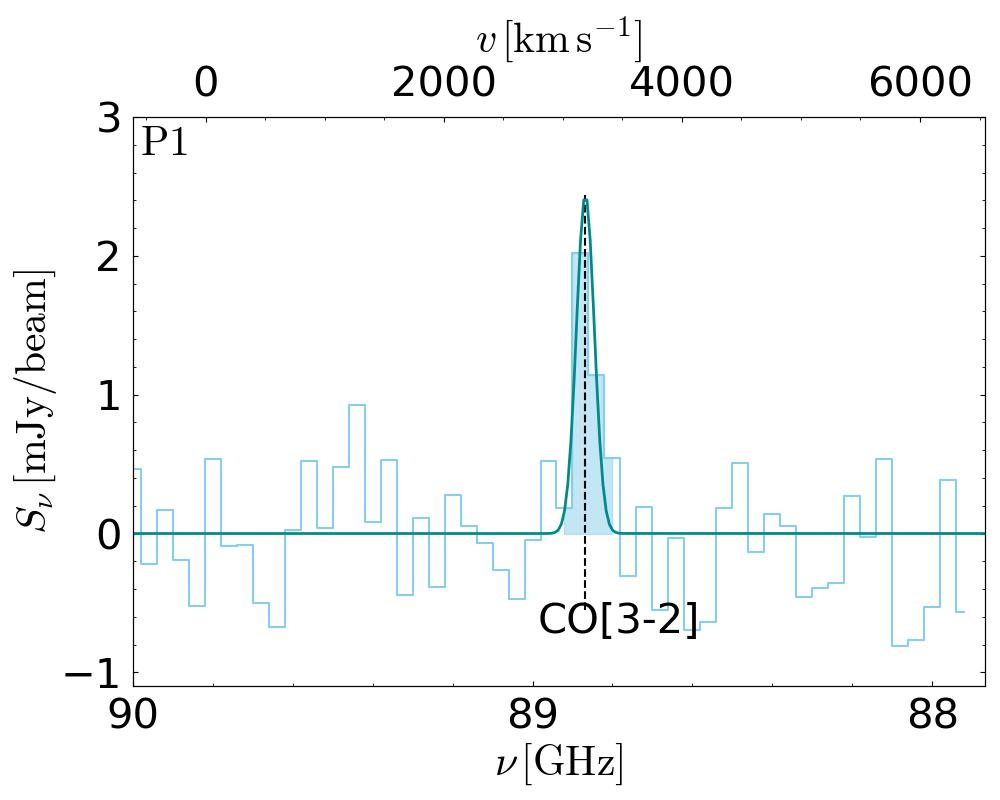} & 
    \includegraphics[width=\width]{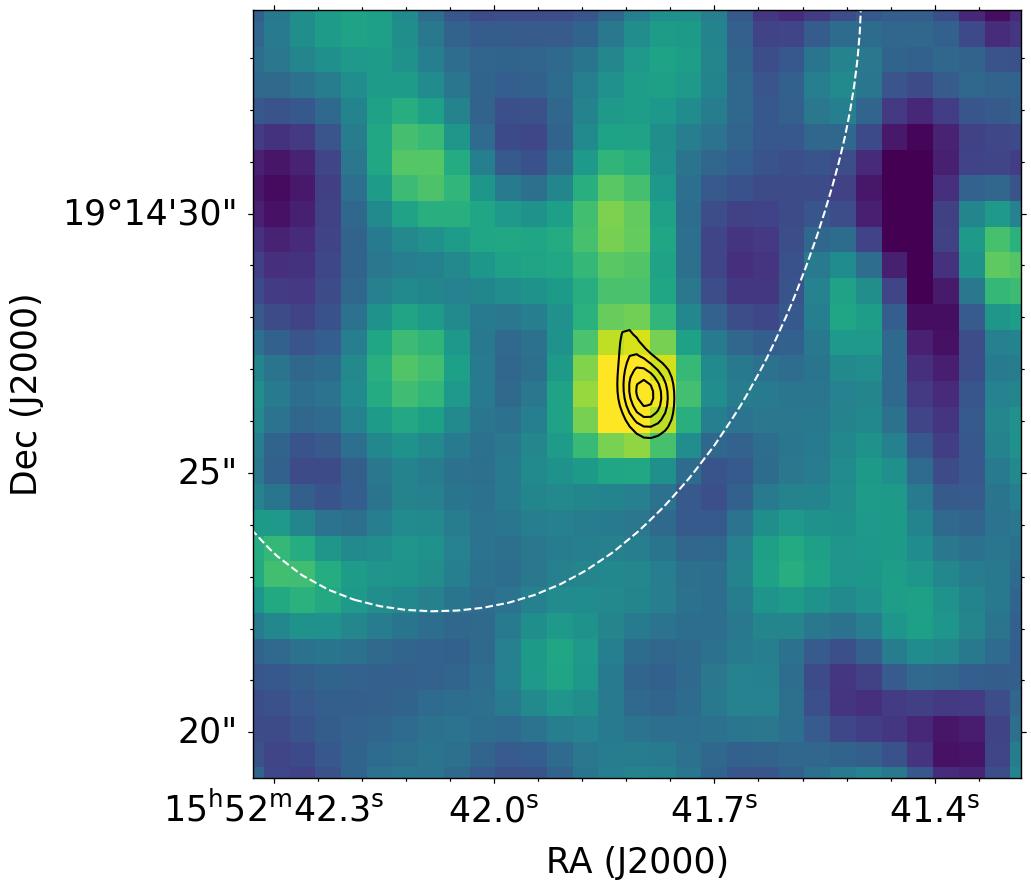} \\
    \includegraphics[width=\width]{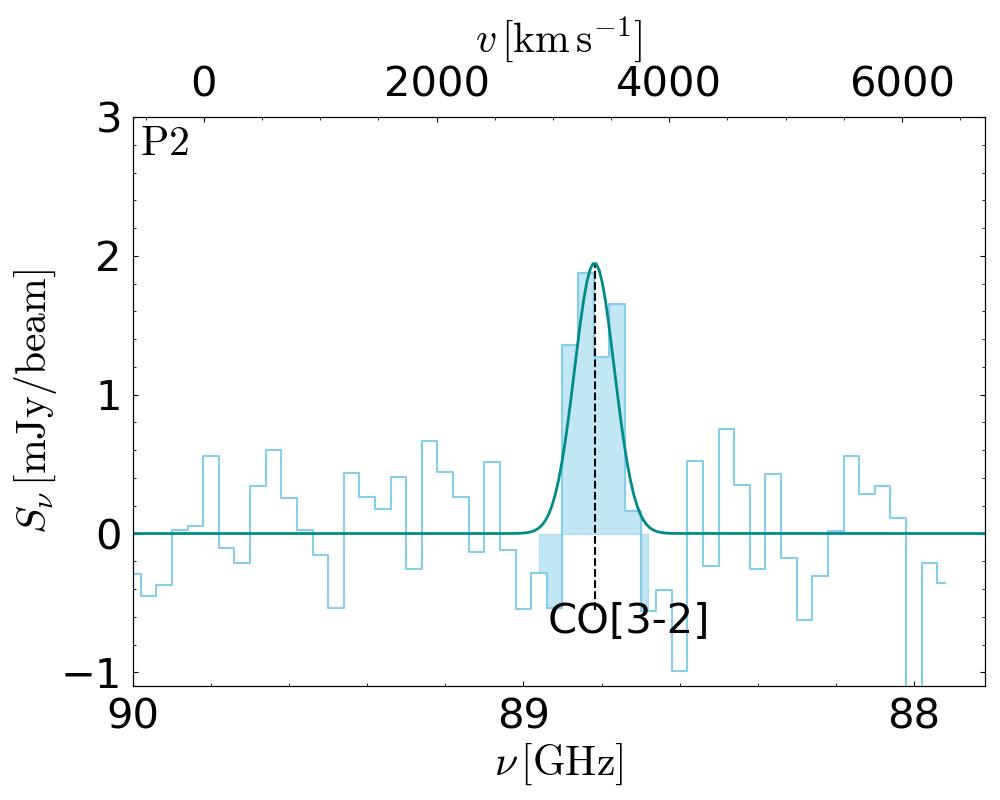} & 
    \includegraphics[width=\width]{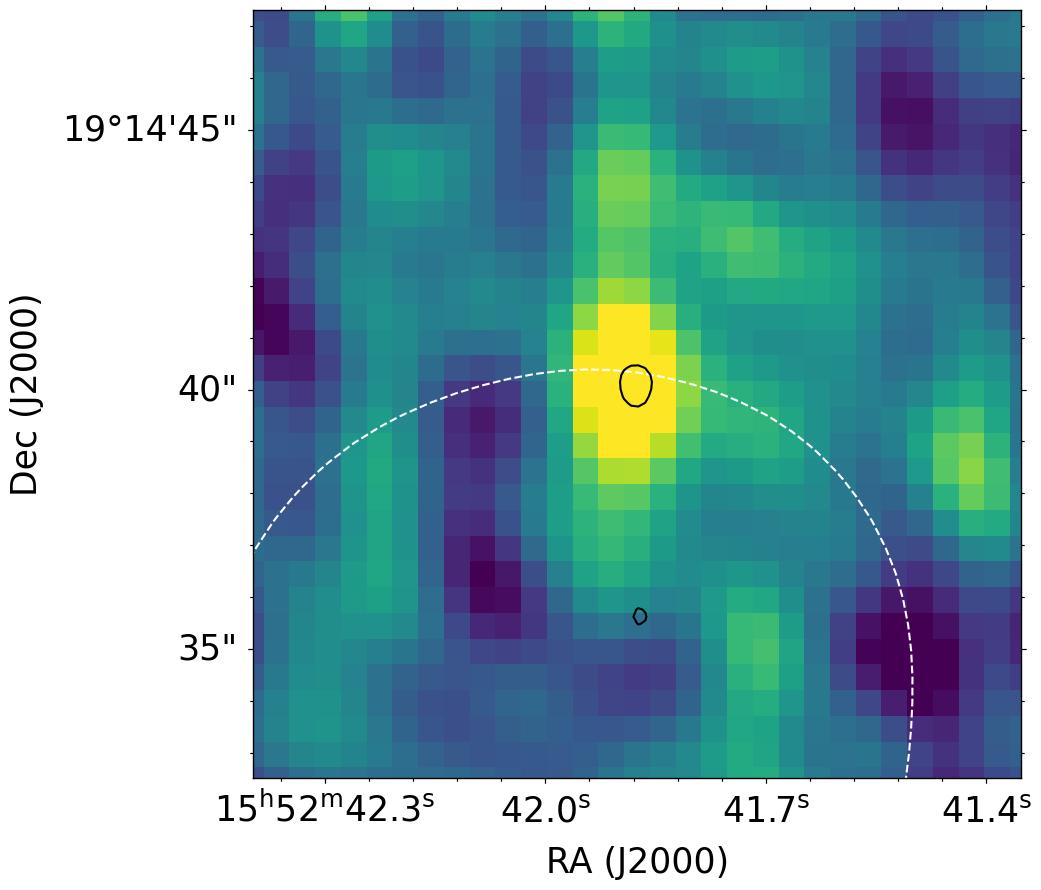} \\
    \includegraphics[width=\width]{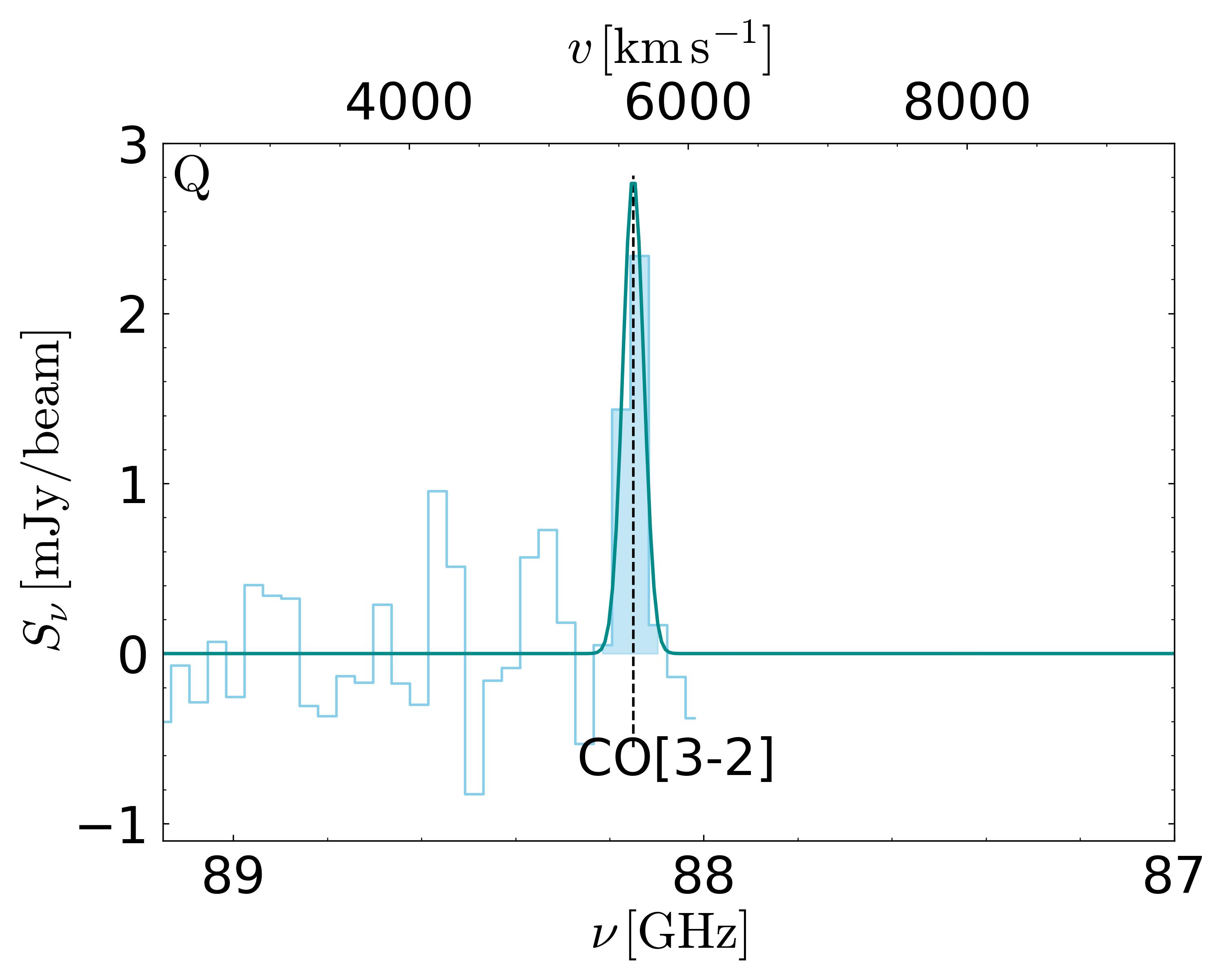} & 
    \includegraphics[width=\width]{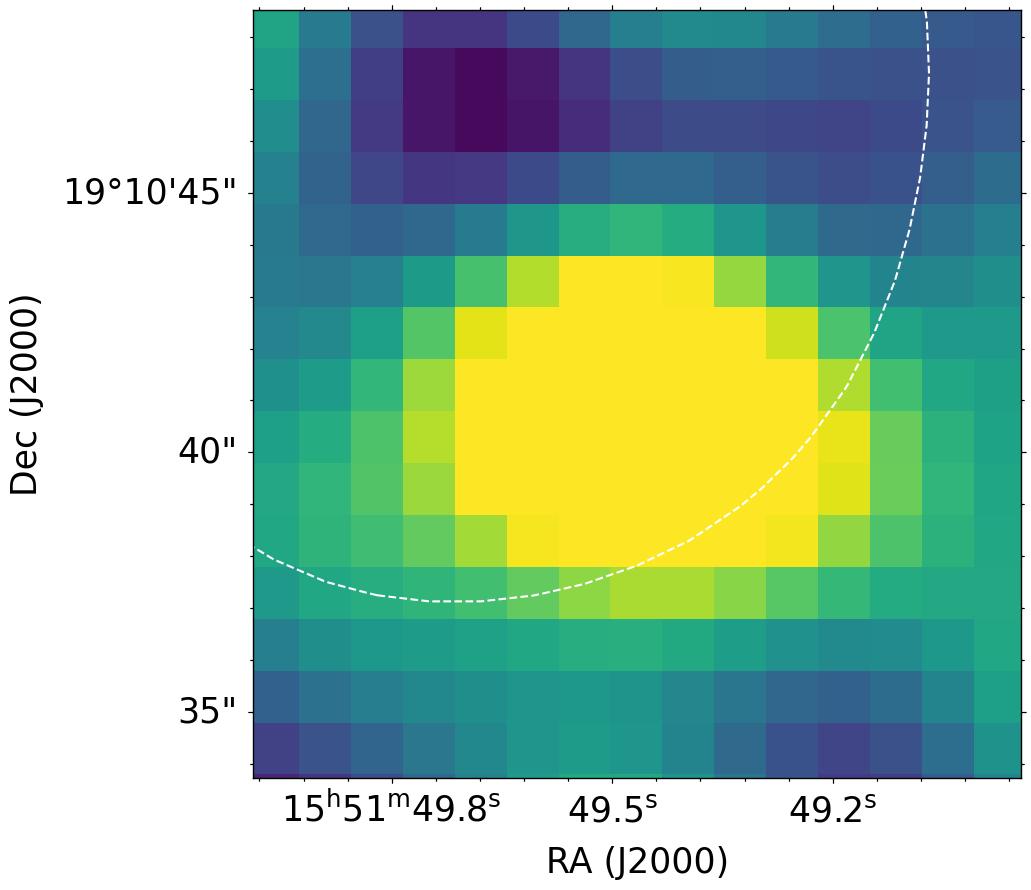} \\
    \end{tabular}
    \caption{CO(3--2) spectra and channel maps, continued.}
\end{figure*}

\begin{figure*}
    \centering
    \begin{tabular}{cc}
    \includegraphics[width=\width]{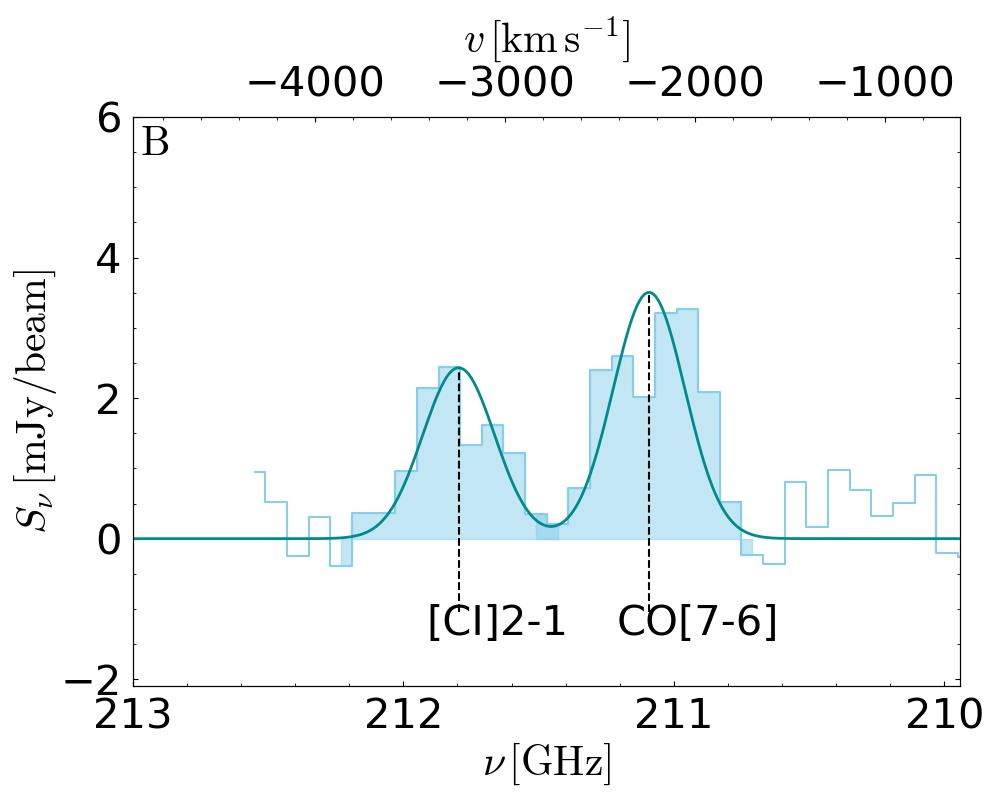} & 
    \includegraphics[width=\width]{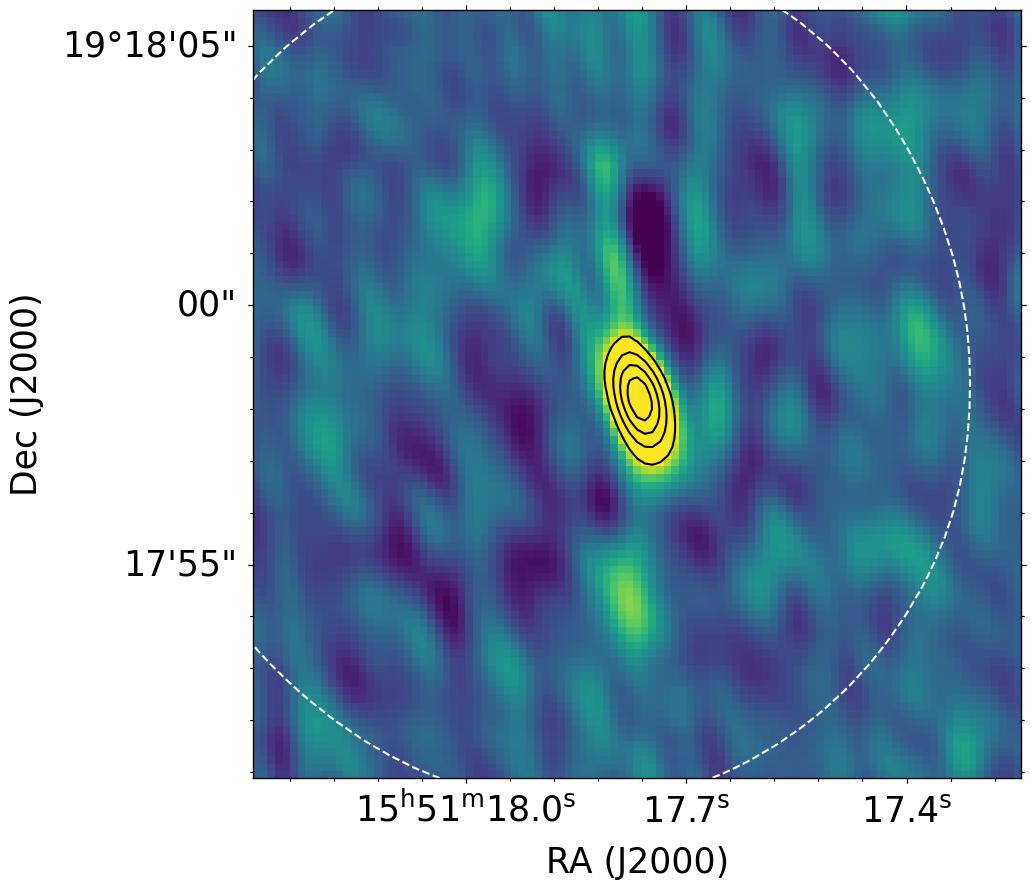} \\
    \includegraphics[width=\width]{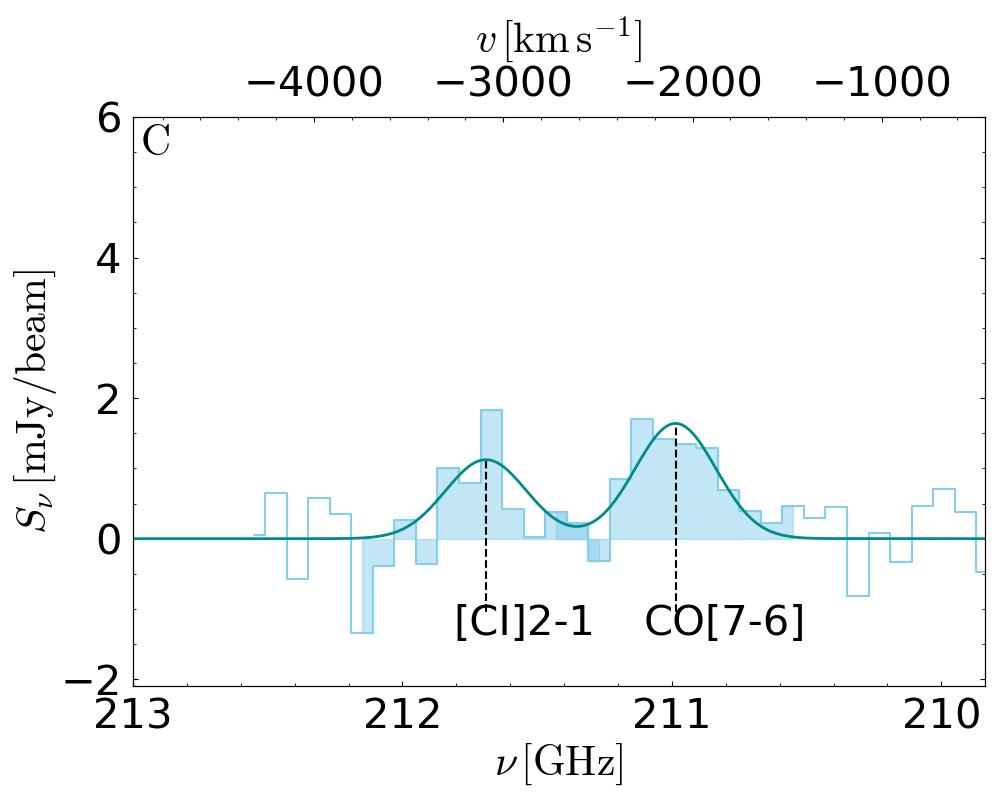} & 
    \includegraphics[width=\width]{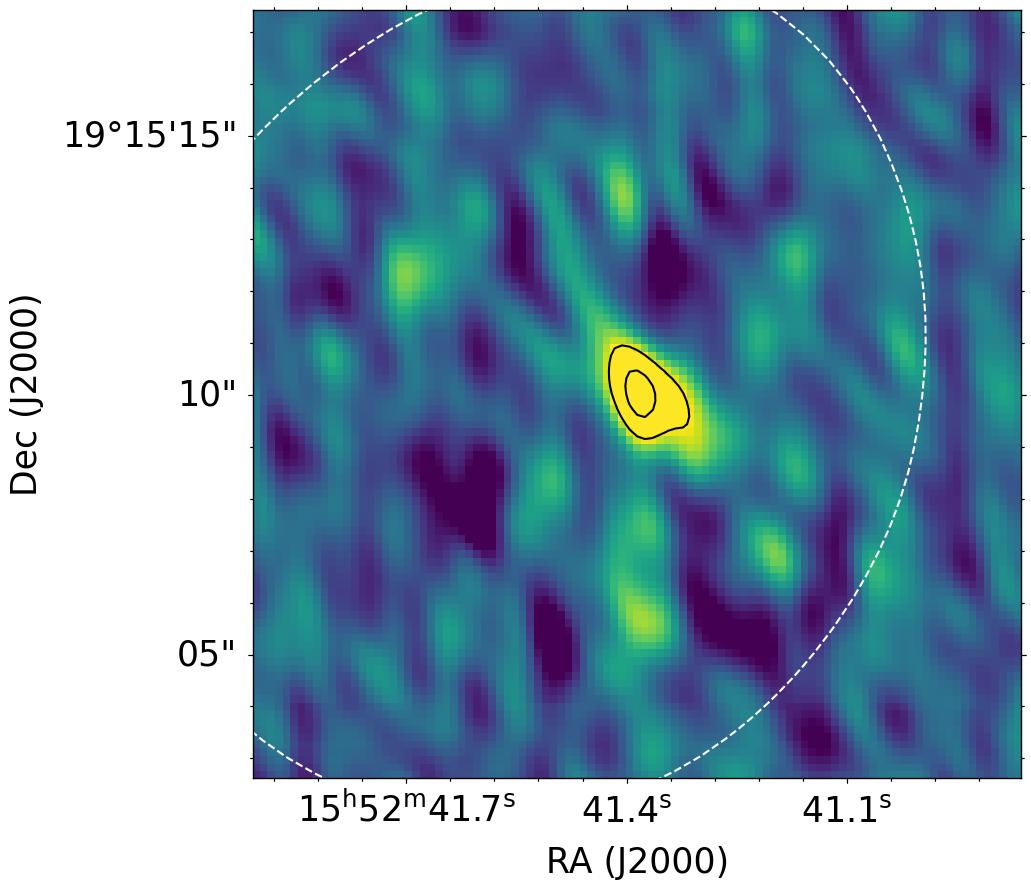} \\ 
    \includegraphics[width=\width]{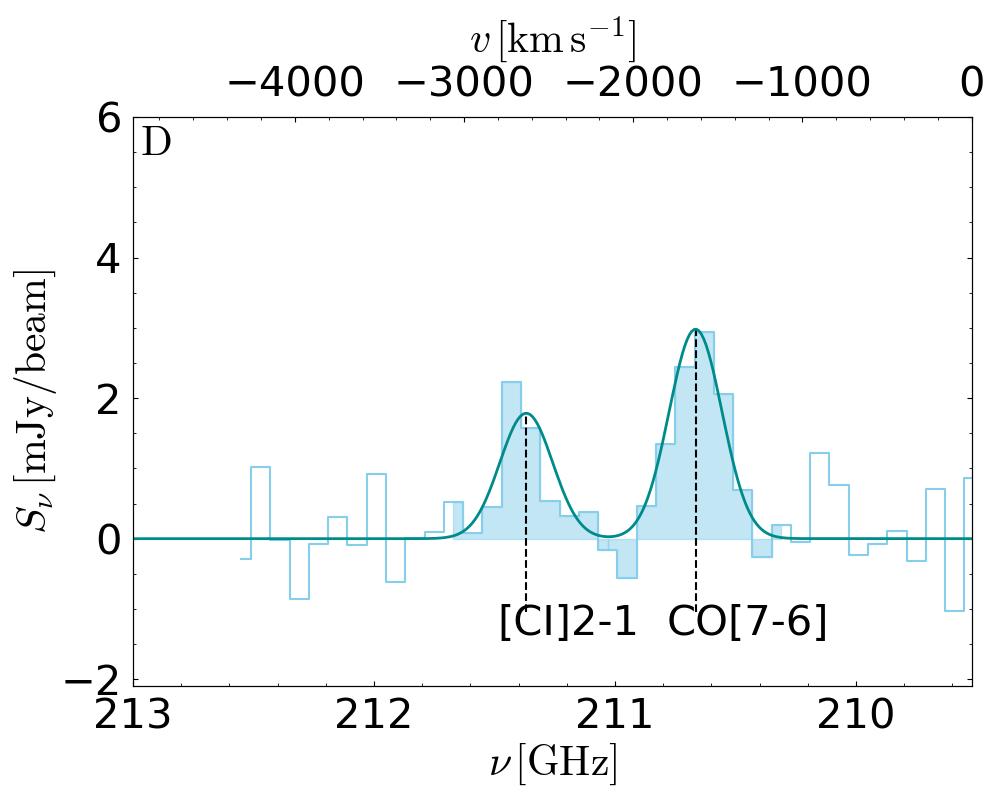} & 
    \includegraphics[width=\width]{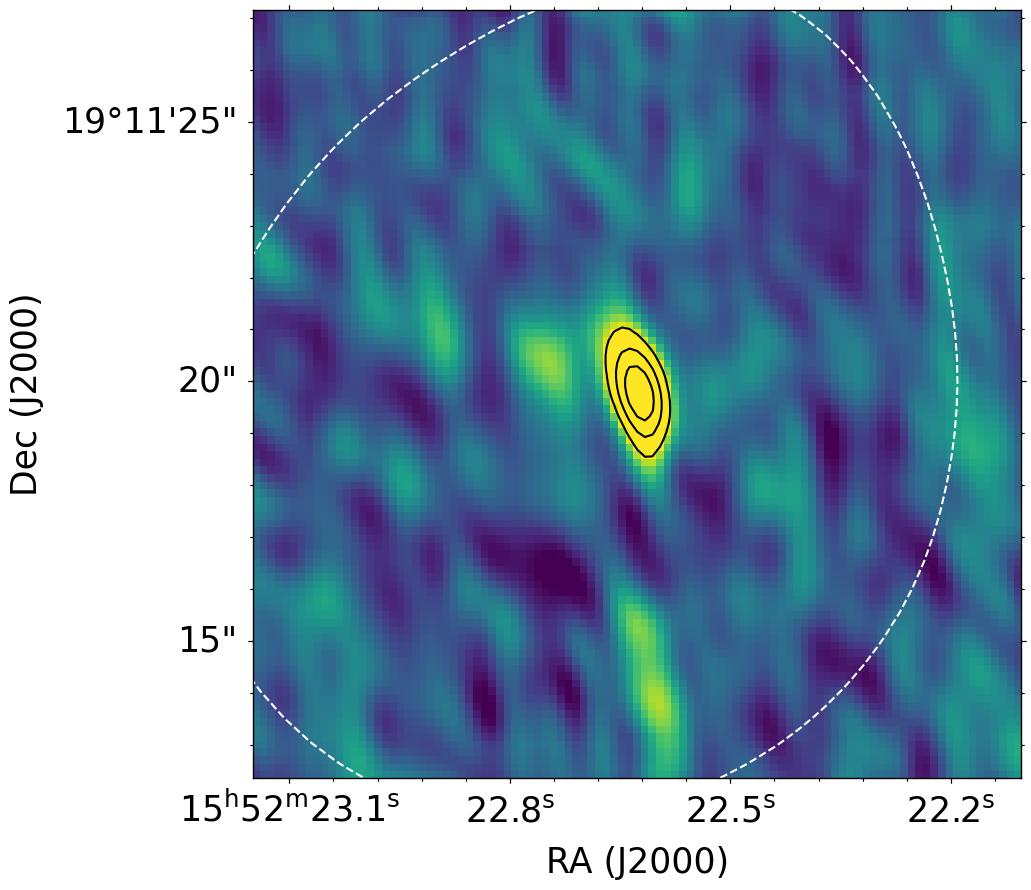} \\
    \includegraphics[width=\width]{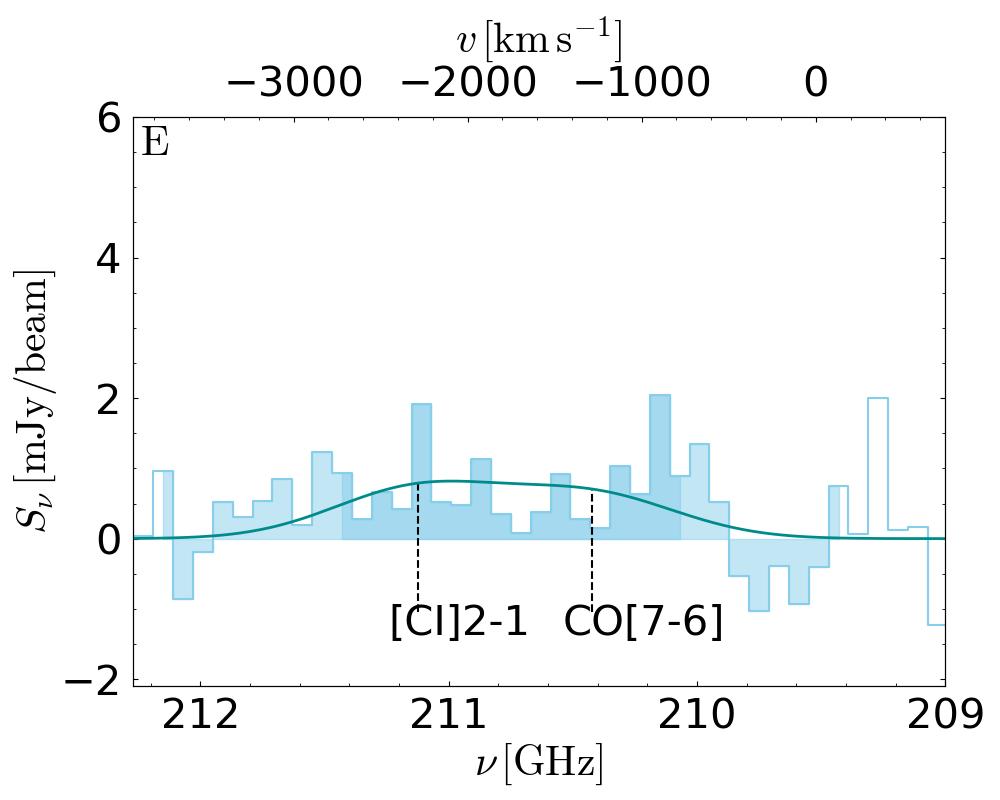} & 
    \includegraphics[width=\width]{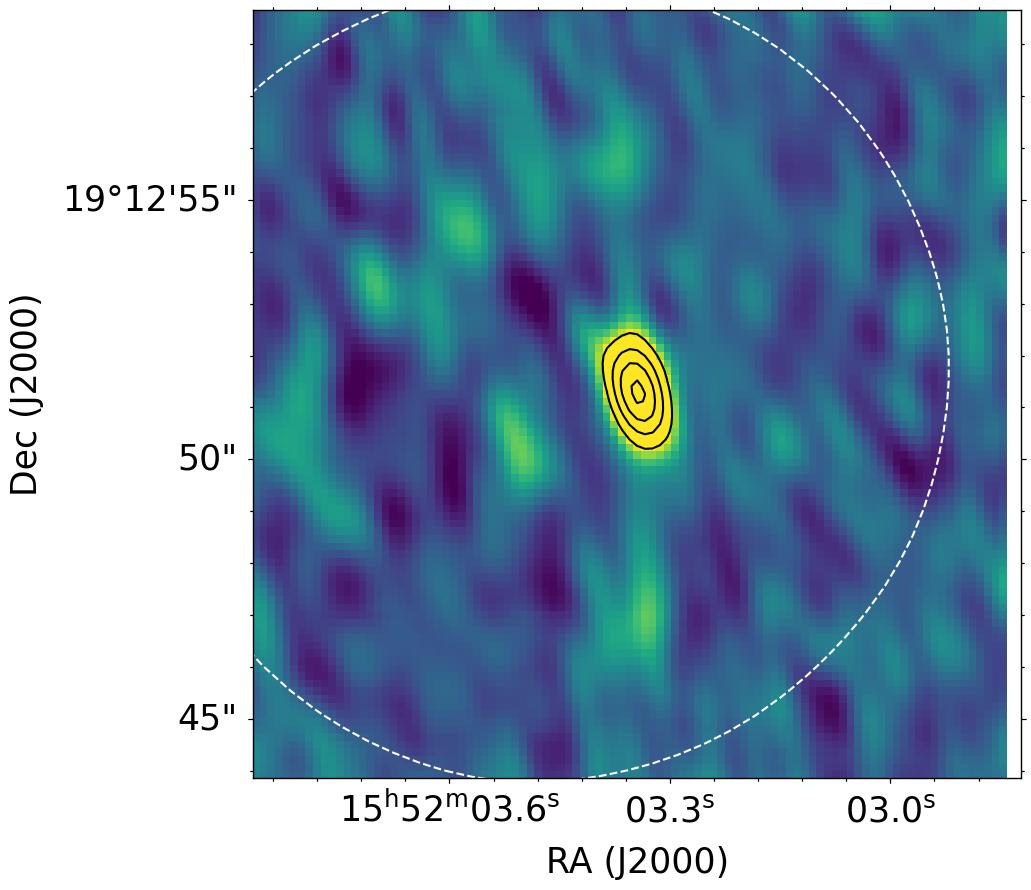} \\
    \end{tabular}
    \caption{CO(7--6)/[CI](2--1) spectra and continuum maps of the sub-sample followed up with NOEMA at 1.4-mm with spectra extracted at the peak pixel of the 1.4-mm continuum. The 850-\um\ SCUBA-2 emission is shown by white contours (70\% of the peak flux of each source). We show the contours of the 1.4-mm map (4, 7, 10, and 13\,$\sigma$). The spectral resolution is 60\,\kms. Double Gaussian fits to each set of line profiles are overlaid. \label{fig: CO7-6}}
\end{figure*}

\addtocounter{figure}{-1}
\begin{figure*}
    \centering
    \begin{tabular}{cc}
    \includegraphics[width=\width]{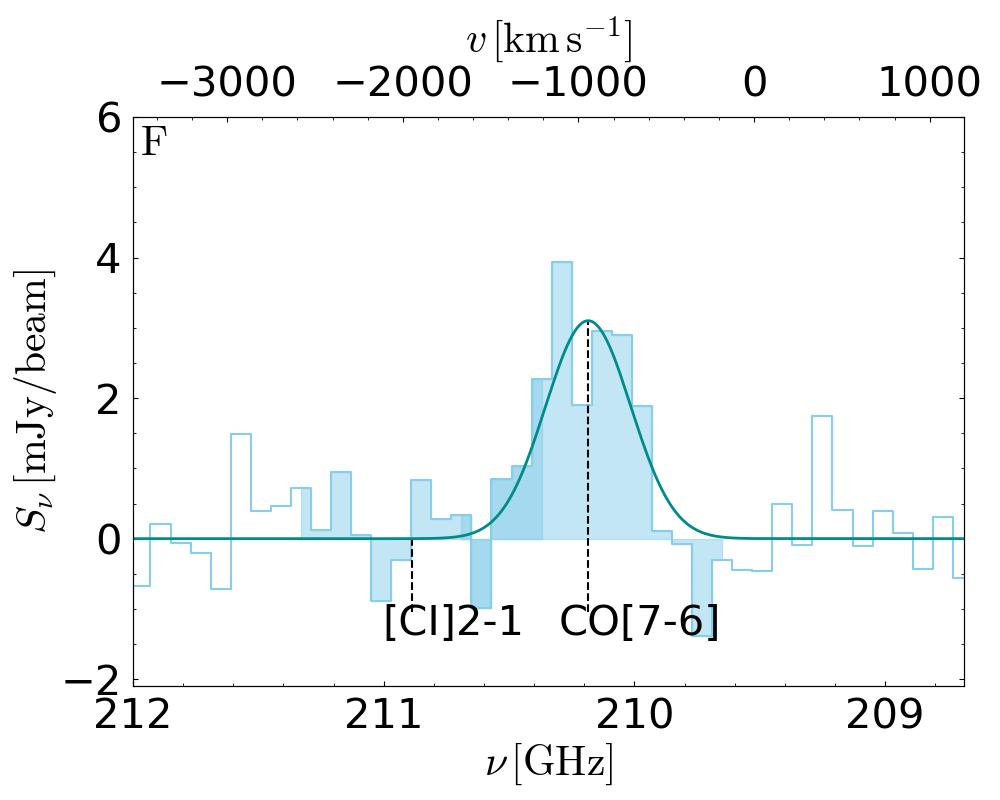} & 
    \includegraphics[width=\width]{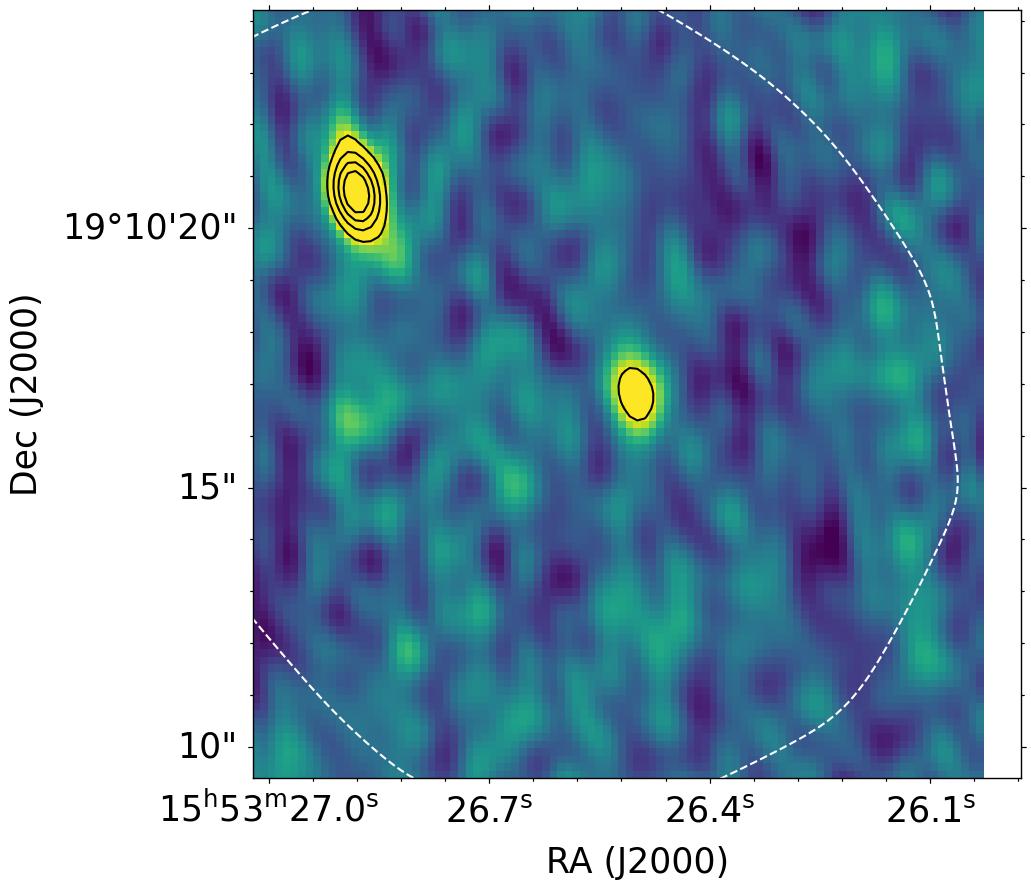} \\
    \includegraphics[width=\width]{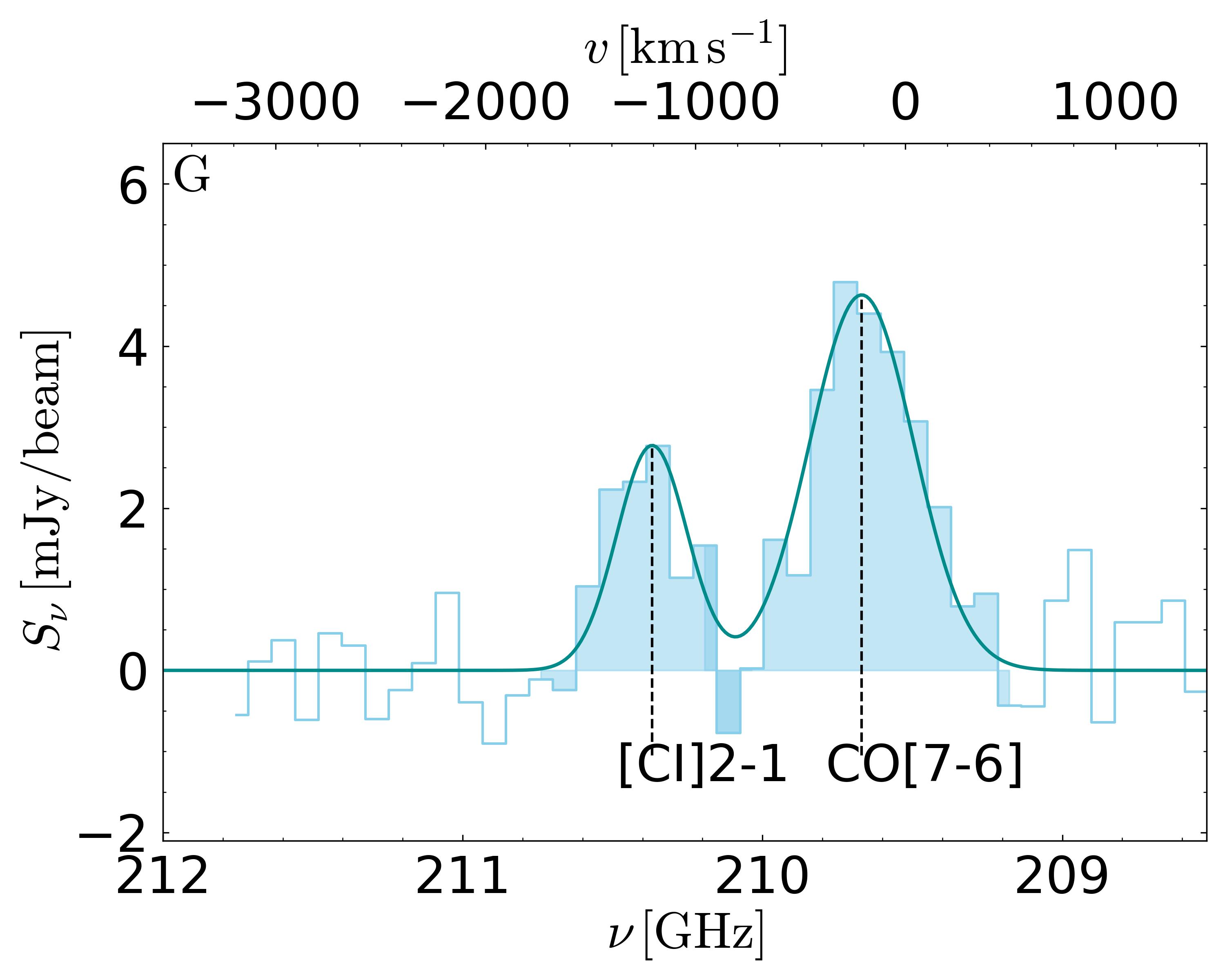} & 
    \includegraphics[width=\width]{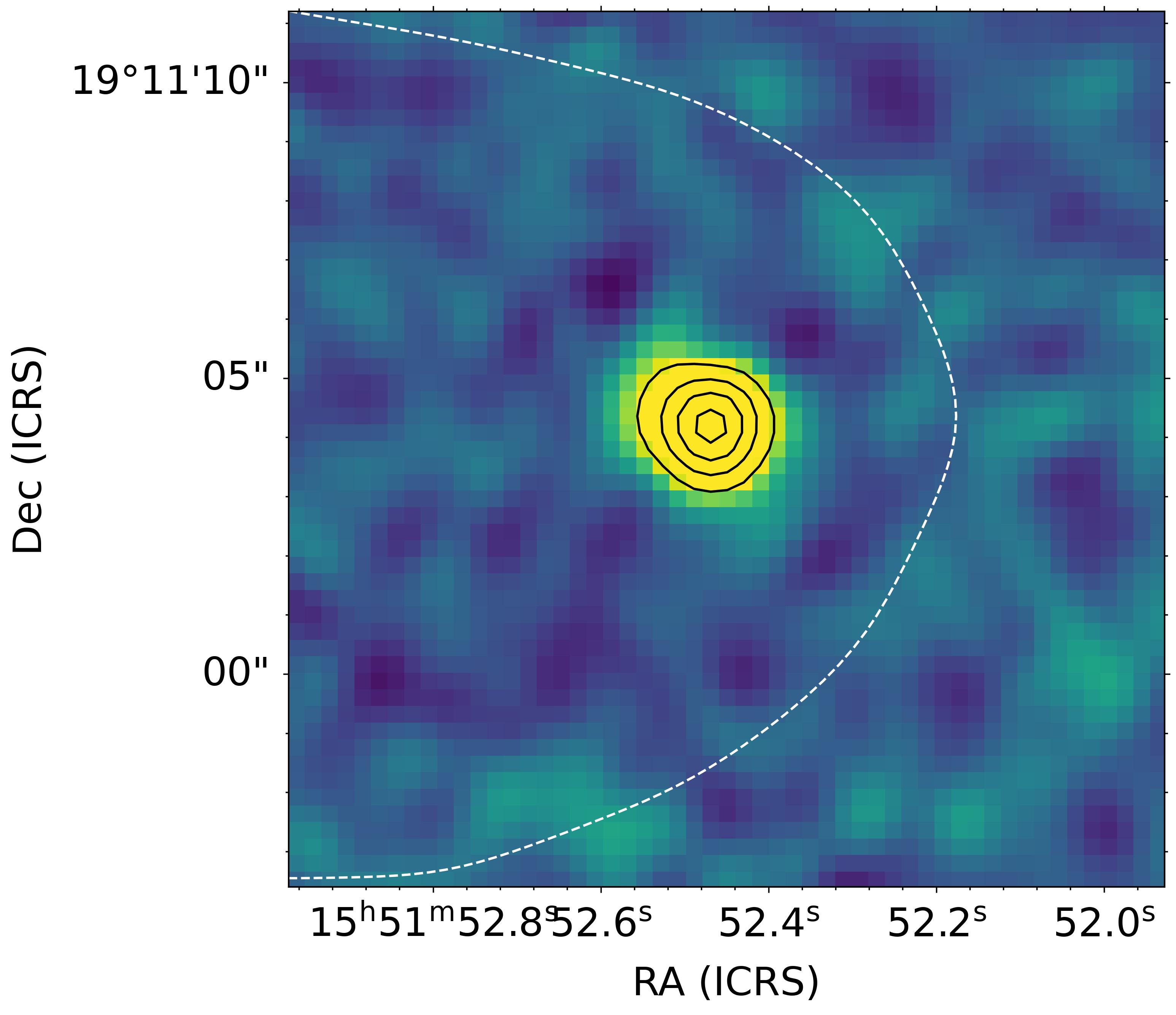} \\ 
    \includegraphics[width=\width]{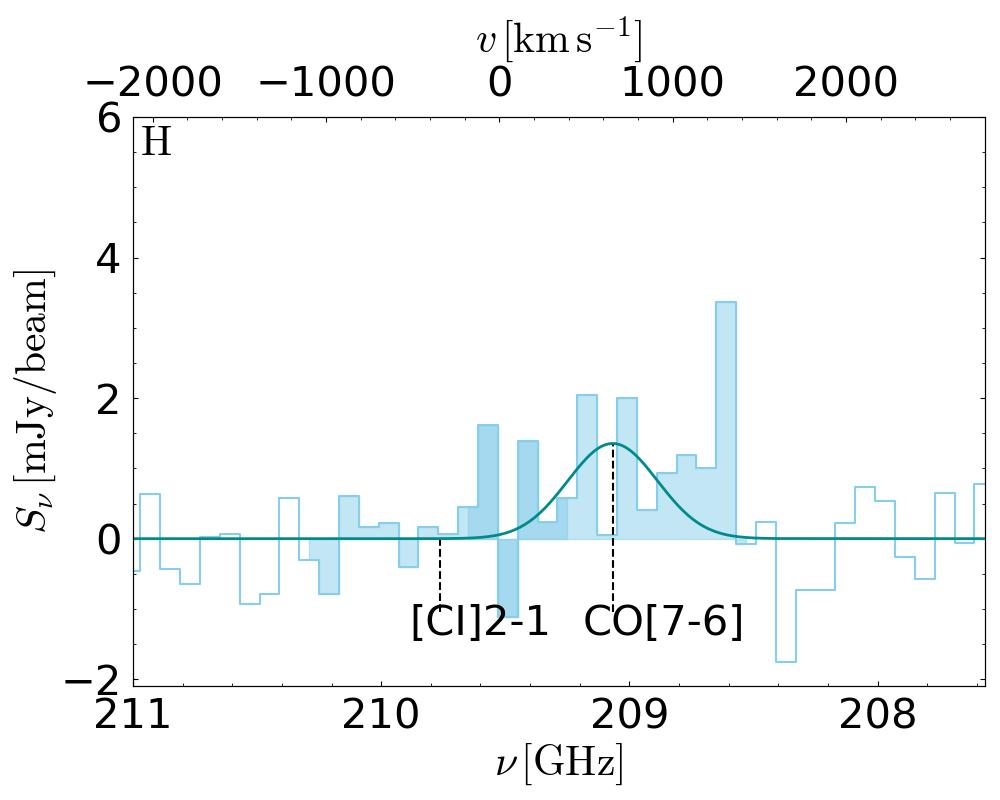} & 
    \includegraphics[width=\width]{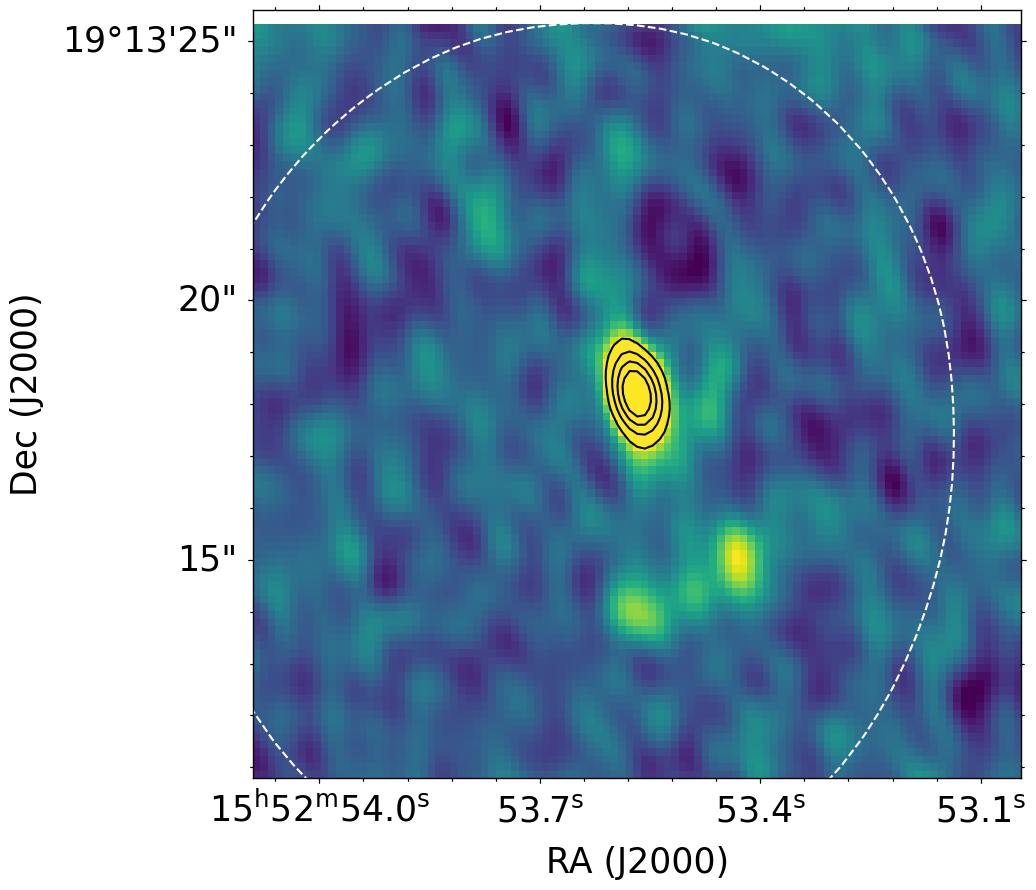} \\
    \includegraphics[width=\width]{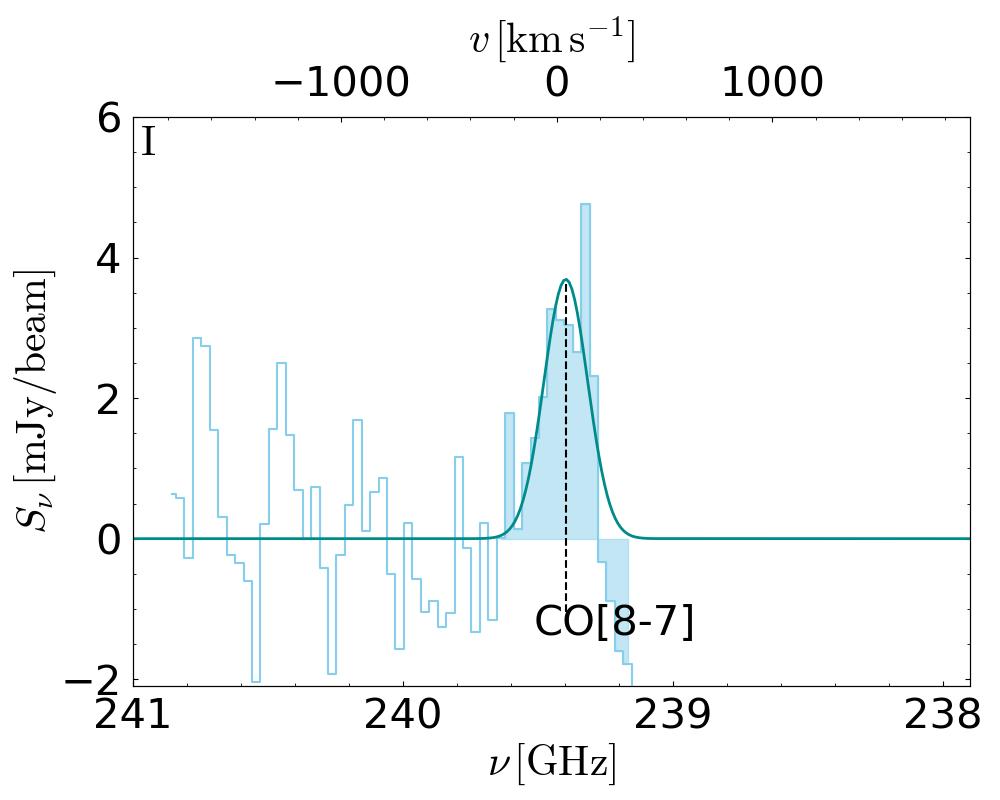} & 
    \includegraphics[width=\width]{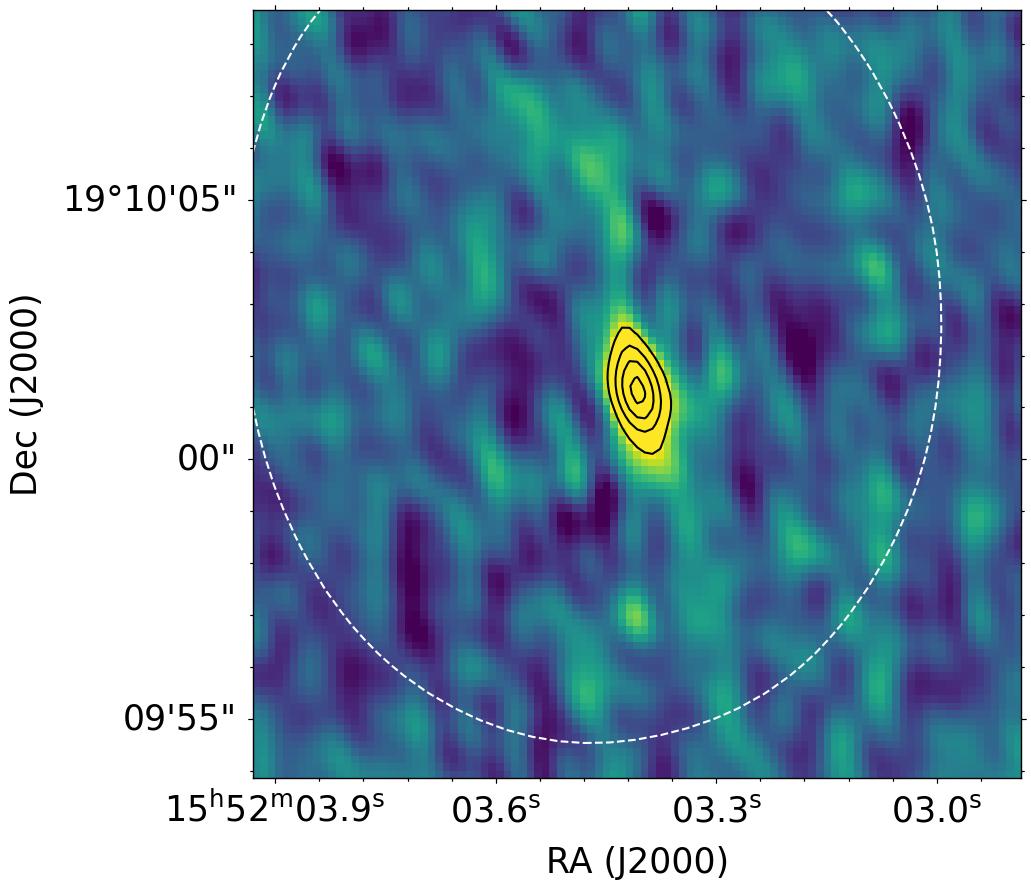} \\
    \end{tabular}
    \caption{CO(7--6)/[CI](2--1) spectra and continuum maps, continued.}
\end{figure*}

\addtocounter{figure}{-1}
\begin{figure*}
    \centering
    \begin{tabular}{cc}
    \includegraphics[width=\width]{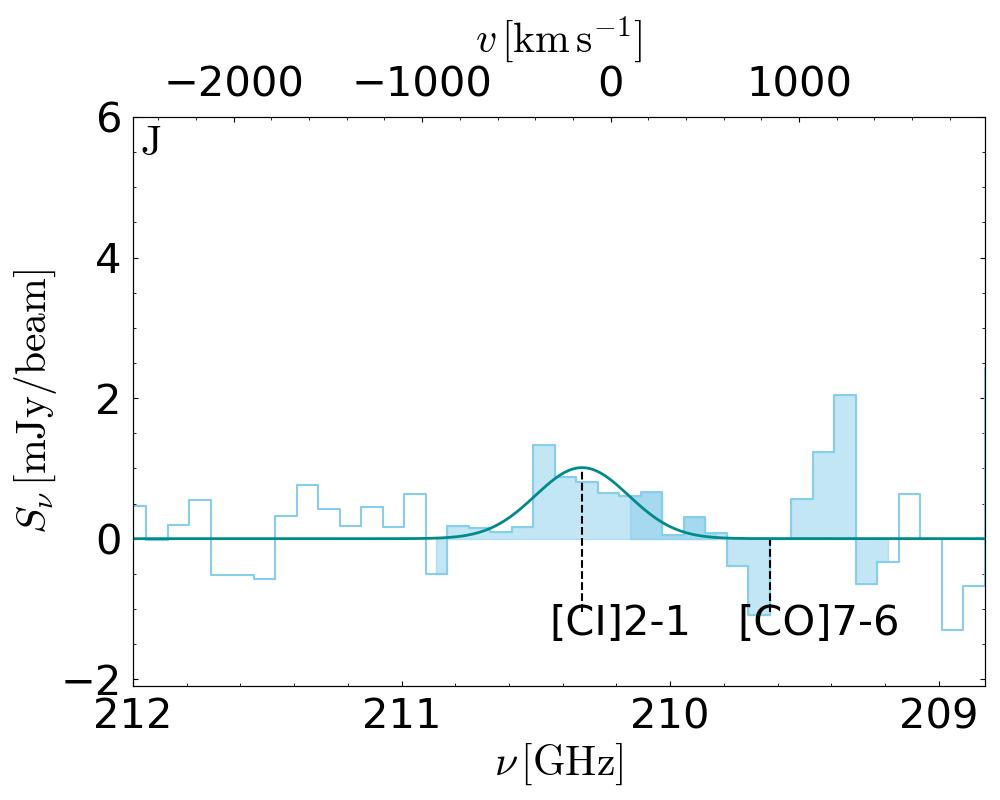} & 
    \includegraphics[width=\width]{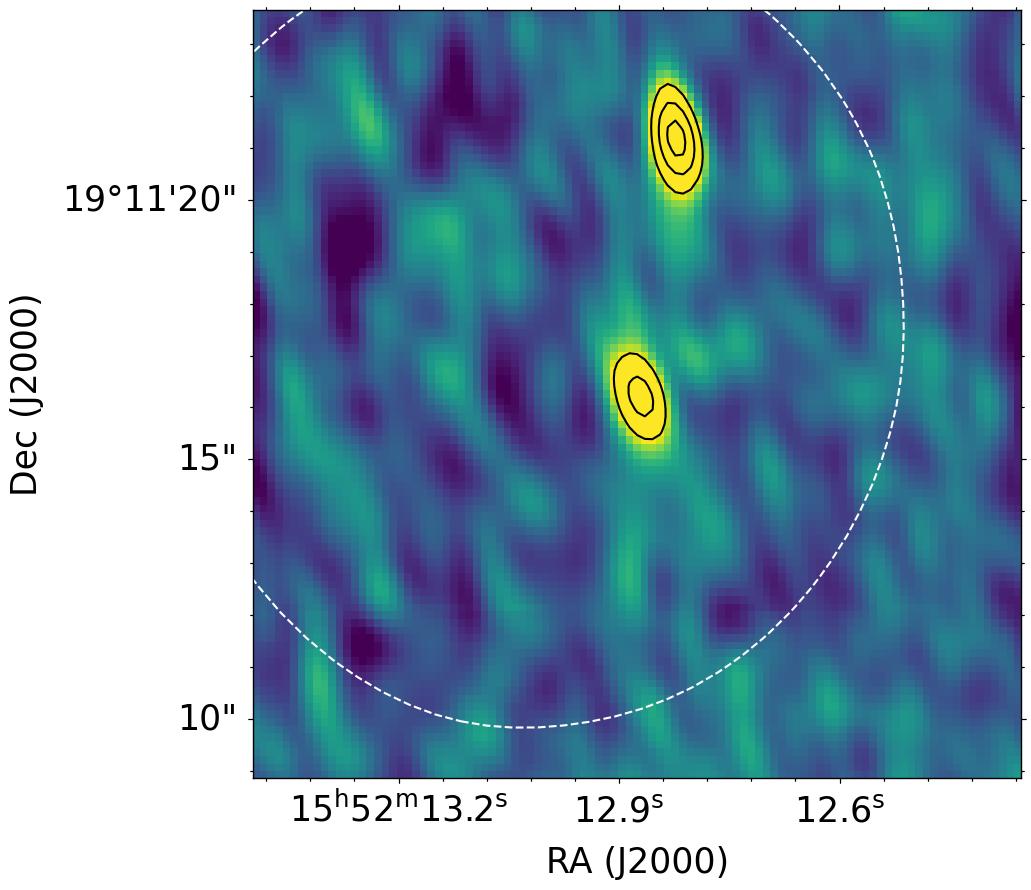} \\
    \includegraphics[width=\width]{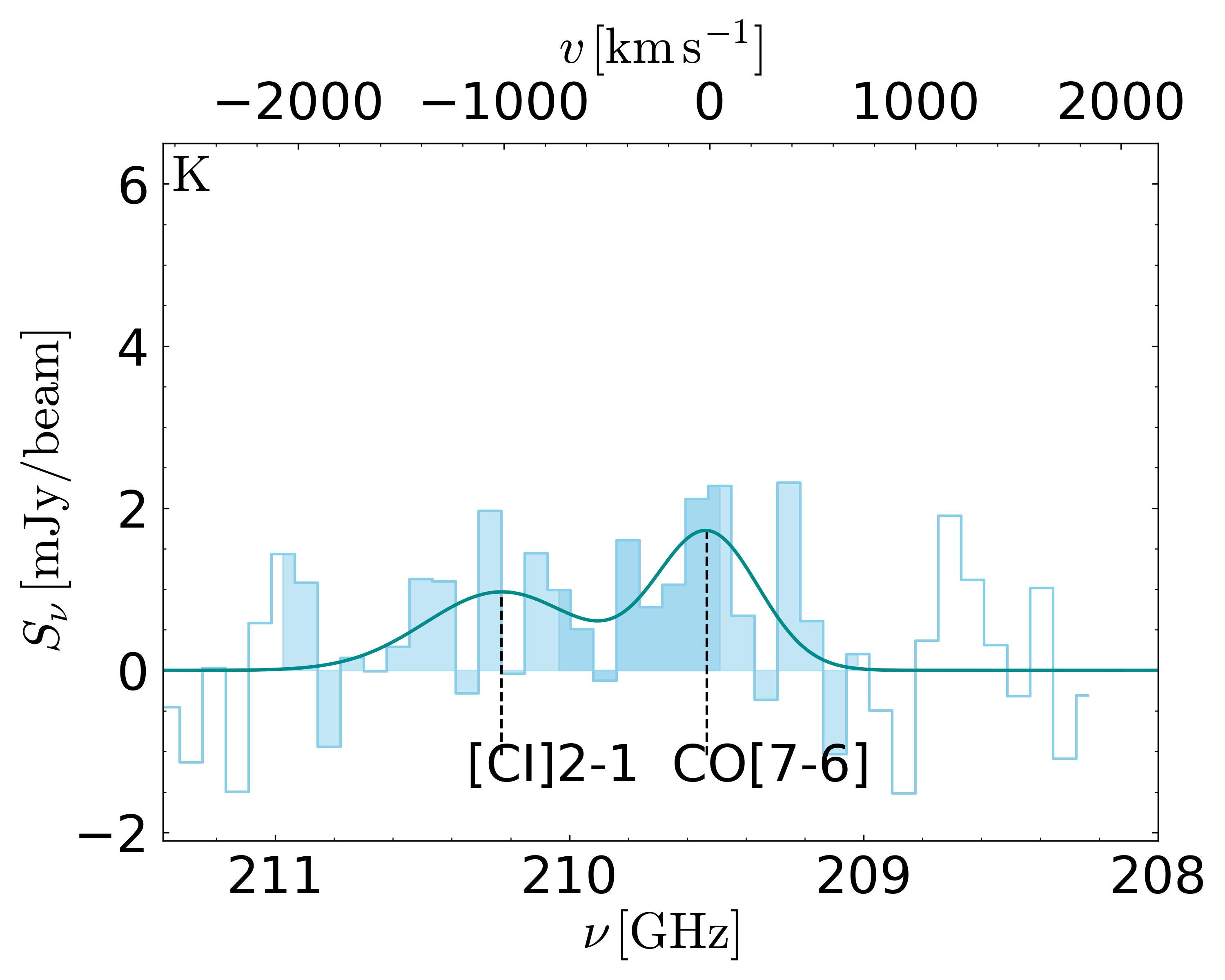} & 
    \includegraphics[width=\width]{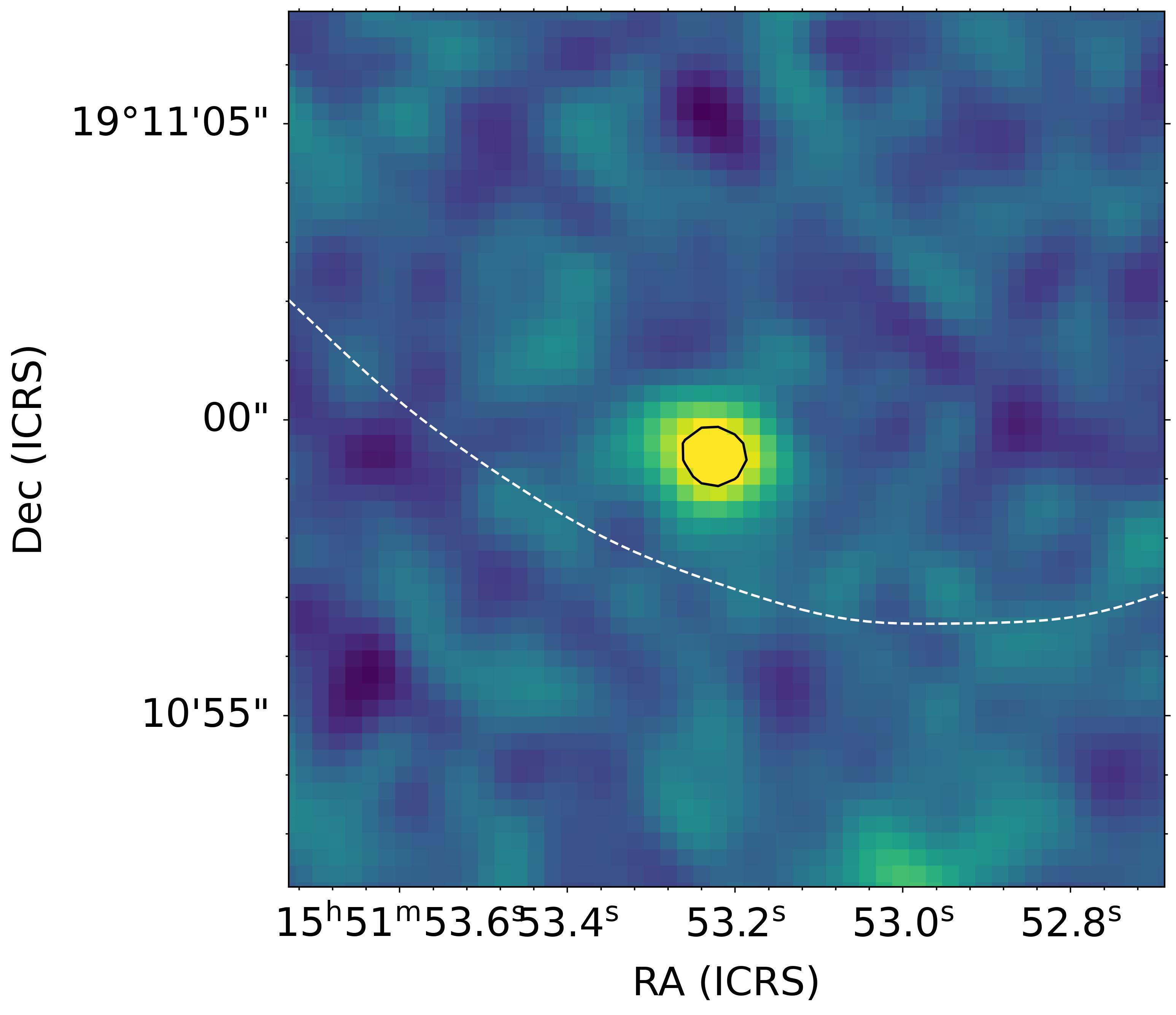} \\ 
    \includegraphics[width=\width]{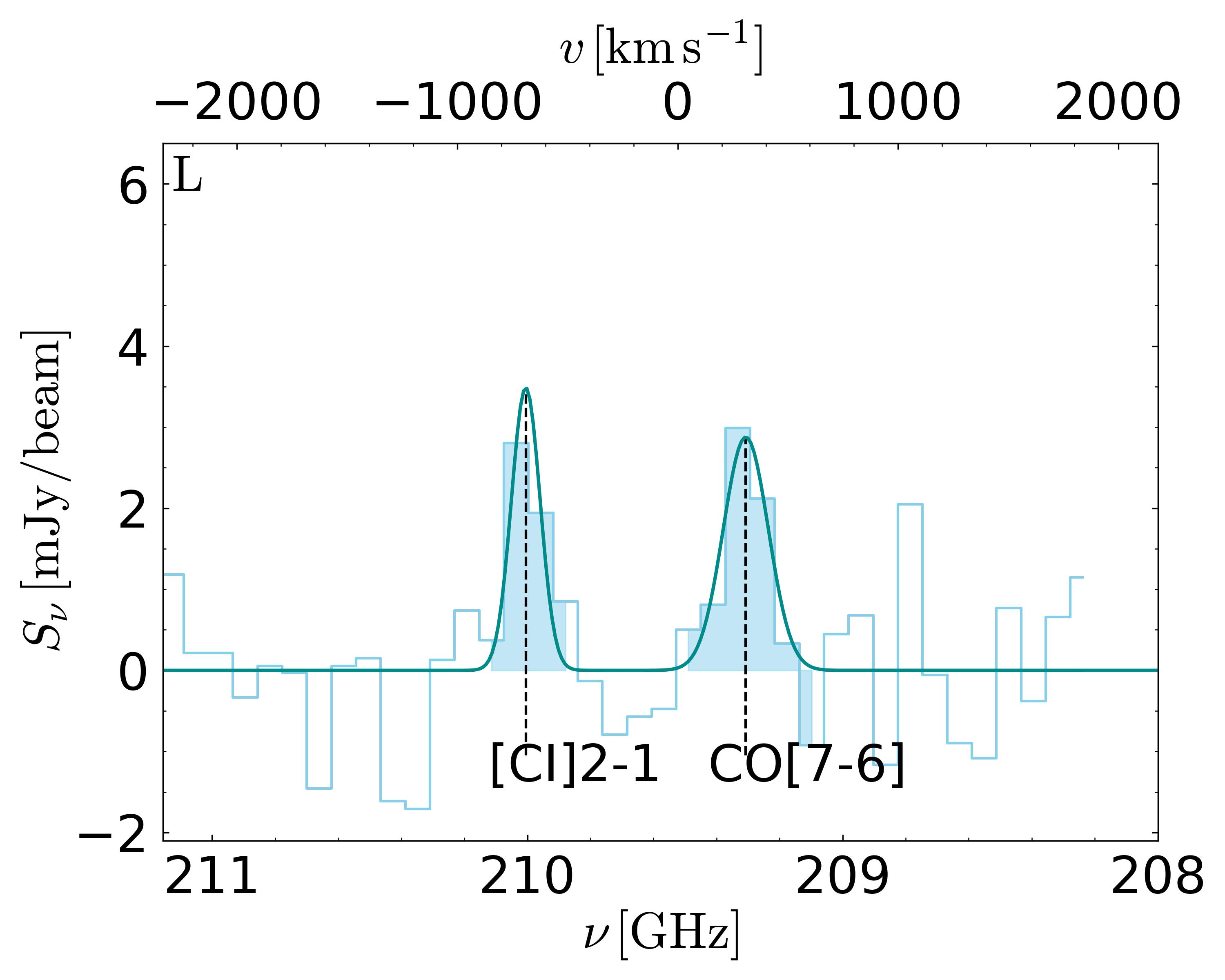} & 
    \includegraphics[width=\width]{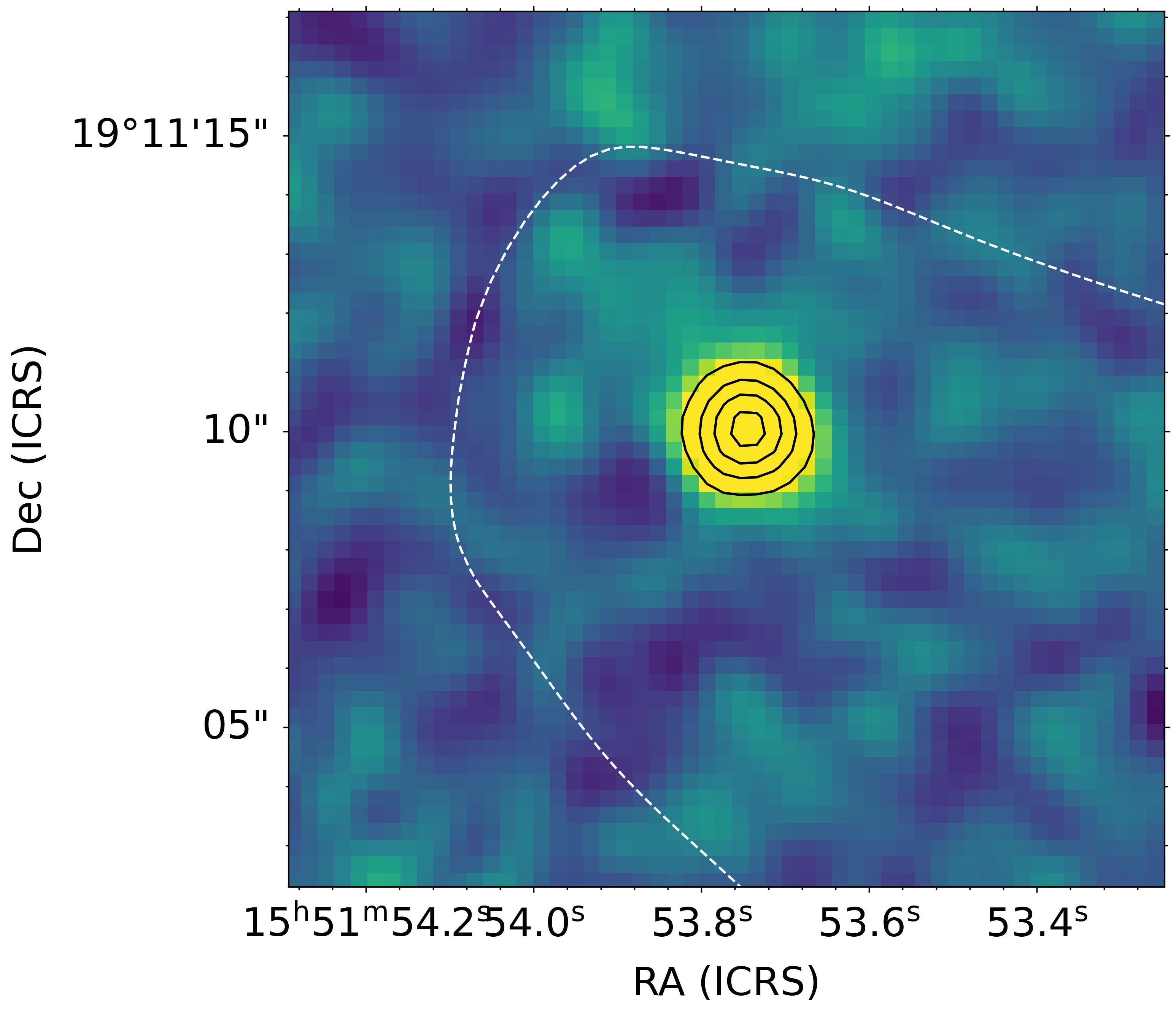} \\
    \includegraphics[width=\width]{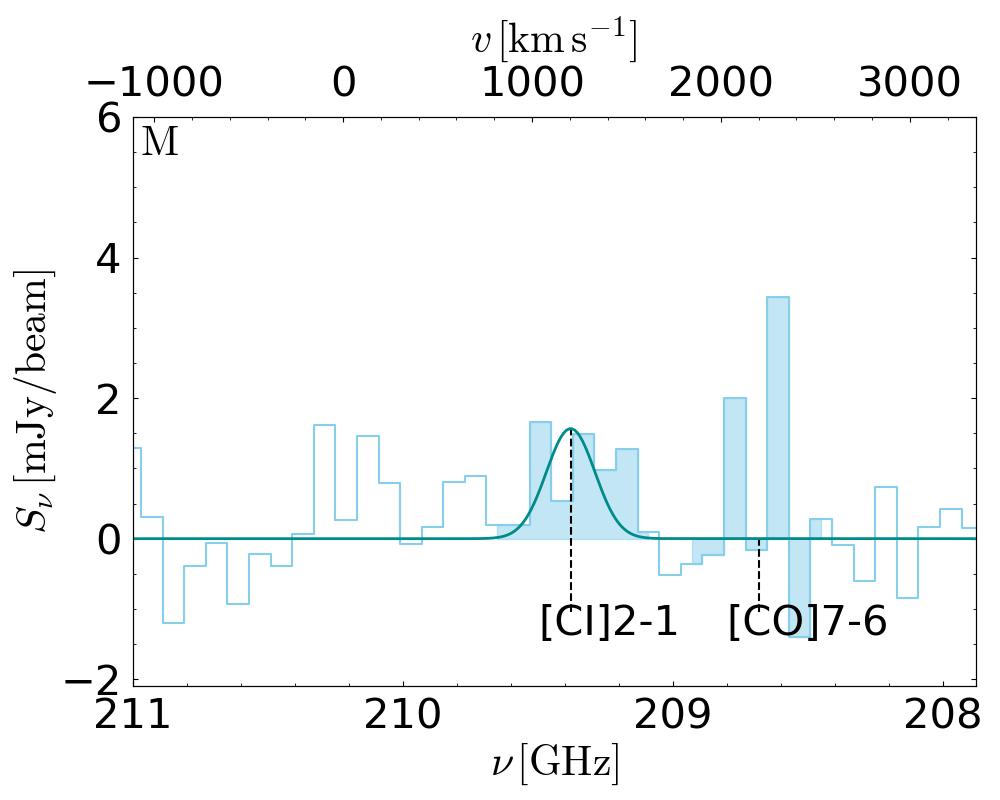} & 
    \includegraphics[width=\width]{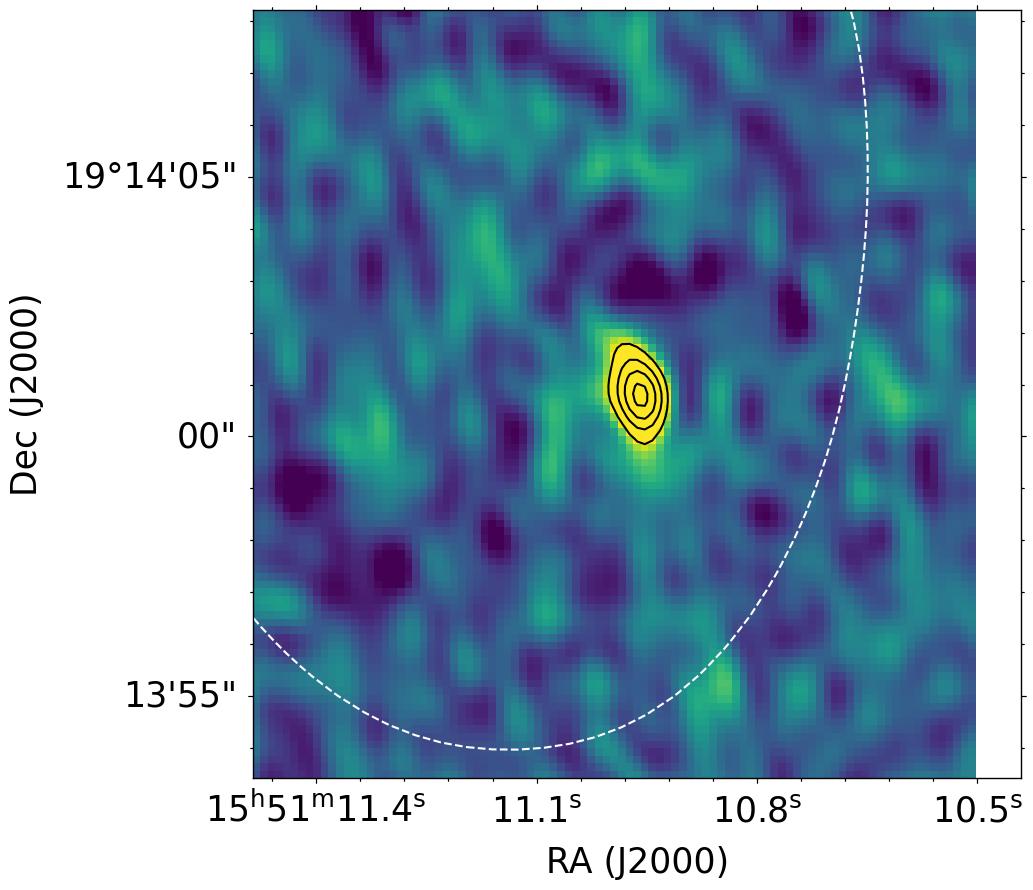} \\ 
    \end{tabular}
    \caption{CO(7--6)/[CI](2--1) spectra and continuum maps, continued.}
\end{figure*}

\addtocounter{figure}{-1}
\begin{figure*}
    \centering
    \begin{tabular}{cc}
    \includegraphics[width=\width]{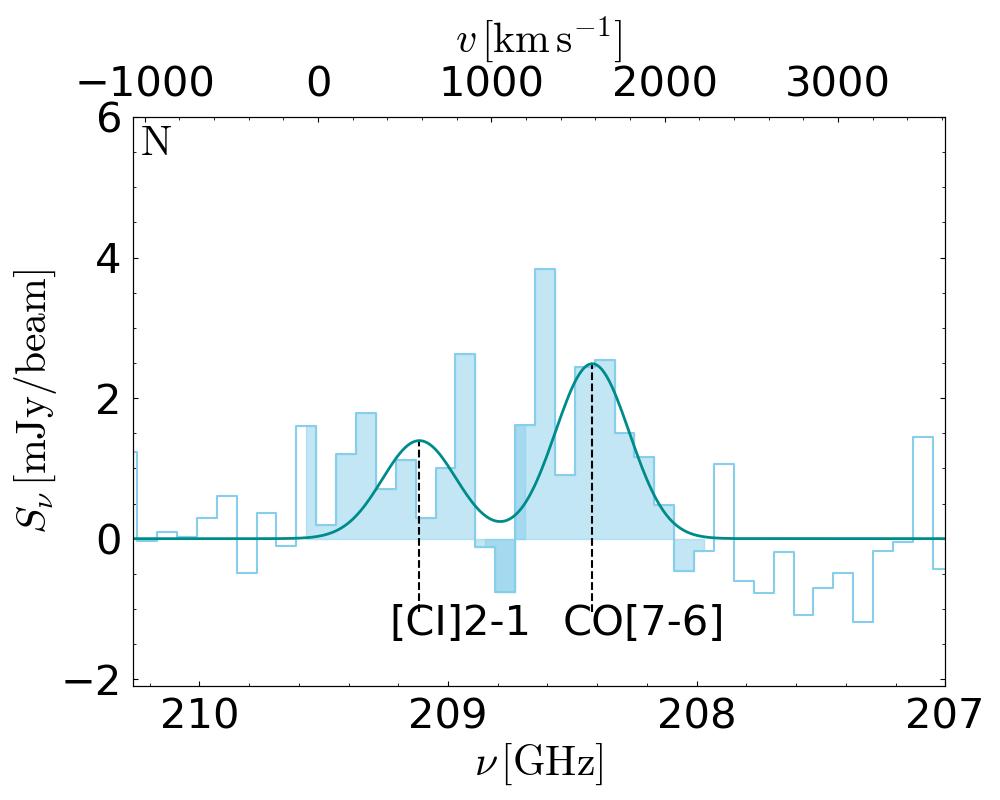} & 
    \includegraphics[width=\width]{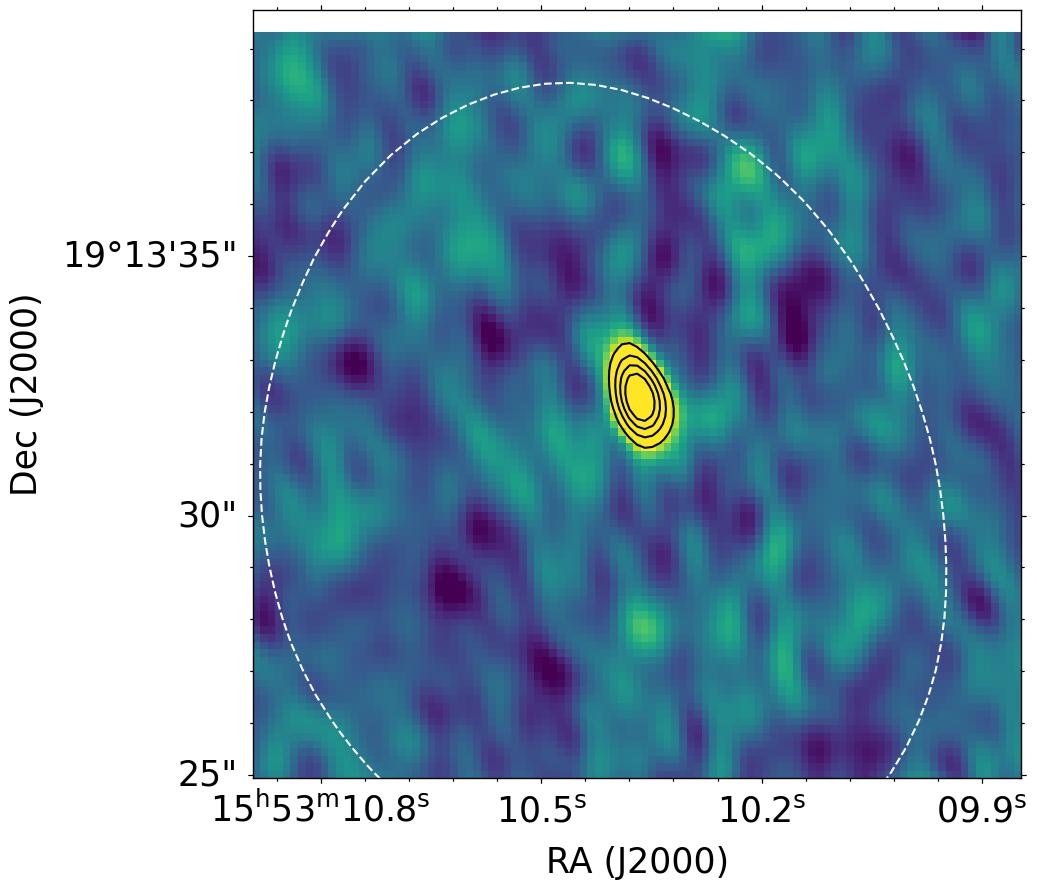} \\
    \includegraphics[width=\width]{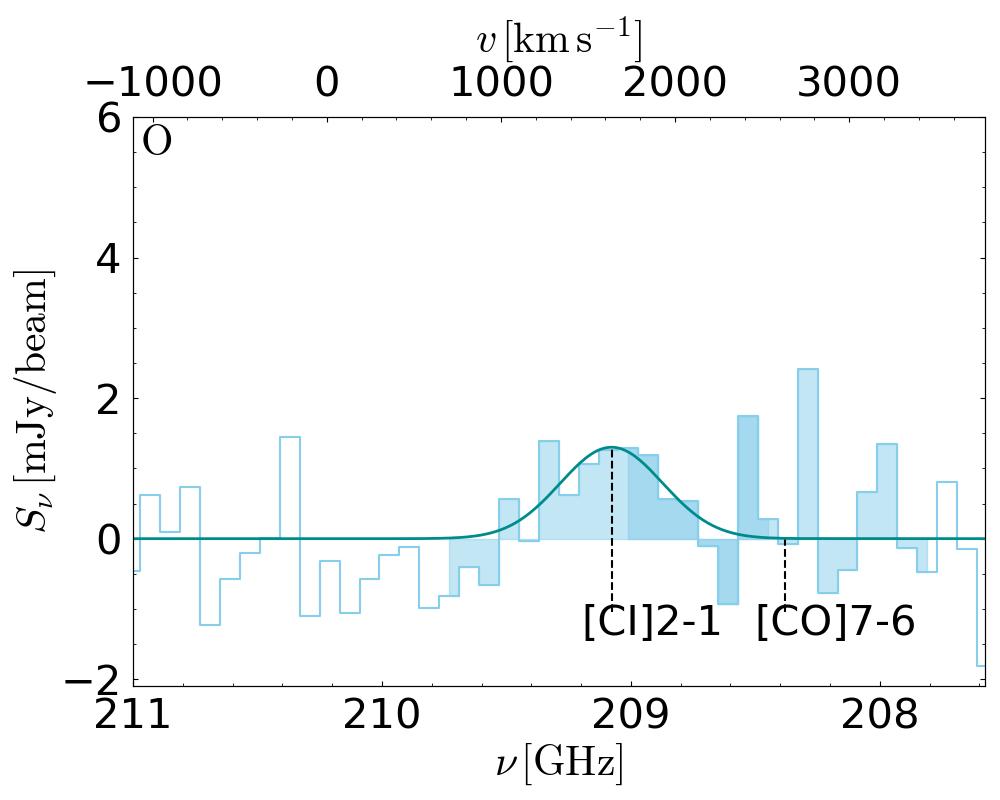} & 
    \includegraphics[width=\width]{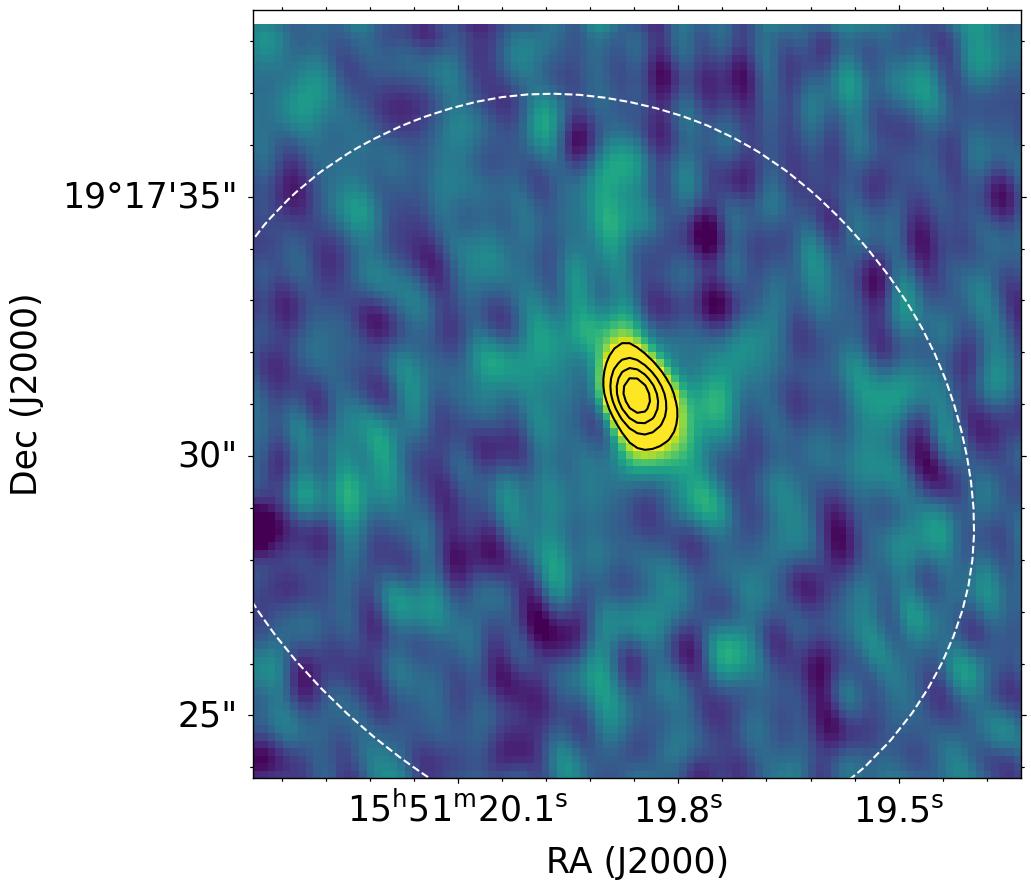} \\ 
    \includegraphics[width=\width]{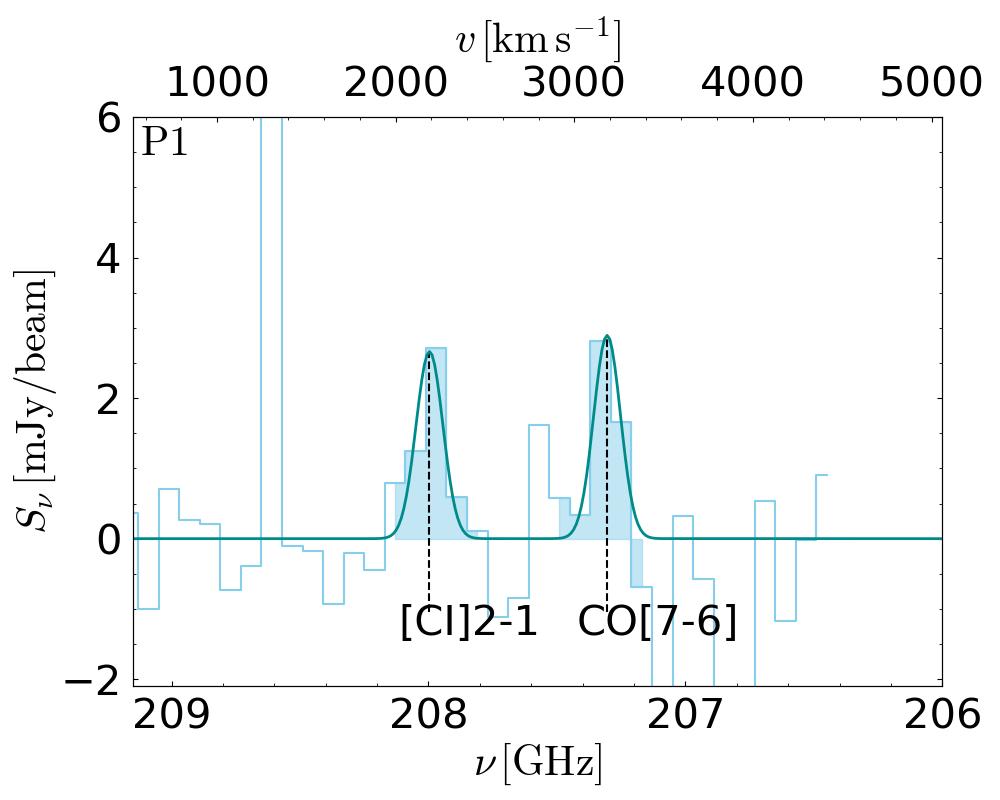} & 
    \includegraphics[width=\width]{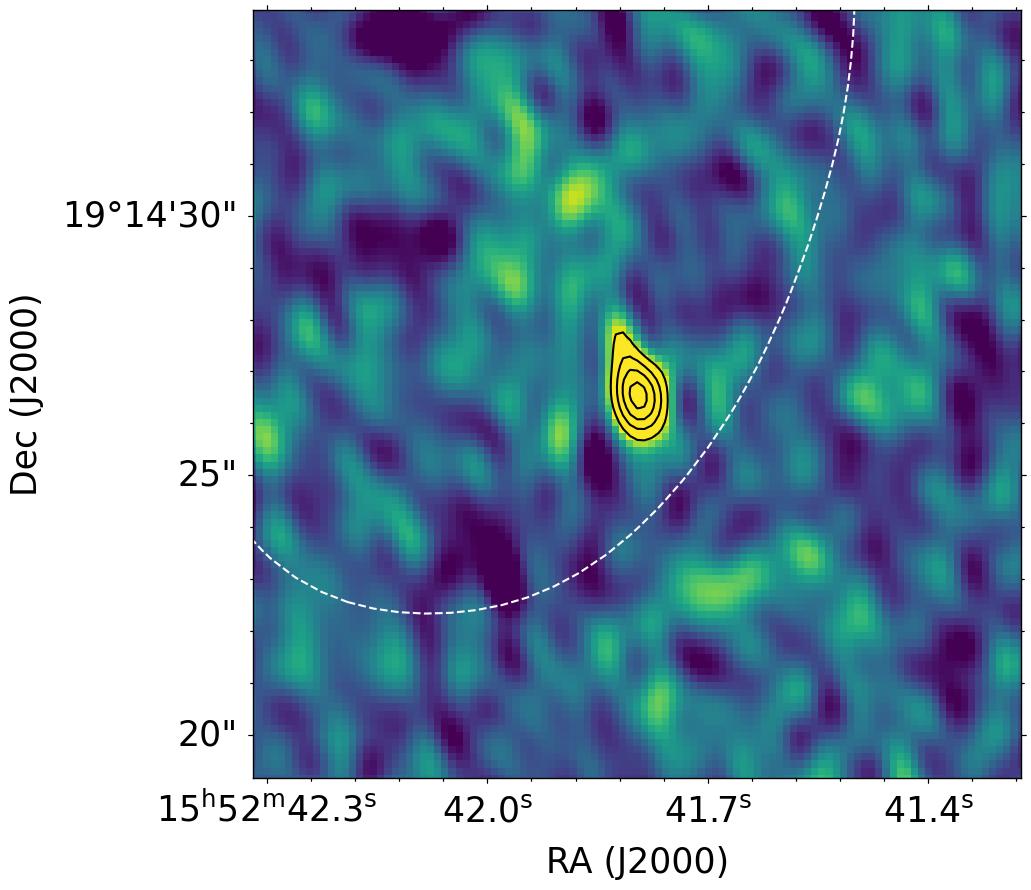} \\
    \includegraphics[width=\width]{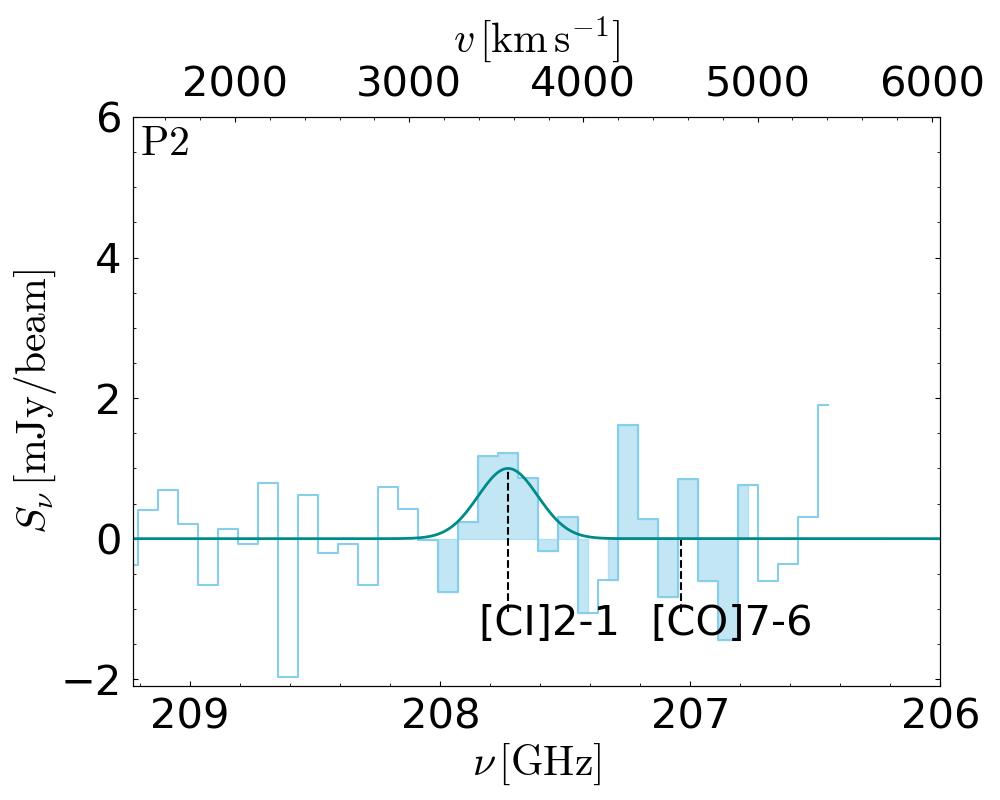} & 
    \includegraphics[width=\width]{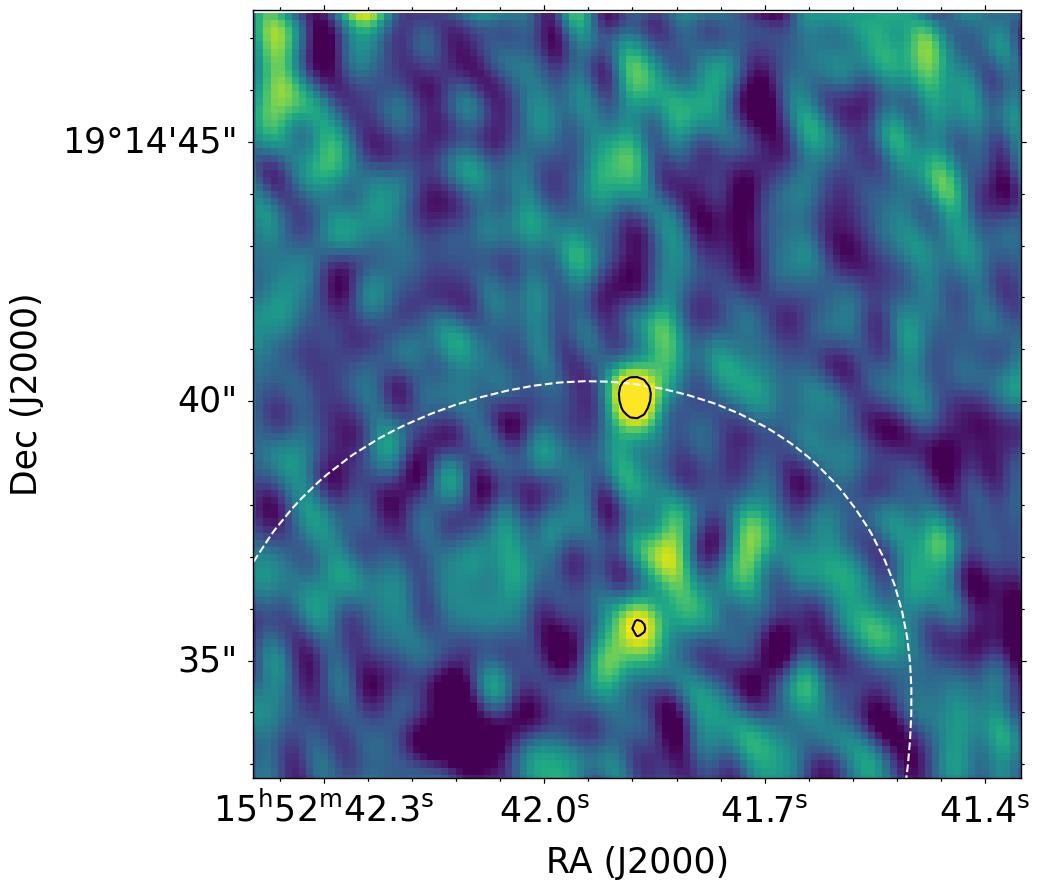} \\
    \end{tabular}
    \caption{CO(7--6)/[CI](2--1) spectra and continuum maps, continued.}
\end{figure*}

\clearpage

\section{Details of protocluster comparison sample} \label{sec: comparison sample}

The GOODS-N overdensity at $z$\,$=$\,1.99 \citep{Chapman2009} spans a roughly 10$^{\prime}$\,$\times$\,$10^{\prime}$ field in the Hubble Deep Field North, containing six SMGs within $\Delta z$\,=\,0.008. The probability of finding this overdensity from fluctuations in the field distribution is $<$\,0.01\%. Interestingly, only a modest overdensity of Lyman-break galaxies (LBGs) is found in this GOODS-N structure. 

The COSMOS $z$\,$=$\,2.5 SMG overdensity \citep[PCL1002]{Casey2015} is similar to the GOODS-N structure in terms of the numbers and luminosities of the component SMGs, the angular size of the system, and the modest overdensity of LBGs associated with it.

The MRC1138 $z$\,=\,2.16 structure was originally discovered as an overdensity of Ly$\alpha$ and H$\alpha$ 
emitters surrounding a radio-loud AGN (known as the ``Spiderweb Galaxy'') that resides in a large Ly-$\alpha$ halo \citep{Kurk2000}. Follow-up observations \citep{Kuiper2011, Dannerbauer2014, Jin2021} revealed the presence of five SMGs, an additional AGN, and 46 CO line emitters. 

The SSA22 protocluster was one of the first protoclusters discovered by observing an overdensity of LBGs \citep{Steidel2000}. It is an extended structure (although less extreme than HS\,1549) at $z$\,=\,3.09, with LAEs spanning over 50\,cMpc \citep{Hayashino2004}. Submm observations of the field have revealed a population of many faint SMGs, five of which have $S_{850}$\,$>$\,8\,mJy \citep{Chapman2001, Chapman2005, Geach2005, Tamura2009, Umehata2015, geach2017, Chapman2023}, one of which is a likely outlier comparable to ``A'' and ``Q'' in HS\,1549. 

The COSMOS $z$\,=\,2.1 protocluster \citep{Hung2016} lacks sufficiently deep 850-\um\ data to characterize the {\it Herschel}-SPIRE sources identified in the structure. We estimate 850-\um\ flux densities by taking their published $L_{\rm IR}$ values (integrated over 3--1100\,\um) and using the SED of Arp 220 to establish the scaling relation that $L_\mathrm{IR}$\,=\,2$\times$\,10$^{12}$\,\lsun\ corresponds to $S_{850}$\,=\,1\,mJy at $z$\,$=$\,2.1. 

While more distant than the protoclusters above, and likely having different characteristics, there have also been detections of SMG overdensities at $z$\,$>$\,4. GN20 in the GOODS-N field shows signs of a protocluster at $z$\,=\,4.05. It was discovered through the serendipitous detection of CO(4--3) from two SMGs \citep{Daddi2009}, with two further SMGs detected subsequently \citep{Daddi2009}. An excess of $B$-band dropouts is also observed in this structure, with several spectroscopically confirmed sources lying at $z$\,$\approx$\,4.05.

The most luminous example at $z$\,$>$\,4 is SPT\,2349$-$56 \citep{Miller2018}, which is characterized by an extremely bright double-lobed LABOCA 870-$\mu$m source, resolved by ALMA into 25 SMGs \citep{Hill2020}. To this day, no satellite SMGs with bound escape velocities have been found beyond this hyper-luminous core region.

Another example is SMMJ\,004224, found from a {\it Herschel}-SPIRE survey \citep{Oteo2018}; it is comparable to SPT\,2349$-$56 but is less concentrated and has a lower total SFR. While an apparent surface overdensity of 870-\um\ sources was found in the field surrounding SMMJ\,004224, most sources lie in the foreground and are not at the protocluster redshift \citep{ivison2020}.

\clearpage

\bibliography{biblio}{}
\bibliographystyle{aasjournal}

\end{document}